\documentclass[a4paper,11pt]{memoir} 
  
\usepackage[latin1]{inputenc}
\usepackage[T1]{fontenc}   
\usepackage{graphicx}  
\usepackage{mathpazo,amssymb,epsf,amsmath}  

\settrimmedsize{297mm}{210mm}{*}  
  
\settypeblocksize{634pt}{448.13pt}{*}
\setulmargins{4cm}{*}{*}
\setlrmargins{*}{*}{1} 
\setmarginnotes{17pt}{51pt}{\onelineskip}
\setheadfoot{\onelineskip}{2\onelineskip}
\setheaderspaces{*}{2\onelineskip}{*}
\checkandfixthelayout
 
\usepackage{amscd} \usepackage[french]{babel} \usepackage{amsmath}
\usepackage{Macros} 
\begin{document}

\newlength{\centeroffset}
\setlength{\centeroffset}{-0.5\oddsidemargin}
\addtolength{\textwidth}{-\centeroffset}
\thispagestyle{empty}

\begin{flushright}
{CPHT-T 051.0805}\\ 
{\ttfamily hep-th/0508242}\\
\end{flushright}

\vspace*{\stretch{1}}
\noindent\hspace*{\centeroffset}\makebox[0pt][l]{\begin{minipage}{\textwidth}
\flushright
{\huge\bfseries D-branes, actions effectives et sym\'etrie miroir}
\noindent\rule[-1ex]{\textwidth}{5pt}\\[2.5ex]
\end{minipage}}

\vspace{\stretch{1}}
\noindent\hspace*{\centeroffset}\makebox[0pt][l]{\begin{minipage}{\textwidth}
{\Large{\bfseries
Pascal Grange}\\[1.5ex]
{\emph{Centre de physique th\'eorique de l'\'Ecole polytechnique,\\
route de Saclay,\\
 91128 Palaiseau Cedex, France}}\\[1ex]
{\ttfamily pascal.grange@polytechnique.org}
} 

\end{minipage}}

\addtolength{\textwidth}{\centeroffset}
\vspace{\stretch{2}}

\def\tr{{\mathrm{tr}}}
\def\d{\partial}

\pagestyle{empty}
\pagestyle{ruled}
 
\begin{center}
 
 {\Large\bfseries ABSTRACT}

\end{center}

\noindent{This} thesis is devoted to derivative corrections to the effective
action of D-branes, and to mirror symmetry with D-branes.\\
Series of derivative corrections first predicted by non-commutative gauge theory are
 completed by couplings between the metric and the gauge
 field. The result is interpreted as a deformation of the
 non-commutative gauge theory, whose structure survives.\\
 The derivation is applied to the tachyon field, whose
 potential is shown to be deformed by the very same
 corrections. Moreover, a prescription is given for the coupling of
 $p$-adic strings to a magnetic field, thus allowing to study $p$-adic
 solitons using non-commutative field-theory techniques.\\
The link with topological D-branes is provided by the non-commutative
 description of D-branes in the B-model. The fibre bundles supported
 by the D-branes are still holomorphic in this
 description. Establishing this property involves the realization of
 D-branes as boundary conditions, within the framework of generalized
 complex geometry.\\
    This geometric framework is then  used to
 describe mirror symmetry with {\hbox{D-branes}} on a Calabi--Yau manifold
 admitting a $T^3$-fibration. Two pure spinors, involved in the
 stability equations for topological {\hbox{D-branes}}, and modified by gauge
 fields, are exchanged, thus unifying Lagrangian
 and non-Lagrangian D-branes of the A-model as mirrors of stable
 holomorphic D-branes of the B-model.
  
\noindent

\begin{flushleft} 
30 ao\^ut 2005 
\end{flushleft}

\vfill\eject 

 \setcounter{page}{1}
 \pagestyle{plain}

\include{page}

\chapter*{Remerciements}

 \`A Ruben Minasian, qui a dirig\'e cette th\`ese, a su me guider avec bienveillance  \`a travers les d\'eveloppements de la th\'eorie des cordes et de la g\'eom\'etrie, m'a encourag\'e \`a diffuser mes r\'esultats, et a r\'eussi \`a faire \'emerger une certaine coh\'erence d'ensemble...\\ 

 \`A Patrick Mora, qui m'a accueilli au Centre de physique th\'eorique, dans des conditions de travail et une ambiance excellentes...\\ 

 \`A Marios Petropoulos, qui coordonne le groupe de cordes et m'a permis de m'y int\'egrer...\\  

  Aux professeurs qui m'ont mis en contact avec la physique th\'eorique contemporaine pendant mes \'etudes \`a l'\'Ecole polytechnique : Christoph Kopper qui m'a expliqu\'e les diagrammes de Feynman, et \'Edouard Br\'ezin qui le premier m'a montr\'e comment les conditions de {\hbox{Dirichlet}} interviennent en th\'eorie des cordes...\\

 \`A Fran\c{c}ois Laudenbach et Jean Lannes, qui m'ont donn\'e, notamment lors du s\'eminaire des \'el\`eves, mes premi\`eres notions de topologie alg\'ebrique, que j'ai retrouv\'ees avec fascination en abordant la th\'eorie des cordes...\\  

 \`A  Marcos Mari$\tilde{\mathrm{n}}$o, dont j'ai eu la chance de suivre les cours sur les cordes topologiques lors de conf\'erences, et qui a accept\'e d'\'ecrire un rapport sur cette th\`ese...\\

 \`A Pierre Vanhove, qui a \'egalement accept\'e la t\^ache de rapporteur, apr\`es  m'avoir prodigu\'e encouragements et r\'eponses au cours de ma th\`ese...\\

 \`A Constantin Bachas et  Massimo Porrati, dont les travaux sur les actions effectives se sont r\'ev\'el\'es si formateurs, et qui ont accept\'e de faire partie du jury...\\

 \`A Jean-Bernard Zuber, qui m'a fait l'honneur de pr\'esider le jury, et dont le travail de lecture a \'et\'e d'une redoutable pr\'ecision...\\

 \`A Luis \'Alvarez-Gaum\'e, qui m'a accueilli pour un stage d'un an au CERN, m'offrant ainsi une premi\`ere exp\'erience de recherche...\\

 \`A Lorenzo Cornalba et Costas Kounnas, qui ont dirig\'e mon m\'emoire de DEA, et m'ont permis, avec leur enthousiasme communicatif, d'appr\'ecier les d\'eveloppement de la th\'eorie des champs non-commutatifs via la physique des D-branes...\\ 

 \`A Christiane Gourber et Jean-Pierre Barani, dont l'enseignement a \'eveill\'e ma curiosit\'e scientifique tout en me conf\'erant de solides bases techniques...\\

 \`A Domenico Orlando, qui a \'ecrit la feuille de style pour cette th\`ese, a r\'esolu tous les probl\`emes informatiques aff\'erents, et a brillamment assur\'e la r\'egie lors de la soutenance...\\

 \`A Sylvain Ribault, qui a endur\'e la lecture de plusieurs versions pr\'eliminaires de ce texte, et sugg\'er\'e tant de n\'ecessaires am\'eliorations avec sa pertinence habituelle...\\

 \`A Claude de Calan, honn\^ete homme dont j'ai eu le privil\`ege de partager le bureau...\\

 Aux camarades dont j'ai si souvent \'eprouv\'e la patience : Luciano Abreu, {\hbox{St\'ephane}} {\hbox{Afchain}},  Pascalis Anastasopoulos, Guillaume Autier, David Marcos Bueno Monge, Fr\'ed\'eric Charv\'e, Julien Guyon, {\hbox{Xavier}} Lacroze, Alexey {\hbox{Yurievich}} Lokhov, Liuba Mazzanti, Chlo\'e Papineau, Szilard Szab\'o, {\hbox{Maria-Cristina}} {\hbox{Timirgaziu}}, Alice-Barbara Tumpach, Yao Yijun...\\   

 \`A l'\'equipe du secr\'etariat et de l'informatique, dont l'efficacit\'e souriante m'a aid\'e \`a mener \`a bien de nombreuses missions...\\  

 Aux organisateurs de s\'eminaires, \'ecoles et conf\'erences, qui m'ont permis d'exposer mes travaux et d'apprendre davantage, en me recevant aux Houches, \`a Caltech, Duke, Hambourg, Porto, {\hbox{Princeton}} et Upenn...\\  

 \`A Mikha\"el Balabane et Alain Bamberger, qui \`a l'\'Ecole des ponts ont accept\'e l'id\'ee de ce sujet de th\`ese, et m'ont assur\'e le soutien du minist\`ere de l'\'Equipement...\\

... j'exprime ici ma gratitude. 

\chapter[Introduction]{Introduction : objets étendus en théorie des cordes, géométrie et dynamique}

\chapterprecistoc{} \chapterprecishere{ Les cordes ouvertes s'appuient
  sur des membranes dites de Dirichlet, induites par les conditions de
  bord dans les équations du mouvement des cordes. Ces membranes sont
  des objets étendus, et correspondent à  des charges. Nous
  commen\c{c}ons par présenter une liste de faits, décrivant tour à 
  tour les membranes de Dirichlet comme sous-variétés de
  l'espace-temps (compatibles avec la T-dualité de la théorie des
  cordes), courants de de Rham (adaptés à  l'étude des actions
  effectives), et charges en K-théorie (adaptées aux fibrés supportés
  par les D-branes, et à  l'annihilation de paires de branes et
  d'anti-branes).  Nous présentons les différentes notions
  géométriques dont nous aurons besoin : non-commutativité, symétrie
  miroir, géométrie complexe généralisée. Les motivations physiques
  correspondantes (non-localité, condensation de tachyons,
  classification des D-branes topologiques) sont annoncées.}

\section{Des cordes aux D-branes}
Une corde fermée est un cercle topologique, qui se propage dans
l'espace-temps en engendrant une surface (sans bord) dite surface d'univers. Or
les surfaces sans bord sont classifiées très simplement par leur
genre. Les modes de vibration de la corde correspondent à un spectre
de particules, dicté par les symétries et la cohérence
interne. La théorie des cordes repose donc sur un développement
topologique de la fonction de partition (amplitude vide-vide),
organisé par la caractéristique d'Euler des surfaces d'univers : entre
un état initial et un état final, on peut interpoler par une série de
surfaces représentant les chemins possibles de la corde, les
différents chemins étant pondérés par l'action de la surface
d'univers. La tension de la corde (ou plutôt son inverse) est notée
$\alpha'$, par allusion à la pente de la relation affine qui lie le
spin $J$ au carré de la masse de la plus légère des particules
de spin $J$. La combinatoire des surfaces d'univers, grâce   au
caractère topologique du développement perturbatif, est beaucoup plus
simple que celle des graphes de Feynman requis pour un traitement
direct de la physique des particules du spectre. Dans le spectre nous
trouvons notamment une métrique $g_{\mu\nu}$, un tenseur antisymétrique $B_{\mu\nu}$, et un
scalaire, le dilaton $D$, dont l'exponentielle de la valeur moyenne, noté $g_s$, n'est
autre que la constante de couplage.\\

En théorie de type IIA (resp. IIB) , les
fermions de Ramond (correspondant à  des conditions aux limites
  périodiques sur la corde fermée, et possédant donc un mode de fréquence
  nulle mimant l'algèbre extérieure) des secteurs droit et
gauche des partenaires supersymétriques des coordonnées sur la surface
d'univers, se combinent pour donner lieu à  des champs de formes
différentielles de degré impair (resp. pair).\\

  Permettre une surface d'univers {\emph{\`a bord}}, c'est inclure un secteur de cordes
ouvertes \cite{TASIPolchinski,lecturesBachas}. Nous aurons donc
affaire à des intégrales de chemin fondées sur une action
classique comportant deux termes : $S_\Sigma=\int_\Sigma
\mathcal{L}_\Sigma$ intégré sur toute la surface d'univers
$\Sigma$, qui représente le couplage à un fond de cordes
fermées, et un terme intégré sur le bord,
$S_{\partial\Sigma}=\int_{\d\Sigma} \mathcal{L}_{\d\Sigma}$. La
fonction de partition $Z$  s'écrira en général comme une intégrale
fonctionnelle par rapport aux coordonnées $X$ du plongement :
$$Z=\int DX \,e^{-S_\Sigma-S_{\partial\Sigma}}.$$ Les équations du
 mouvement pour la surface d'univers demandent
 l'annulation des variations du terme intégral bidimensionnel et du terme de
 bord, donnant ainsi à  la fois les équations du mouvement et les
 conditions de bord, qui définissent les points de l'espace-temps
 auxquels les extrémités des cordes ouvertes sont attachées. L'ensemble de
 ces points forme une membrane, dite de Dirichlet, ou D-brane. En se
 propageant dans la direction temporelle, une D-brane de dimension
 $p$, ou D$p$-brane, engendre un volume d'univers, de dimension
 $p+1$.\\

Un tel objet possède une tension, il représente une certaine
concentration d'énergie. Tel est le principe du point de vue dit
d'espace-temps sur les D-branes : une D-brane est une solution étendue
d'une certaine action effective. Les conditions de bord, qui sont à 
la base du point de vue de surface d'univers présenté plus haut,
donnent accès à  cette action sous forme d'intégrales de chemin.\\

  La T-dualité, invariance vis-à-vis de
 l'inversion du rayon d'une dimension compacte, est organiquement liée
 au phénomène de la symétrie miroir. C'est aussi un moyen de calcul et
 un puissant facteur de cohérence interne : elle peut notamment \^etre
 combinée à  l'invariance de jauge pour contraindre les actions
 effectives.\\

\section{Charges de Ramond--Ramond et champs de jauge}

\subsection{Champs de formes différentielles} 
Comme nous l'avons vu plus haut, les secteurs de Ramond  engendrent des champs de formes différentielles dans le spectre des supercordes. Or ces champs (de Ramond--Ramond) 
  ne sont pas couplés au dilaton : leurs transformations de jauge sont
simplement les changements de représentant au sein d'une classe
d'équivalence en cohomologie. Ils sont donc d'essence non-perturbative.  La
considération de conditions de Dirichlet a culminé avec
l'identification par Polchinski des D-branes comme charges de Ramond--Ramond
\cite{Polchinski}. La tension d'une D$p$-brane  vaut
$$
T_p=\frac{1}{g_s\sqrt{\alpha'}}\frac{1}{(2\pi\sqrt{\alpha'})^p},$$
qui est
inversement proportionnelle à  la constante de couplage $g_s$, ce qui signe le
caractère non-perturbatif des D-branes.

\subsection{Les D-branes portent  une charge en théorie des supercordes}
  En première
approximation, les membranes de Dirichlet ne sont que des
sous-variétés de l'espace-temps. Si nous prenons le point de vue
de l'action effective et du couplage aux champs de formes
différentielles du secteur de Ramond--Ramond, ce sont des courants
de de Rham, c'est-à-dire des objets le long desquels on peut
intégrer des formes différentielles.  En effet, étant donné un
champ de formes $C^{(p)}$ de rang $p$, et un volume d'univers $X^p$ de
dimension $p$ occupé par une D-brane, le geste naturel produisant
une action consiste à intégrer le champ de formes :
$$S=\int_{X^{p}} C^{(p)}.$$

\subsection{Les cordes ouvertes portent des champs de jauge}

Pour une pile de $N$ D-branes nous avons un champ de jauge et un
scalaire transverse, portant chacun deux indices, désignant la paire
de D-branes reliées par une corde ouverte, et se transformant sous
l'action du groupe de jauge $U(N)$ :

$$
(A_\mu)_i^j, (\phi^\mu)_i^j.$$
La séparation des $N$ D-branes brise la
symétrie $U(N)$ en $U(1)^N$, et en particulier la théorie de jauge abélienne
correspond à  des conditions de bord pour les cordes ouvertes comportant une
seule D-brane. Cette symétrie de jauge appara\^it
 lorsqu'une corde ouverte s'appuie sur une D-brane :
pour chaque direction {\emph{longitudinale}}, nous avons un reparamétrage possible du
volume d'univers de la D-brane, redondance dans laquelle on reconnaît celle des champs de
jauge. En revanche, une fluctuation d'un scalaire {\emph{transverse}} modifie la
forme de la D-brane. La T-dualité échange directions longitudinales et
transverses, champs de jauge et scalaires transverses.

 Le spectre des cordes ouvertes contient donc un champ de jauge, sous
lequel les extrémités des cordes ouvertes sont chargées. Ce fait
induit une théorie de jauge le long du volume d'univers des D-branes. En d'autres termes
le volume d'univers d'une D-brane est la base d'une structure fibrée,
la fibre étant isomorphe au groupe de jauge. La courbure $F$ du fibré
est le tenseur du champ de jauge, et peut être consid\'er\'ee comme un champ de
formes différentielles sur la base, autorisant un couplage entre le
champ de jauge et des formes de Ramond--Ramond de rang faible. {\hbox{L'action}} effective comporte donc des termes couplant les champs de jauge
aux potentiels de Ramond--Ramond, via des quantités topologiques, d'où
leur nom de termes de Wess--Zumino ou de Chern--Simons
\cite{Chern--Simons}. Les diverses puissances permises pour $F$ sont
des composantes du caractère de Chern, et induisent des charges moins
étendues à l'intérieur de la D-brane principale, Ces charges de
dimension plus faible sont tenues ensemble par le champ de jauge,
selon l'image mise au point par Douglas \cite{within} :
$$ S=\sum_{k, p-2k\geq 0}\int_{X^p}
C^{(p-2k)}\land\frac{1}{k!}F^k=\int_{X^p} C\land e^F. $$ Nous
observons que l'action effective des D-branes fait apparaître des
couplages entre les secteurs de cordes ouvertes et fermées. Les
corrections dérivatives que nous calculerons au chapitre 3 couplent le
tenseur antisymétrique $B_{\mu\nu}$ au champ de jauge. De plus, il est
manifeste que les puissances $F^k$ issues du développement du
caractère de Chern mesurent des charges de D(p-2k) branes le long du
volume d'univers de la D-brane ambiante.  Elles
sont nombreuses, comme les fils de la trame d'un tapis.  En
particulier, elles engendrent une symétrie de jauge non-abélienne,
et leurs scalaires transverses ne commutent pas, même si la D-brane
ambiante ne porte que des champs de jauge abéliens.  Des termes mis en
évidence par Myers \cite{Myers}, couplant la D$p$-brane à  des champs
de Ramond--Ramond de degré plus élevé, sont demandés par la symétrie
de jauge et la T-dualité, et permis par une multiplication intérieure
entre scalaires transverses et champs de formes différentielles :

$$S=\int \left(e^{\iota_{[\phi, \phi]}}C\right)\land e^F.$$

 Ils contraignent l'inclusion des champs de jauge dans les quantités
 géométriques échangées par symétrie miroir en présence de D-branes,
 comme nous le verront au dernier chapitre. Le commutateur
$$[\phi^\mu, \phi^\nu]$$
est aux charges de D$(p+2)$-branes ce que le tenseur du champ de jauge
$F_{\mu\nu}$ est aux charges de D$(p-2)$-branes. Nous exploiterons cette
conséquence de la T-dualité dans le dernier chapitre, dans l'idée d'étendre
nos relations de symétrie miroir à des champs de jauge non-abéliens.

\subsection{Les charges de Ramond--Ramond ont valeur dans la
  K-théorie de l'espace-temps}  Les charges de Ramond--Ramond ont un
caractère algébrique : il doit exister des anti-branes pouvant s'annihiler
mutuellement avec les D-branes. Le simple fait qu'une D-brane soit la base
d'un fibré requiert donc une notion de soustraction des fibrés : une D-brane
stable doit pouvoir être vue comme le résultat d'un processus d'annihilation
entre une D-brane et une anti-brane. Mathématiquement, il nous faut donc une
structure de groupe définie à  partir de fibrés. C'est la notion de
K-théorie, qui est adaptée aux charges de Ramond--Ramond, ainsi que l'ont
montré Minasian et Moore \cite{MMK}. Le caractère de Chern est un objet
naturel en K-théorie. Quant au processus physique d'annihilation, il a été
relié par Witten \cite{WittenK} à  la condensation de tachyons, transition de
phase conjectur\'ee par Sen,  qui tient à la structure non-perturbative de la théorie des cordes.\\

\subsection{Le potentiel de tachyons gouverne les phases de la théorie des cordes} Certaines techniques de théorie des champs non-commutatifs
donnent accès, dans des modèles simplifiés exactement solubles, à  des
actions effectives qui induisent une dynamique des scalaires instables
vérifiant la conjecture de Sen \cite{Senconjecture}. Les D-branes
possédant une mauvaise dimension en théorie des supercordes (dimension
impaire en type IIA, paire en type IIB) ne correspondent pas à  une
charge conservée, puisqu'il n'y a pas de champ de Ramond--Ramond dont
le degré lui permette de s'intégrer le long du volume d'univers avec
un résultat non nul. Il en va de même pour toutes les D-branes de la
théorie des cordes bosoniques. Un système brane-antibrane est
également instable, et un scalaire de masse carrée négative est
présent dans le spectre.

\section{Géométries}

La séquence d'objets mathématiques décrite ci-dessus (sous-variétés,
courants de de Rham, K-théorie) est organiquement liée à  la physique des
cordes ouvertes (les cordes ouvertes s'appuient sur des D-branes, qui sont chargées sous
les champs de Ramond--Ramond, et elles présentent des champs de jauge dans
leur spectre). Toutefois, d'autres structures mathématiques peuvent être
adaptées aux D-branes, de manière à  pouvoir les étudier plus facilement.
Disposer d'une structure supplémentaire pertinente peut en effet fournir des
résultats physiques (actions effectives, conditions de stabilité) par la
vertu des contraintes de cohérence interne. Les principales structures
utilisées dans cette thèse sont les théories de jauges non-commutatives, les
structures complexes généralisées, et les twists topologiques qui localisent
la géométrie des D-branes sur les objets complexes ou symplectiques.

\subsection{Non-commutativité}

\subsubsection{Non-localité et objets étendus}
Non-localité et non-commutativité doivent génériquement
intervenir dans la dynamique des objets étendus.
La notion même de point étant secondaire dans la théorie des
cordes, il est naturel d'espérer une réalisation physique d'objets
géométriques non-commutatifs en théorie des cordes, même dans
des situations très simples. C'est en effet le cas dans une limite de
la théorie des cordes, où en particulier la tension de corde tend vers
zéro, avec une loi d'échelle particulière. Une partie des
travaux exposés dans cette thèse vise à  étendre au-delà  de
cette limite des résultats concernant les actions effectives. Une
telle extension permet de rendre justice au caractère bidimensionnel
de la surface d'univers, qui constitue une motivation géométrique
de ces développements, et était pourtant atténué dans la
limite considérée.

\subsubsection{Algèbre des observables}

Un espace peut être décrit par l'algèbre des fonctions vivant sur cet
espace. Cette algèbre n'est pas en général commutative. Prendre l'algèbre
des observables comme point de départ constitue le principe de la géométrie
non-commutative. Les rotations en dimension sup\'erieure \`a deux, la
mécanique quantique, les niveaux de Landau, sont autant d'exemples
d'interventions de la non-commutativité en physique, dès avant le
développement de la théorie des cordes.\\

Nous nous limiterons à  la notion la plus simple de non-commutativité,
justiciable d'une description par des opérateurs de coordonnées qui ne
commutent plus :
$$[x^\mu,x^\nu]=i\theta^{\mu\nu}.$$ L'approximation du commutateur par
 les crochets de Poisson associés à  $\theta$ est analogue à 
 l'approximation semi-classique de la mécanique quantique.  La théorie
 de Yang--Mills a été étudiée par Connes, Douglas et Schwarz
 \cite{CDS} sur un tore non-commutatif. La quantification formelle et
 les produits non-commutatifs sont apparus dans le travaux de Chu et
 Ho \cite{Chu--Ho} et d'Alekseev, Recknagel et Schomerus
 \cite{Schomerus,Schomerusquant}. Les arguments de dualité en faveur d'une origine en
 théorie des cordes, dans une certaine limite, ont été développés par
 Seiberg et Witten \cite{SW}.

\subsection{Symétrie miroir}
\subsubsection{T-dualité}
La symétrie miroir, phénomène violent et d'un intérêt considérable sur le
plan mathématique puisqu'il apparie deux variétés de topologies différentes,
 provient d'une dualité entre théories superconformes \cite{observation},
échangeant modules complexes et modules de Kähler. L'établissement de la
symétrie miroir par des arguments physiques repose toujours sur une certaine
forme de T-dualité\footnote{T est mis pour {\emph{target space}} ;
  l'expression symétrie miroir sera souvent employée en lieu et place du nom
  de T-dualité, surtout dans le contexte des cordes topologiques. Le mot
  miroir fait référence à une symétrie par réflexion du tableau de Hodge qui
  code les dimensions des cohomologies complexes des variétés échangées par
  symétrie miroir. L'argument
  de Strominger, Yau et Zaslow justifiant cette assimilation sera passé en
  revue au chapitre 7.}, fondée sur l'inversion du rayon d'une dimension compacte :
$$ R\mapsto \alpha'/R.$$
Cette transformation affecte les D-branes, notamment en  échangeant
directions longitudinales et transverses. L'interprétation comme parité
asymétrique sur la surface d'univers sera présentée au prochain chapitre. L' effet de la symétrie
miroir sur les D-branes et un domaine de recherche fécond sur le plan des
dualités et de la géométrie.

\subsubsection{Variétés de Calabi--Yau}
 Mes travaux en rapport avec la théorie des champs non-commutatifs
 supposent un espace-cible plat. Dans la suite, il sera question de
 variétés de Calabi--Yau de dimension complexe trois, ou plutôt de
 paires de telles variétés, formées par la symétrie miroir. Donnons
 quelques mots de motivation pour cette géométrie de Calabi--Yau. Les
 conditions de cohérence interne et de symétrie (invariance
 relativiste, symétrie conforme, supersymétrie), donnent lieu, et
 c'est là  un trait tout à  fait remarquable de la théorie des cordes,
 à  des contraintes sur la géométrie de l'espace-temps, à 
 commencer par la dimension critique, 26 pour les cordes bosoniques et
 10 pour les supercordes. Le cahier des charges à  remplir par une
 variété riemannienne\footnote{Le graviton faisant partie du spectre
 de masse nulle des cordes fermées, nous héritons d'une métrique sur
 l'espace-cible} pour permettre la propagation des cordes et une
 physique d'espace-temps supersymétrique, comprend la condition de
 Kähler et la condition de Calabi--Yau. Une variété de Calabi--Yau
 est une variété complexe de Kähler, admettant une métrique plate au
 sens de Ricci. La conjecture de Calabi,
 démontrée par Yau, est un théorème d'existence pour
 une telle métrique sur une variété complexe de Kähler $X$, moyennant
 la condition topologique d'annulation de la première classe de Chern
 :

$$[c_1(X)]=0.$$      
 
Il suffit donc de calculer un invariant topologique pour décider si la
condition est ou non vérifiée, ce qui est considérablement plus simple que
la résolution de l'équation différentielle qu'est la condition de Ricci. On
peut caractériser  l'annulation de la première classe de Chern par
l'existence d'une forme différentielle holomorphe de degré maximum sur $X$.\\

Les travaux présentés dans le chapitre sur le comportement des
D-branes sous la symétrie miroir supposent un espace-cible de
dimension complexe trois vérifiant la condition de Calabi--Yau. La
dimension complexe trois, naturelle pour la compactification des supercordes, est de plus  critique pour le couplage à la gravité
topologique \cite{gravitetopologique}, qui consiste à intégrer sur
l'espace de toutes les métriques permises.\\

\subsubsection{Localisation et D-branes topologiques}
 Les versions topologiques de la théorie des cordes
permettent d'étudier séparément certains faits
géométriques. 
C'est le principe de la localisation. On espère que les D-branes topologiques, moyennant une condition de
stabilité, correspondent à  d'authentiques objets étendus de la
théorie des cordes. Dans le contexte de la symétrie miroir, reliant
deux versions topologiques A et B, le modèle A est sensible à  la
géométrie symplectique (via des D-branes lagrangiennes ou
coisotropes), et le modèle B à  la géométrie complexe (via des
{\hbox{D-branes}} équipées de fibrés holomorphes).\\ 

Il y a une notion de stabilité, développée en géométrie pour les fibrés
\cite{LeungPhD}, et réalisée ensuite en théorie des cordes \cite{stability}
comme une condition de supersymétrie sur les solitons (objets étendus
concentrant de l'énergie) que sont les D-branes. Cette approche
d'espace-temps utilise une action effective et a conduit aux équations
 de Mari$\tilde{\mathrm{n}}$o, Minasian, Moore et
Strominger \cite{MMMS}, dont nous donnerons une version non-commutative en
utilisant une approche de surface d'univers.

\subsection{La géométrie complexe généralisée}
 Introduite par Hitchin \cite{HitchinGCY}, la géométrie complexe
 généralisée est adaptée aux théories supersymétriques
 $\mathcal{N}=(2,2)$. Un modèle sigma non-linéaire supersymétrique
 $\mathcal{N}=1$ peut en effet avoir une seconde supersymétrie,
 linéairement indépendance de la première, pourvu que l'espace-cible
 possède deux structures complexes donnant lieu à  deux structures
 hermitiennes. De plus, si les champs de fond
 sont indépendants de certaines des coordonnées, au nombre de $d$,
 alors l'action effective hérite d'une action du groupe d'isométries
 $O(d,d)$ \cite{Meissner--Veneziano}. Deux
 copies de l'algèbre superconforme sont alors disponibles.\\

 Les développements récents de la géométrie complexe généralisée
permettent d'unifier les deux modèles A et B comme deux cas particuliers
d'une structure vivant sur la somme de l'espace tangent et de l'espace
cotangent. L'ambition de cette g\'en\'eralisation consiste en une interpolation
entre les twists topologiques A et B. Il s'est avéré que des
techniques de surface d'univers permettent d'obtenir des objets
étendus en imposant des conditions de bord à un espace dédoublé. \\

\section{Plan de ce mémoire}

J'ai donc présenté les différents personnages qui ont animé mes
travaux. Le fil conducteur de cette thèse consiste en une
réalisation de différentes géométries par des conditions de
bord sur la surface d'univers. Nous aurons donc affaire à
différents appariements entre des objets géométriques et des
propriétés des D-branes : non-commutativité et action du champ
de jauge, non-commutativité et potentiel de tachyons, géométrie
complexe généralisée et B-branes topologiques non-commutatives,
spineurs purs et symétrie miroir.\\

Le chapitre 2 sera consacré à  l'étude des conditions de bord
définissant les D-branes. Nous mettrons progressivement en place les
outils utilisés dans la suite. Nous discuterons des propagateurs pour
les champs de coordonnées sur la surface d'univers, en présence d'une
D-brane placée dans un espace plat dans lequel règne un champ de fond
antisymétrique. Les cordes et D-branes topologiques seront ensuite
présentées, avec une illustration transparente des propriétés de
localisation complexe et symplectique évoquées plus haut. La symétrie
miroir sera illustrée sur l'exemple simple de la T-dualité le long
d'un cercle sur le tore $T^2$, et l'inclusion d'un champ
antisymétrique complexifiant la forme de Kähler apparaîtra
naturellement.\\

 La théorie de jauge non-commutative est présentée au chapitre 3, avec
 son origine en théorie des cordes. Afin de motiver mes travaux sur
 les actions effectives (qui font l'objet du chapitre 4), j'illustrerai
 le fait que les actions effectives de cordes ouvertes peuvent
 s'écrire de beaucoup de manières. La versatilité de ces expressions
 devient un moyen de calcul, via le fait que l'expression
 non-commutative est sévèrement contrainte par la richesse même de sa
 structure.  Le lien intime entre les champs de Ramond--Ramond et la
 redéfinition des champs de jauge de la théorie non-commutative
 abélienne de Yang--Mills, ou transformation de Seiberg--Witten,
 invite à  penser que les D-branes topologiques doivent être
 justiciables de certaines techniques de théorie de jauge
 non-commutative. Cette analogie avec les couplages de Ramond--Ramond
 se retrouvera au dernier chapitre avec nos spineurs purs modifiés, et
 sera rendue plus naturelle par la géométrie complexe généralisée.\\
 
 Au chapitre 4, je confirmerai les corrections dérivatives dictées par la
 non-commutativité, en effectuant un calcul d'intégrales de chemin en
 théorie {\emph{commutative}} des cordes. Je prolongerai ensuite ces
 prédictions au-delà de la limite, dite de Seiberg--Witten, dans
 laquelle elles ont été faites. Le fait de ne pas avoir utilisé
 d'argument d'invariance de jauge non-commutative dans les calculs
 permettra {\emph{in fine}} de prolonger la notion de théorie de jauge
 non-commutative au-delà  de la limite de Seiberg--Witten sur le
 volume d'univers de la D-brane.\\
 
Le chapitre 5 concernera les actions effectives des champs de
tachyons. La non-commutativité est prise en compte a priori en
K-théorie, puisque la K-théorie de l'algèbre des fonctions sur un
espace est encore une notion bien définie si l'algèbre n'est pas
commutative. Je débute le chapitre par un exposé sur la robustesse du
produit de Moyal vis-à-vis de l'altération qui conduit au modèle
exactement soluble des cordes $p$-adiques.  La condensation de
tachyons dans la théorie ordinaire (archimédienne) est bien sûr une
motivation cruciale dans le développement de ce mod\`ele simplifi\'e. Des tests de la conjecture de Sen ont
été effectués \cite{KMM1,KMM2,tachyonCornalba,Okuyama}, souvent dans la limite
de Seiberg--Witten, et l'interprétation en termes de tachyons
non-commutatifs \cite{DMR} reliée à  la théorie des champs de cordes
\cite{Wnctachyons,overview}. L'universalité du potentiel de tachyons
semble survivre au-delà  de la limite de Seiberg--Witten, en
conséquence assez immédiate (et heureuse) des calculs du chapitre
3. L'objet naturel codant simultanément les champs couplés au bord
dans les chapitres 3 et 4 est une superconnexion. Je reprendrai des
propositions antérieures à  cette thèse pour l'écriture des actions
effectives en termes de superconnexions, et j'argumenterai en leur faveur
à  l'aide des corrections d'ordre élevé calculées au cours de cette
thèse.\\

 La géométrie complexe généralisée sera introduite {\emph{in situ}} au
 chapitre 5, dans une fécondation croisée avec la non-commutativité ; elle
 permettra d'obtenir une notion de fibré stable holomorphe non-commutatif, qui
 avait été conjecturée  en 1999 dans  l'approche d'espace-temps. Je donnerai à 
 la fin de ce chapitre quelques notions plus formelles sur cette géométrie,
 d'après Hitchin et Gualtieri. Les spineurs dits purs, vides d'une algèbre de
 Clifford naturelle impliquant champs de vecteurs et formes différentielles,
 peuvent appartenir soit au domaine symplectique du modèle A, comme
 l'exponentielle de la forme de Kähler, soit au domaine complexe du modèle B,
 comme la trois-forme holomorphe.\\

 Ces notions, et en particulier celle de spineur pur, nous serviront à 
 calculer au chapitre 6 les modifications par les D-branes de la
 formule de symétrie miroir qui échange la trois-forme holomorphe et
 l'exponentielle de la forme de Kähler ; nous supposerons l'existence
 d'une fibration en tores $T^3$, et nous expliquerons comment la
 symétrie miroir peut être identifiée à  une T-dualité de long des
 fibres.  Nous terminerons ce chapitre par quelques spéculations sur la
 notion de K-théorie adaptée à  la géométrie complexe généralisée. Les
 projets de recherche associés sont énumérés et motivés dans le
 dernier chapitre.\\

La table suivante indique quelles notions sont abordées dans les
différents chapitres, afin de permettre au lecteur de choisir un fil
conducteur.

\begin{table}[htbp]
  \centering
  \begin{tabular}{|c|c|c|c|c|c|}
    \hline {} & Non-commutativité & D-branes & Tachyons & Symétrie
    miroir & Spineurs purs \\\hline {Ch. 1} & $\ast$ & $\ast$ & $\ast$
    & $\ast$ & $\ast$ \\ \hline {Ch. 2} & {} & $\ast$ & {} & $\ast$ &
    {} \\ \hline {Ch. 3} & $\ast$ & $\ast$ & {} & {} & {} \\ \hline
    {Ch. 4} & $\ast$ & $\ast$ & {} & {} & {} \\ \hline {Ch. 5} &
    $\ast$ & $\ast$ & $\ast$ & {} & {} \\ \hline {Ch. 6} & $\ast$ &
    $\ast$ & {} & {} & $\ast$ \\ \hline {Ch. 7} & {} & $\ast$ & {} &
    $\ast$ & $\ast$ \\ \hline

  \end{tabular}
  \caption{Thèmes abordés}
  \label{tab:e-uno}
\end{table}


\chapter{Des conditions de bord aux D-branes topologiques}
\chapterprecistoc{} \chapterprecishere{Les conditions de bord donnant lieu à
  des D-branes sont présentées. Les deux thèmes qui seront étudiés
  dans cette thèse, à savoir les actions effectives induites par le spectre
  de cordes ouvertes le long du volume d'univers d'une D-brane, et les
  D-branes topologiques, sont abordés. La symétrie miroir relie les deux
  différents twists topologiques.}

\section{Les D-branes sont  requises par la T-dualité}

Nous allons effectuer quelques calculs pour acclimater différents
aspects de la T-dualité qui seront utiles au cours de ce mémoire. Pour
une introduction plus systématique, le lecteur pourra consulter les
références \cite{GPR,bigbook}. Tout d'abord un calcul des variations de
l'aire d'une surface d'univers avec bord donnera les équations du
mouvement pour les coordonnées du plongement, avec les conditions de bord
définissant les D-branes comme le lieu sur lequel le bord des surfaces
d'univers (les extrémités des cordes ouvertes) peuvent
s'appuyer. Puis l'évaluation de la fonction de partition d'une
corde fermée, avec pour espace-cible un cercle, rendra manifeste la symétrie par
inversion du rayon de compactification ; ce sera le
premier exemple de T-dualité. L'impulsion du centre de masse et le
nombre d'enroulement seront échangés, ce qui pour les D-branes induira l'échange des
directions longitudinales et transverses par T-dualité. Cette transformation est
également une parité asym\'etrique sur la surface d'univers, 
agissant sur le secteur droit des
coordonnées. Passant à un tore $T^2$, nous aurons une structure
complexe et une forme de Kähler, échangées par T-dualité, première
indication de la symétrie miroir. Une présentation approfondie des
liens physiques et géométriques entre T-dualité et symétrie
miroir se trouve dans le volume dirigé par Hori et ses collaborateurs
\cite{mirrorbook}.

\subsection{Conditions de Neumann et de Dirichlet}
Envoyons la surface d'univers (à bord) $\Sigma$ d'une corde ouverte dans un
espace plat $\mathbf{R}^d$ ; nous allons discuter de la partie
bosonique de l'action de surface d'univers :

$$S:=\int_\Sigma d^2\sigma\, \partial_\alpha X^\mu \partial^\alpha X_\mu,$$
$$
X^\mu : \Sigma \to \mathbf{R}^d.$$
L'équation du mouvement
obtenue en demandant la stationnarité vis-à-vis d'une variation
infinitésimale $\delta X^\mu$ des scalaires dans l'intérieur de la surface
d'univers est l'équation de Klein--Gordon pour un champ scalaire sans masse.
Quant aux variations portées par le bord, elles donnent lieu via le théorème
de  Stokes à la condition de bord 
$$\delta X_\mu \partial_n X^\mu|_{\partial\Sigma}=0,$$
dans laquelle le
symbole $\partial_n$ désigne la dérivée dans la direction normale au bord de
la surface d'univers. Dans chaque direction $\mu$ de l'espace-cible, nous
avons le choix pour le champ scalaire porté
par le bord  entre une condition de Neumann 
$$ \partial_n X^\mu|_{\partial\Sigma}=0,$$ 
et une condition de Dirichlet 
$$\delta X^\mu|_{\partial\Sigma} =0.$$ Autrement dit, le bord de la
surface d'univers peut s'appuyer sur un sous-espace de l'espace-cible,
engendré par les directions selon lesquelles une condition de Neumann
a été choisie. Les directions normales à cet espace sont celles selon
lesquelles une condition de Dirichlet a été choisie. Ce sous-espace
est naturellement appell\'e membrane de Dirichlet, ou D-brane, ou
D$p$-brane en signalant sa dimension $p$ (\'egale au nombre de conditions de Neumann).\\

En munissant la surface d'univers de coordonnées complexes $\sigma±
i\tau$, o\`u $\sigma$ désigne l'abscisse curviligne le long de la corde
ouverte, et $\tau$ la direction temporelle, nous arrivons à la forme qui
s'adaptera à l'inclusion des champs de jauge sur l'espace tangent \`a la
D-brane :
$$\partial X^\mu=+\bar{\partial} X^\mu
    {\text{(direction tangente  ou de Neumann)}}, $$
    $$\partial X^\mu=-\bar{\partial} X^\mu
        {\text{(direction normale  ou de  Dirichlet)}}.$$
 Nous écrirons les conditions de bord à l'aide
        d'une matrice dite de réflexion not\'ee $R$ :
        $$\partial X= R(\bar{\partial}X).$$ Le sous-espace propre
        de $R$ associé à la valeur propre $-1$ déterminera le fibré
        normal à la D-brane. Les directions tangentes portent à la
        fois le fibré tangent et le fibré principal correspondant
        à la théorie de jauge. Nous verrons comment la métrique
        et les champs de jauge sont mélangés par la matrice de
        réflexion.

\subsection{Un cercle pour espace-cible : inversion du rayon}

Compactifions l'une des dimensions de l'espace-cible. Pour simplifier
les notations, nous pouvons aussi bien ignorer les autres dimensions, et considérer un cercle $S^1$ de
rayon $R$ comme l'espace-cible. Le champ de coordonnées est donc
simplement un scalaire périodique de période $R$ :
$$X : \Sigma\mapsto S^1.$$
Nous
considérons des cordes fermées.  Il y a un double effet topologique
dans cette description\footnote{le calcul est effectué dans des
unités telles que $\alpha'=1$.}. Ainsi, dans le développement du
champ de coordonnées, scindé en parties droite et gauche

$$X(\sigma,\tau)= X(\tau-\sigma) + \tilde{X}(\tau+\sigma),$$
$$X(\tau-\sigma)=\frac{x_0}{2}+\frac{1}{\sqrt{2}} p_0
(\tau-\sigma)+\frac{1}{\sqrt{2}}\sum_{n\neq 0}\frac{1}{n} \alpha_n
e^{-in(\tau-\sigma)}, $$
$$
\tilde{X}(\tau+\sigma)=\frac{x_0}{2} + \frac{1}{\sqrt{2}}\tilde{p}_0
(\tau+\sigma)+\frac{1}{\sqrt{2}}\sum_{n\neq 0}\frac{1}{n} \tilde{\alpha}_n
e^{-in(\tau+\sigma)},$$
$$p_0=\frac{m}{R}-{w}{R},\;\;\;\;\;\;\;\;\;\;\;\;\;\;\tilde{p}_0=\frac{m}{R}+{w}{R},$$
 deux entiers apparaissent : l'un, noté $m$, quantifie la quantité de
 mouvement du centre de masse en unités $R^{-1}$, de manière à assurer
 le caractère monovalué des ondes planes, et l'autre, noté $w$,
 quantifie l'enroulement de la corde fermée autour du cercle cible en
 unités $R$, puisque deux coordonnées différant d'un multiple de $R$
 sont identifiées.\\
 
La propagation de la corde fermée dans une amplitude vide-vide identifie
deux extrémités d'un cylindre (ayant pour longueur la durée de propagation
$\tau_1$), après leur avoir imprimé une rotation relative mesurée par
l'angle $\tau_2$. Pour construire la surface d'univers, nous avons donc besoin d'un patron de tore avec une structure
complexe donnée par $\tau=\tau_1 +i \tau_2$.\\ 


  Calculant la fonction de
partition comme dans un gaz de Bose pour les hamiltoniens des secteurs droit
et gauche, affectés des constantes d'ordre normal donnés par des valeurs de
la fonction z\^eta
$$H= \frac{1}{2}\left( \frac{m}{R}-{w}{R} \right)^2 + \sum_{n\geq
  1}\alpha_{-n}\alpha_n -\frac{1}{24},$$
$$\tilde{H}=\frac{1}{2}\left( \frac{m}{R}+{w}{R}\right)^2 + \sum_{n\geq
  1}\tilde{\alpha}_{-n}\tilde{\alpha}_n -\frac{1}{24},$$
nous obtenons un produit de fonctions modulaires\footnote{La
fonction \^eta de Dedekind appara{î}t, et ses transformations
  modulaires
$$\eta(\tau+1)=e ^{i\pi/12}\eta(\tau),\;\;\;\;
\eta(-1/\tau)=(-i\tau)^{1/2}\eta(\tau),$$
permettent de conclure {à}
l'invariance modulaire de la fonction de partition~:
$$Z(\tau,\bar{\tau})=\frac{1}{\sqrt{\Im\tau}}\frac{1}{|\eta(\tau)|^2},$$
$$\eta(\tau)= q^{1/24}\prod_{n\geq 1}(1-q ^n).$$}, multiplié par une
double somme sur les deux entiers $m$ et $w$ :

$$Z(\tau,\bar{\tau})= ({q\bar{q}})^{-1/24}\frac{1}{\sqrt{\tau_2}}\prod_{n\geq
  1}\frac{1}{1-q^n}\frac{1}{1-\bar{q}^n}\sum_{m,w \in
  \mathbf{Z}}q^{\frac{1}{4}\left( \frac{m}{R}-{w}{R} \right)^2
}\bar{q}^{\frac{1}{4}\left( \frac{m}{R}+{w}{R} \right)^2 },$$
$$q:=e^{2i\pi\tau}.$$ 

 Nous lisons sur cette expression la symétrie de
 T-dualité, qui échange le rayon et son inverse, ainsi que les nombres
 d'impulsion et d'enroulement:
 $$ R\longleftrightarrow 1/R, \;\;\;\;\;\;\;\;\;\;\;\;\;
 m\longleftrightarrow w.$$ Revenons aux expressions des secteurs droit
 et gauche des coordonnées, et donnons une interprétation de la
 T-dualité valable pour les cordes ouvertes. Ces derni\`eres ne donnent pas lieu à
 un nombre d'enroulement, ce qui nous prive \`a premi\`ere vue de l'un des effets topologiques. Nous voyons que la T-dualité agit comme une
 parité asymétrique :

$$ (p_0,\tilde{p}_0)\longleftrightarrow (-p_0,\tilde{p}_0)$$
$$
(\alpha_n, \tilde{\alpha}_n) \longleftrightarrow (-\alpha_n,
\tilde{\alpha}_n).$$
Quant aux dérivées, elles s'échangent suivant la loi
$$\partial+\bar{\partial}\longleftrightarrow \partial-\bar{\partial},$$
ce
qui modifie la matrice de réflexion, et fournit l'interprétation de la
T-dualité pour les sous-variétés supportant les D-branes : une opération de
T-dualité dans une direction transverse (resp. longitudinale) la transforme
en direction longitudinale (resp. transverse).
 
\subsection{Échange entre structure complexe et structure de Kähler sur le tore}
Prenons maintenant pour espace-cible le produit de deux cercles de
  rayons $R_1$ et $R_2$ :
$$X : \Sigma\to S^1_{R_1}\times S^1_{R_2}.$$.
 Génériquement, c'est-à-dire pour deux rayons
  différents, la symétrie du problème est héritée de la
  symétrie par T-dualité que présente chacun des deux cercles.\\

Nous avons cependant un peu plus de structure \`a notre disposition 0: la forme du patron du tore
rectangulaire équivalente à la donnée du rapport $R_1/R_2$, correspond à la donnée
d'une structure complexe sur le tore. Quant à la taille du tore $R_1 R_2$,
elle correspond à la donnée d'une forme de Kähler, puisqu'une telle forme
en dimension deux n'est autre qu'une forme volume.\\

Or la T-dualité le long de l'un des cercles agit en permutant le quotient et
le produit des deux rayons :
$$ R_1 R_2\longleftrightarrow \frac{R_1}{R_2}.$$ Il s'agit de la
première apparition d'une telle permutation entre géométrie complexe
et géométrie de Kähler. Elle est cependant imparfaite, puisque nous
n'avons considéré que des patrons de tore rectangulaires, au lieu des
formes de parallélogramme générique que donne une structure complexe
générique. L'espace des structures complexes doit en effet être décrit
par une affixe complexe, comme nous venons de le voir lors du calcul
de la fonction de partition sur le cercle ; l'échange avec la forme de
Kähler a  lieu lorsque celle-ci est complexifiée par un champ
antisymétrique (dont la donnée nécessite exactement un nombre réel
dans le cas présent bidimensionnel).

\section{Cordes ouvertes dans un champ électromagnétique}

Ajoutons un champ de fond antisymétrique constant, conformément à
l'opération à laquelle nous avons fait allusion pour réaliser
l'échange des géométries complexe et kählérienne. Avant d'interpréter
cet échange comme la symétrie miroir, ce que nous ferons dans le
contexte des cordes topologiques, nous allons inclure le champ de fond
antisymétrique dans les conditions de bord, ce qui mélange les
conditions de Neumann et de Dirichlet. Considérons une corde ouverte,
se propageant sur fond de champ magnétique constant $B_{\mu\nu}$ dans
un espace-cible plat. Elle engendre une surface d'univers $\Sigma$,
avec un bord.  L'action contient un terme couplant le champ magnétique
à la surface d'univers :
$$S=\int_\Sigma dX^\mu\land\ast dX_\nu+\int_{\partial\Sigma}
\frac{1}{2}B_{\mu\nu} X^\mu \frac{dX^\mu}{dt}.$$ Les champs de fond
représentent un certain état cohérent de cordes fermées, auquel se
couple la surface d'univers $\Sigma$ des cordes ouvertes.  Le
tenseur antisymétrique est lié au spectre des cordes ouvertes par une
symétrie de jauge : en présence d'une D-brane, un champ de fond
antisymétrique constant $B_{\mu\nu}$ (qui est une deux-forme exacte)
est équivalent à un champ magnétique :
$$\int_\Sigma B= \int_{\partial\Sigma} \frac{1}{2}B_{\mu\nu} X^\mu
\frac{dX^\nu}{dt},$$
$$\int_\Sigma B +\int_{\partial\Sigma} A =\int_\Sigma
\left(B - d\Lambda \right)+\int_{\partial\Sigma} \left(A+\Lambda\right).$$

Autrement dit, on ne peut se débarrasser de l'influence du champ $B$
sans hériter d'un photon sur la D-brane. Il sera souvent instructif
d'adopter successivement le point de vue des cordes ouvertes et des
cordes fermées pour l'interprétation des résultats concernant le champ
magnétique de fond.  Au cours des calculs, la combinaison invariante
$B+F$ sera souvent considérée comme l'expression de fluctuations des
champs de jauge autour de la valeur moyenne $B$. Cette démarche est à
la base de la résolution des équations de Seiberg--Witten de la
théorie de jauge non-commutative par utilisation du lemme de Darboux
\cite{Cornalba,Liu}, que nous évoquerons au prochain chapitre.\\

Nous étudierons d'abord dans ce mémoire un fond de métrique plate et
de deux-forme à coefficients constants $B_{\mu\nu}$, puis les cordes
sur un espace compactifié sur une variété de Calabi--Yau de dimension complexe 
trois, ce qui induira, par l'indépendance de la physique
quadridimensionnelle vis-à-vis des dimensions supplémentaires, des
symétries $O(6,6)$ qui motiveront le recours à la géométrie complexe
généralisée.\\
 
Puisque la surface d'univers possède un bord, les équations du
mouvement doivent à nouveau être assorties de conditions de bord,
correspondant à l'annulation des termes de bord qui apparaissent lors
de la variation infinitésimale des coordonnées :
$$S=\int_\Sigma g_{\mu\nu} dX^\mu\land\ast dX_\nu+\int_\Sigma B,$$
$$\delta S=\int_\Sigma d\sigma d\tau\,( \delta
X^\mu)(\partial_\sigma^2+\partial_\tau^2)X_\mu +\int_{\partial\Sigma}
g_{\mu\nu}( \delta X^\mu)(\partial+ \bar{\partial}) X^\nu.$$
 Ainsi les conditions de bord se lisent comme une compensation entre la
force élastique entre les deux extrémités de la corde et les
forces électromagnétiques :
$$(g_{\mu\nu}-B_{\mu\nu})\partial
X^\nu=-(g_{\mu\nu}+B_{\mu\nu})\bar{\partial} X^\nu.$$

Au cours de ce m\'emoire, nous allons faire deux usages principaux de
cette équation.\\ {\emph{1. Usage comme moyen de calcul : cette
équation autorise à calculer le propagateur des champs scalaires
$X^\mu$ et à calculer des intégrales de chemin en utilisant le
formalisme des états de bord. Ce genre de calcul confirme et précise
les prédictions de la théorie des champs non-commutatifs,que nous
exposerons au chapitre 3.\\
 2. Usage comme moyen de description
géométrique des D-branes dans les théories des cordes topologiques (le
lien entre ces théories topologiques et les états de bord est apparu
dans l'article~\cite{OOY} par Ooguri, Oz et Yin).\\}}

Il sera utile, lors de la discussion des D-branes topologiques, d'adopter des
coordonnées complexes $z=\sigma+i \tau, \bar{z}=\sigma- i\tau$ sur la
surface d'univers. Nous avons montré que les conditions de bord pour les
coordonnées du plongement s'écrivent
$$\partial_z \phi^\mu-R^\mu_\nu \partial_{\bar{z}}\phi^\nu =0,$$
avec la
matrice de réflexion $R$ donnée pour des coordonnées longitudinales (ou dans
le cas d'une D-brane remplissant tout l'espace), par l'expression :
$$R^\mu_\nu=\left(\frac{1}{g-B}\right)^{\mu\rho}(g+B)_{\rho\nu}.$$
En incluant les directions transverses, ou en étudiant une D-brane de
dimension éventuellement plus petite, supportée par une
sous-variété $Y$ de l'espace-cible, nous trouvons la décomposition
suivante adaptée au fibré normal $NY$ et au fibré tangent $TY$,
suivant la formule plus intrinsèque :
$$R=(-\mathrm{Id})_{NY}\oplus \left(\frac{1}{g-B} (g+B)\right)_{TY},$$
où la restriction au fibré tangent est automatique si l'on considère
$B$ comme un champ magnétique \`a valeur de distribution. En l'absence de ce champ, les
directions normales à la D-brane sont des vecteurs propres associés à
la valeur propre $-1$, et les directions tangentes des vecteurs
propres associés à la valeur propre $+1$.

\section{Propagateurs et mélanges Neumann--Dirichlet}
Les conditions aux limites écrites plus haut se prêtent à des calculs
d'action effective pour les D-branes considérées comme des états étendus  
(\emph{états de bord} notés $|B\rangle$), définis par l'action algébrique des
opérateurs contenus dans les conditions aux limites :
$$ \left(\partial \phi^\mu-R^\mu_\nu
\bar{\partial}_{\bar{z}}\phi^\nu\right) |B\rangle=0 .$$ C'est le
formalisme utilisé dans les travaux exposés dans le chapitre 4, qui
concerne les corrections dérivatives. Une introduction efficace est
donnée dans les notes de Di Vecchia et Liccardo \cite{dVL}. Ce type
de calcul ne fait pas usage de la notion de théorie de jauge
non-commutative. Il nous donnera un moyen de valider (et de compléter)
dans le langage ordinaire les méthodes issues de la
théorie de jauge non-commutative.\\

En présence d'un champ magnétique, les conditions de bord définissant
la position des D-branes, qui sont les premières équations imposées à
ces objets, mélangent les conditions de Dirichlet et de Neumann, ce
qui ouvre des perspectives naturelles de recherche, en relation avec
la T-dualité qui dicte l'incorporation des D-branes en théorie des
cordes. En effet, la T-dualité n'est autre qu'une parité n'agissant
que sur le secteur droit des coordonnées. Les D-branes
plongées dans un champ magnétique, avec leur mélange de conditions de
bord, doivent pouvoir
être décrites à partir d'une rotation effectuée sur les conditions de
bord. C'est en tout cas une façon de calculer le propagateur des
champs scalaires \cite{Blumenhagen}, dont il est fait grand usage dans
la suite de ce mémoire.\\

Pour simplifier les notations, nous nous plaçons sur un tore $T^2$, ce
qui revient à diagonaliser par blocs le champ de fond $B$. La notation
$B$ désignera donc provisoirement dans les calculs la valeur propre
antisymétrique correspondant à un bloc
$$\begin{pmatrix}
0 & B\\
-B & 0\\
\end{pmatrix}.
$$
 Nous partons d'une D0-brane et nous
allons reconnaître le résultat d'une action asymétrique du groupe des
rotations $SO(2)$ sur cette D0-brane dans la séquence suivante de trois
opérations élémentaires : T-dualité dans la direction $1$, rotation
de la D1-brane résultante, T-dualité dans la direction $2$. En
d'autres termes, le diagramme suivant commute :

\begin{equation*}
  \begin{CD}
    IIA @> \text{T-dualité} >> IIB \\ @VV{\text{rotation}}V
    @VV{\text{rotation asymétrique}}V \\ V_1 @> \text{T-dualité} >>
    V_2
  \end{CD}
\end{equation*}

 De plus,
l'action asymétrique donne précisément les conditions mixtes de
Dirichlet et Neumann, moyennant une identification de $B$ à une
certaine fonction trigonométrique de l'angle de rotation.\\

Les parties holomorphe et anti-holomorphe du champ de coordonnées
bosoniques dans l'espace-cible
$$
Z_L:=X^1(z)+ i X^2(z), \;\;\;\;\;\;{Z}_R:=X^1(\bar{z})+ i X^2(\bar{z}),$$
se transforment par action d'un sous-groupe. La prise en compte
    d'un sous-groupe discret de $U(1)$ (via une phase rationnelle $\varphi$,
    ce qui autorise la discussion de l'équivalence de Morita \cite{AGB}), disons
$\mathbf{Z}_N$, induit la loi 
$$
[Z_L,Z_R]\mapsto [e^{i\varphi}Z_L, e^{i\varphi}{Z}_R],$$
alors que la T-dualité
dans la direction $1$ agit comme une parité asymétrique. Ceci implique que
la séquence des trois opérations décrites plus haut agit de façon
asymétrique sur les secteurs droit et gauche :
$$[Z(z),Z(\bar{z})]\mapsto [e^{-i\varphi}Z(z),e^{i\varphi}Z(\bar{z})].$$

Effectuons ces trois transformations sur les conditions de bord. Nous
partons d'une D0-brane :
$$
\begin{cases}
\partial_\tau X^1= 0\\
\partial_\tau X^2= 0,
\end{cases}
$$
qui se transforme par T-dualité en une D1-brane dans la direction $1$ :
$$
\begin{cases}
 \partial_\sigma X^1= 0\\
\partial_\tau X^2= 0,
\end{cases}
$$
qui subit une rotation de manière géométrique, 
$$
\begin{cases}
\cos \varphi \,\partial_\sigma X^1 -\sin\varphi \,\partial_\sigma X^2=0\\

\sin\varphi \,\partial_\tau X^1+\cos\varphi \,\partial_\tau X^2=0.
\end{cases}
$$
La
T-dualité suivant la direction $1$ donne précisément les conditions mixtes
induites par le tenseur antisymétrique, moyennant l'identification
$$ B:= \cot\varphi,$$

$$ 
\begin{cases}
\partial_\sigma X^1- B \,\partial_\tau X^2=0\\
\partial_\sigma X^2+ B \,\partial_\tau X^1=0.
\end{cases}
$$
Nous avons donc parcouru trois côtés du diagramme commutatif.\\

En appliquant des rotations d'angles opposés $\varphi$ et $-\varphi$
aux secteurs droit et gauche d'une fonction harmonique du demi-plan
complexe supérieur, nous déduisons le propagateur des coordonnées
dans le champ de fond antisymétrique $B$, avec deux nouveaux termes,
l'un fonction paire de $\varphi$, et l'autre fonction impaire :
\begin{align}
\langle X^\mu(z) X^\nu(z') \rangle &= -\alpha' \delta^{\mu\nu}\ln|z-z'| \nonumber\\
& -\alpha' \epsilon^{\mu\nu} \sin\varphi\cos\varphi
\ln\left(\frac{z-\bar{z}'}{\bar{z}-z'}\right)-\alpha'
\delta^{\mu\nu}(\sin^2\varphi-\cos^2\varphi) \ln|z-\bar{z}'|.
\nonumber
\end{align}

En prenant en compte toutes les directions en dimension plus grande, nous
retrouvons le propagateur de \cite{CLNY} en fonction des champs de fond. Au
chapitre suivant nous l'utiliserons pour le calcul du développement en
produits d'opérateurs de deux ondes planes, ce qui motivera l'introduction
de la limite de Seiberg--Witten et de la non-commutativité. Le propagateur
complet sera utilisé dans le chapitre 4, concernant les corrections dérivatives
au-delà de la limite de Seiberg--Witten.

\section{Cordes topologiques}

\subsection{La R-symétrie et les twists topologiques}

Introduisons les cordes topologiques, obtenues en faisant jouer à une
symétrie interne de la supersymétrie $\mathcal{N}=2$ le rôle d'une symétrie
d'espace-temps. Nous allons voir qu'il existe deux façons distinctes
d'effectuer cette modification, et que l'une et l'autre conduisent à de
remarquables propriétés de localisation, donnant  aux
D-branes topologiques une riche structure géométrique.  Le point de départ
est la supersymétrie $\mathcal{N}=2$, ce qui suppose quatre courants
conservés : le tenseur énergie-impulsion $T$, le R-courant $J$ correspondant
à la R-symétrie entre les deux générateurs de supersymétrie, et les deux
super-courants $G^+$ et $G^-$, dans lesquels le signe désigne la R-charge.
Ces courants étant définis sur la surface d'univers, l'algèbre de
supersymétrie existe naturellement sous la forme de deux copies, droite et
gauche :
$$ (T_L,J_L, G^+_L, G^-_L),$$ 
$$({T}_R,{J}_R,{G}^+_R,{G}^-_R).$$ 

Le courant $J$ peut \^etre incorpor\'e dans les symétries d'espace-temps
  par modification du tenseur énergie-impulsion, ce qui altère
les spins de tous les champs chargés sous la R-symétrie. En
particulier, un nouvel opérateur de spin 1 est produit à partir de
l'une des super-charges si la modification du tenseur
énergie-impulsion est affectée d'un facteur $1/2$ :
$$ T\mapsto T± \frac{1}{2}\partial J.$$ L'indétermination dans le
signe conditionne le nouveau courant de spin 1, et le complexe BRST
associé (la charge conservée donnée par l'intégrale du courant de spin
1 possède un spin 0, c'est donc un scalaire, qui peut vivre sans
donnée supplémentaire sur une géométrie courbe). C'est à ce choix de
signe, bénin semble-t-il, que nous devons les deux modèles de cordes
topologiques. Ceux-ci seront reli\'es, avec des g\'eom\'etries
d'espace-cible tr\`es diff\'erentes, par la sym\'etrie miroir. Witten
a donn\'e dans \cite{WittenTFT} une description des liens entre la
sym\'etrie miroir et les th\'eories topologiques.\\

 La liste des courants conservés doit être complétée par l'opérateur
 de flot spectral $e^{i\varphi}$, responsable de la supersymétrie
 d'espace-temps, et lié à la bosonisation des R-courants et à la forme
 holomorphe, dans une variété de Calabi--Yau de dimension complexe
 $d$, par les relations
$$ e^{i\phi}=\Omega_{i_1\dots i_d}\psi^{i_1}\dots\psi^{i_d},$$
$$J=i\partial \phi,$$
 
La condition de recollement
entre les deux tenseurs énergie-impulsion modifiés possibles définit les
conditions de bord\footnote{Les conditions sur les flots spectraux sont
  écrites à une phase près ; la phase du modèle A peut être cachée dans une
  redéfinition de la 3-forme holomorphe, mais elle a d'autre part été identifiée comme
  l'image par symétrie miroir de la phase du mod\`ele B. Nous reviendrons sur ce point au chapitre 7.} des D-branes des modèles A :

$$
\begin{cases}
J_L=-J_R,\\
e^{i\phi_L}= e^{i\alpha} e^{-i\phi_R},\\
G_L^+=± G_R^-,\\
\end{cases}
$$
et B :

$$
\begin{cases}
J_L=J_R,\\
 e^{i\phi_L}= (± 1)^d e^{i\alpha} e^{i\phi_R},\\
G_L^+=± G_R^+,\\
\end{cases}
$$
Ces conditions sont vérifiées moyennant l'existence d'une matrice
 de réflexion orthogonale, notée $R$.

\subsection{D-branes topologiques}
Les conditions de bord correspondent à l'envoi par une application $\Phi$ des bords
de la surface d'univers $\Sigma$
sur une sous-variété $Y$ de l'espace-cible
$X$:
 $$ \Phi(\partial\Sigma)\subset Y.$$ La contribution des champs de
 jauge viendra progressivement corriger cette image strictement
 géométrique. Cette contribution est contenue dans la matrice $R$,
 orthogonale par rapport à la métrique de l'espace-temps:
 
 $$\partial_z \phi^\mu= R^\mu_\nu (\phi,g,F) \partial_{\bar{z}}
 \phi^\nu.$$ Cette matrice de réflexion contient toujours {\emph{a priori}} la
 contribution des champs de jauge, ce qui peut corriger, en les
 rendant moins symétriques, certaines conclusions de géométrie
 différentielle\footnote{par exemple en permettant des A-branes
 non-lagrangiennes}.\\
 
Un vecteur propre $v$ de la matrice $R$ associé à la valeur propre
(-1) correspond à une condition de Dirichlet 
$$ v_\mu\d_\tau X^\mu|_{\partial\Sigma}=0 ;$$
 il est donc normal
à la D-brane. Lorsque le tenseur du champ de jauge $F$ est nul, la
matrice $R$ est symétrique et les vecteurs propres de $R$ associés
à la valeur propre (+1) correspondent à une condition de Neumann
et engendrent en tout point l'espace tangent à la D-brane. 
Les fermions font également l'objet de conditions de bord, 
$$\psi^\mu_+= R^\mu_\nu\psi_-^\nu.$$ Ces conditions ne préservent pas
l'entier de la symétrie superconforme $\mathcal{N}=(2,2)$, et
restreignent la géométrie des D-branes de différentes façons selon le
twist topologique. Nous allons écrire des conditions de bord
(définissant des D-branes dans l'approche de surface d'univers) pour
des théories superconformes $\mathcal{N}=2$ qui préservent la
supersymétrie de la surface d'univers, et la moitié de la
supersymétrie $\mathcal{N}=2$ d'espace-temps.\\

Considérons pour commencer le modèle A. En
utilisant des coordonnées holomorphes associées à une structure
complexe $J$ sur l'espace-cible, le twist topologique implique que les
coefficients de la matrice $R$ des conditions de bord ne relient les
directions holomorphes qu'aux directions anti-holomorphes :
$$ R^\mu_\nu= R^{\bar{\mu}}_{\bar{\nu}}=0.$$
Autrement dit, nous avons dans le modèle A la relation 
$$R^t J R=-J.$$ De plus, faisons l'hypothèse que le tenseur du champ
de jauge $F$ est nul\footnote{Une décomposition par blocs de la
matrice de réflexion a conduit Kapustin et Orlov a mettre en
évidence la possibilité de A-branes non-lagrangiennes,
coisotropes, portant une courbure de jauge non nulle. Nous donnerons un exemple
sur le tore au dernier chapitre. Nous nous servirons des symétries
de cette situation non-lagrangiennes pour déduire une relation de
T-dualité en théorie des cordes ouvertes entre deux spineurs
purs}. La matrice $R$ est alors symétrique. Soit un vecteur $v$
tangent à la D-brane. C'est un vecteur propre de $R$ associé à la
valeur propre $+1$.  La relation précédente implique que le vecteur
$Jv$ est un vecteur propre associé à la valeur propre $-1$. Il est
donc orthogonal à la D-brane. Comme la structure complexe $J$ est
inversible, la dimension de la D-brane est égale à la moitié de la
dimension de l'espace-cible. De plus, l'orthogonalité entre les
vecteurs $v$ et $Jv$ s'écrit en utilisant la forme de Kähler $\omega$ : 
$$\omega=g_{\mu\rho} J^{\rho}_\nu dx^\mu\land
 dx^\nu,$$
$$ \omega(u,Jv)=0.$$ En libérant les vecteurs tangents $u$ et $v$, nous obtenons le
fait suivant : {\emph{les D-branes du modèle A, lorsqu'elles supportent un fibré à courbure nulle, sont des
sous-variétés lagrangiennes de l'espace-cible}}.\\

L'étude des cas de courbure non nulle  par des
méthodes symplectiques, sous l'hypothèse d'une fibration en tores
$T^3$, sera l'objet du
chapitre 7, mais nous allons exposer ici les arguments de théorie
supersymétrique $\mathcal{N}=2$ qui conduisent aux conditions
définissant les D-branes topologiques dans les deux modèles (on parlera de A-branes et de B-branes pour d\'esigner les D-branes des mod\`eles topologiques A et B).\\

Dans le modèle B, la décomposition de l'espace tangent en somme directe du
fibré tangent et du fibré normal au cycle supportant la D-brane, permet de
définir une structure complexe sur ce cycle. En présence d'un champ de
jauge, le fibré associé est holomorphe. Quant à la condition de stabilité,
elle exprime la phase $ e^{i\alpha}$ en fonction de la forme de Kähler et
du champ de jauge \cite{Kapustin--Li}, exprimant la non-dégénérescence d'une
certaine forme différentielle : $$
e^{i\alpha}=
\frac{(\omega+F)^p}{\mathrm{vol}|_Y}.$$

Passons en revue la déduction de ce résultat à partir du twist du modèle B. Les conditions sur
les courants conservés sont
$$ Q^{±}=\bar{Q}^{±},\;\;\;\;\;\;\;\;\;\;\;\;\;\;\;\;J=\bar{J}.$$
Grâce à l'orthogonalité de la matrice de réflexion $R$, il suffit pour
vérifier cette équation d'avoir la relation supplémentaire
$$R^t\omega R=\omega.$$
 Avec la structure presque complexe 
$$J:= G^{-1}\omega,$$ la condition d'orthogonalité et la condition
liant $R$ à $\omega$ donnent lieu à une relation de commutation :
$$
R^{-1}JR=J.$$
Cette relation implique que les sous-espaces propres de $R$
associés aux valeurs propres $-1$ et $+1$ sont stables par action de $J$. Le
fibré normal est donc stable par $J$. Sur le fibré tangent, cette stabilité
induit une structure complexe, et un fibré holomorphe. C'est la localisation
sur le géométrie complexe dans le modèle B.\\

Nous énonçons donc le résultat-slogan de cette section  :\\

\emph{les A-branes sont définies par des conditions de géométrie
  symplectique induites par la forme de Kähler, et les B-branes sont
  définies par des conditions de géométrie complexes : elles portent des
  fibrés holomorphes}.\\

Le problème de la complexification de la forme de Kähler par le
champ antisymétrique $B$ discuté brièvement dans l'exemple du tore,
ainsi que les rapports étroits entre ce champ et le tenseur du champ
de jauge, motivent l'étude de la symétrie miroir en présence de
D-branes, qui sera abord\'ee au  chapitre 7, et mettra en évidence
une intervention du champ de jauge qui ne pouvait être prévue, pour
des raisons dimensionnelles, dans l'exemple simple du tore $T^2$.


\chapter{Théories de jauge non-commutatives sur les D-branes}

\chapterprecistoc{} 

\chapterprecishere{Dans ce chapitre, je présente la notion de théorie
  de jauge non-commutative qui émerge dans une limite d'échelle de la
  théorie des cordes, lorsqu'une D-brane est suspendue dans un fond de
  cordes fermées avec une métrique plate et un champ antisymétrique
  $B_{\mu\nu}$. Les actions effectives gouvernant la dynamique des
  D-branes, actions dites de Born--Infeld et de Wess--Zumino (ou de
  Chern--Simons) peuvent être écrites de différentes façons, dont
  l'une utilise une théorie de jauge non-commutative.  Ces actions
  offrent en outre une voie physique vers la solution des problèmes de
  cohérence interne (invariance de jauge et non-localité) de la
  théorie de jauge non-commutative.}

\section{Théorie de jauge non-commutative et symétrie de dualité}

\subsection{Argument heuristique : champ magnétique et échelle 
  de non-commutativité} Reprenons la situation introduite au début du
 chapitre précédent : une D-brane est présente dans l'espace plat,
 ainsi qu'un champ $B_{\mu\nu}$ constant,
 équivalent à un champ magnétique le long de la D-brane. L'action est
 la somme de deux termes, l'un provenant de l'action de Polyakov  de la surface
 d'univers $\Sigma$ , et l'autre de l'intégrale de $B$. Dans la limite
 où les cordes sont très tendues, moyennant une limite d'échelle, on
 peut imaginer que ce dernier terme gouverne à lui seul la dynamique
 de la corde:
$$S_{\Sigma}\sim\int_{\partial
\Sigma} d\tau \,B_{\mu\nu}X^\mu \partial_\tau X^\nu.$$ Ainsi les
équations du mouvement deviennent des relations de commutation sur
l'algèbre des coordonnées le long du volume d'univers de la
 D-brane:
$$[X^\mu, X^\nu]=i\theta^{\mu\nu},$$ avec la définition, qui suppose
que le tenseur $B_{\mu\nu}$ est inversible, autrement dit qu'il 
induit une forme symplectique sur le volume d'univers de la D-brane:
$$\theta^{\mu\nu}:= (B^{-1}) ^{\mu\nu}.$$ Dans leur article de 1999
 sur la géométrie non-commutative en théorie des cordes \cite{SW},
 Seiberg et Witten ont argumenté en faveur de ces relations de
 commutation entre les coordonnées, et identifié une limite de
 théorie non-commutative des champs le long des D-branes.

\subsection{Limite de Seiberg--Witten et non-commutativité}

 Le propagateur des champs scalaires en théorie des supercordes peut
 être évalué en présence d'une D-brane. Afin de simplifier les
 notations, nous consid\'ererons une D-brane remplissant tout
 l'espace. Le calcul se g\'en\'eralise imm\'ediatement aux dimensions
 plus faibles par restriction aux directions longitudinales. Nous
 avons donc une D9-brane, ou des conditions de Neumann dans toutes les
 directions. Le propagateur des champs scalaires est un tenseur à deux
 indices
$$\langle X^\mu(\sigma_1) X^\nu(\sigma _2)\rangle=\alpha'G^{\mu\nu}
\ln|\sigma_1-\sigma_2|^2
+\frac{i}{2\pi}\Theta^{\mu\nu}\epsilon(\sigma_1-\sigma_2),$$
$$G^{\mu\nu}:=\left(\frac{1}{g+2\pi\alpha' B}\right)\Large{|}_{sym}^{\mu\nu},$$
$$\Theta^{\mu\nu}:=\left(\frac{1}{g+2\pi\alpha'
B}\right)\Large{|}_{antisym}^{\mu\nu},$$
 dont la partie symétrique est liée  à
 des dimensions anormales dans les développements en produits
 d'opérateurs. Cette partie symétrique a le même rôle que la
 métrique dans le produit d'opérateurs, 
 ce qui doit être traduit
 dans un langage adapté  à notre secteur de cordes ouvertes sur fond de champ
 électromagnétique couplé  à la ligne des extr\'emit\'es des cordes. Le
 tenseur symétrique en question mérite donc le nom de métrique de cordes
 ouvertes. On le note $G$, alors que son partenaire antisymétrique se note
 $\Theta$, et ceci en général, avant de prendre une quelconque limite.
\'Ecrivons la modification du développement en produit
d'opérateurs de deux ondes planes, induite par le champ
antisymétrique. Il s'agit d'une simple phase, qui sera la signature
des théories non-commutatives, comme facteur de vertex dans les
règles de Feynman
$$e^{ik_1 X(\sigma_1)}e^{ik_2 X(\sigma_2)}\sim
   |\sigma_1-\sigma_2|^{\alpha'k_{1,\mu} G^{\mu\nu}k_{2,\nu}}
   e^{\frac{i}{2}k_{1,\mu}\theta^{\mu\nu}k_{2,\nu}}e^{i(k_1+k_2)
   X(\sigma_1)}.$$

Il est possible d'identifier une limite de théorie des champs
non-commutative telle que le raisonnement heuristique du début
(consistant  à négliger le terme de tension superficielle dans l'action
de la surface d'univers pour ne retenir que le couplage de bord),
donne effectivement le bon résultat. Ce que nous demandons  à cette
 limite tient  à la régularité  à courte distance des développements
en produit d'opérateurs : pour retrouver les relations algébriques
entre les coordonnées évoquées plus haut, nous sommes contraints
d'écrire une limite d'échelle, dans laquelle la métrique de cordes
fermées $g$ tend vers 0 plus vite que l'inverse de la tension de corde :
$$g\sim \epsilon,$$
$$\alpha '\sim \epsilon ^2.$$ Nous avons donc affaire  à des cordes
très tendues, mais dont la structure confère  à la théorie des champs
une structure non-commutative. Le tenseur antisymétrique $\Theta$ se
réduit bien  à l'inverse du champ magnétique dans cette limite,
$$\Theta\mapsto \theta:= B^{-1},$$ et le propagateur ne comporte plus
qu'un terme, ce qui rendra plus simple le calcul des intégrales de
chemin par contraction de Wick dans cette limite d'échelle, dite de
Seiberg et Witten~\cite{SW}:
$$\langle X^\mu(\sigma_1) X^\nu(\sigma_2)\rangle=
\theta^{\mu\nu}\epsilon(\sigma_1-\sigma_2).$$ Il est  à noter que cette
notion de non-commutativité, invariante par translation, a motivé
l'introduction de la limite d'échelle.  Ce point de vue géométrique
sur l'existence d'une formulation non-commutative de la théorie des
champs est massivement utilisé dans \cite{SW}, et le problème de la
transformation des champs de jauge est posé par la cohérence interne
avec ce point de vue géométrique.  Cet usage n'est cependant pas
obligatoire : une fois la limite d'échelle identifiée, rien n'interdit
(et ce sera le propos du prochain chapitre) de garder un point de vue
ordinaire (au sens de commutatif\footnote{{Les termes {\emph{ordinaire
}} ou {\emph{commutatif}} sont employés dans ce contexte pour
qualifier les lois de multiplication des champs, par opposition au
terme {\emph{non-commutatif}} ; pour parler de la structure du groupe
de jauge, on utilisera les termes {\emph{abélien}} et
{\emph{non-abélien}}.}}), et d'évaluer des actions effectives  à
l'ordre du disque 
$$Z[A]=\int DX\, e ^{-S[X,A]},$$ par contractions de Wick à l'aide du propagateur
antisymétrique, sans référence  à l'interprétation
géométrique des opérateurs non-locaux qui ne manqueront pas
d'émerger.\\

 Faisant l'économie des considérations
géométriques préalables, nous n'aurons pas l'usage des
symétries de dualités, ce que nous paierons par une combinatoire
plus lourde au cours des calculs. Néanmoins, les résultats, pour
autant qu'ils puissent se comparer  à des prévisions pour les
actions effectives issues du point de vue non-commutatif, auront
valeur de test en théorie des cordes pour un argument de symétrie
émis au niveau de la théorie non-commutative des champs.\\

 Afin d'appr\'ecier cet argument, présentons les
conséquences les plus générales de la
non-commutativité sur les actions effectives et sur la description
des champs de jauge.

\subsection{Non-commutativité, non-localité et actions effectives}
\subsubsection{Les produits}
Les coordonnées du volume d'univers de la D-brane étant assujetties
aux relations de commutation
$$[X^\mu,X^\nu]=i\theta^{\mu\nu},$$ c'est toute l'algèbre des
fonctions de ces coordonnées qui doit être modifiée. En particulier,
pour des champs scalaires, la loi de multiplication ordinaire est
remplacée par le produit dit de Moyal, ou star-produit,
$$f\ast g(x):=f(x) \exp \left(
\frac{i}{2}\theta^{\mu\nu}\partial_\mu\partial'_\nu\right)
g(x')\Big{|}_{x'=x},$$ ce dont on peut se convaincre en séparant les
modes de Fourier les uns des autres. Nous reviendrons sur la question
de la non-localité induite par le nombre infini de dérivées dans ce
produit, car il sera difficile d'écrire des couplages invariants de
jauge en remplaçant tous les produits ordinaires par cet
opérateur. Néanmoins nous pouvons faire quelques observations
concernant les théories des champs scalaires. L'action s'obtient  à
partir de celle de la théorie ordinaire \cite{Komaba} en
remplaçant la multiplication ordinaire par un produit de
Moyal\footnote{Quant  à l'intégrale, il s'agit en fait d'une trace sur
l'espace de Hilbert associé aux relations de commutation dans une
interprétation  à la Heisenberg, par la prescription dite de Weyl, qui
associe un opérateur $\hat{f}$  à chaque champ $f$ de manière
compatible avec le produit de Moyal:
$$\hat{f}\hat{g}=\widehat{f\ast g}.$$ Nous continuerons  à écrire des
intégrales par rapport  à des coordonnées non-commutatives, mais
elles sont  à prendre en un sens formel.}, si bien que nous avons  à
notre disposition une théorie non-commutative avec ses règles de
Feynman induites par la simple correspondance :
$$ S=\int dx(\dots +\phi ×\dots×\phi+\dots)\;\rightarrow\; \hat{S}=\int
dx(\dots +\phi\ast\dots\ast\phi+\dots).$$  À propos des règles de
Feynman, d'intéressantes observations ont été faites par
Minwalla, Van Raamsdonk et Seiberg~\cite{NPD}. La modification de la
théorie par déformation du produit induit une modification des
règles de Feynman par une simple inclusion de facteurs de phase gaussiens aux
vertex, ce qui est manifeste via l'écriture du produit de Moyal dans
l'espace de Fourier :

$$\widetilde{f\ast g} (k)=\delta(k-k_1-k_2) f(k_1)\exp \left(
\frac{i}{2}\theta^{\mu\nu}k_{1,\mu}k_{2,\nu}\right) g(k_2).$$ Tout se
passe comme si le nouveau paramètre dimensionné $\theta$ régularisait
la théorie, rendant possible l'intégration sur l'espace des
impulsions. Le problème de l'ordre des limites engendre celui du
mélange entre les régions infra-rouge et ultra-violette, vestige de
l'origine de la non-commutativité en théorie des cordes, qui pose la
question du potentiel phénoménologique des théories
non-commutatives~\cite{AGVM}, à commencer par l'électrodynamique
récrite avec des produits de Moyal. Pour poursuivre l'étude des
D-branes, nous avons besoin de champs de jauge, c'est-à-dire d'une
structure fibrée qui doit réagir au nouveau caractère non-commutatif
de l'espace qui lui sert de base.

\subsubsection{Les champs de jauge}

 Seiberg et Witten ont donné dans \cite{SW} un argument général de
 théorie quantique des champs, qui dédramatise l'introduction de la
 non-commutativité. Celle-ci apparaît en effet naturellement via les
 commutateurs associés  à une régularisation par séparation ponctuelle
 des opérateurs. Une théorie commutative est en revanche adaptée  à une
 régularisation  à la Pauli--Villars.\\

Puisque la géométrie du volume d'univers de la D-brane peut être
commutative ou non-commutative selon la régularisation adoptée, cette
géométrie, au moins pour des valeurs suffisamment génériques des
champs de fond, n'est rien d'autre qu'un mode de description relevant
de notre choix. Le choix de la description non-commutative peut être
motivé, dans des cas que nous exposerons en détail, par la richesse de
la structure géométrique associ\'ee, qui fournit à bon marché des
résultats jusqu'alors inaccessibles aux calculs effectués dans le
cadre commutatif. Dans une situation invariante de jauge, la
description non-commutative a un effet inattendu : même dans le cas
abélien, des commutateurs de champs de jauge apparaissent, qui sont
nuls dans la théorie ordinaire de groupe de jauge $U(1)$. On parle
alors de th\'eorie de jauge non-commutative ab\'elienne de
Yang--Mills.  Or la notion d'orbite de jauge est, elle, topologique et
intrinsèque, indépendante de notre choix d'un mode de
régularisation. Elle doit \^etre pr\'esente dans l'une et l'autre des
descriptions. Il doit donc exister un champ de jauge non-commutatif
$\hat{A}$, défini de manière certainement non-linéaire à partir du
champ de jauge ordinaire $A$, tel que les transformations de jauge,
elles-mêmes redéfinies à l'aide d'un paramètre $\hat{\lambda}$,
respectent les orbites de jauge :
$$ \delta A_\mu= \partial_\mu \lambda, $$ 
$$ \hat{\delta}_{\hat{\lambda}} \hat{A}_\mu= \partial_\mu
\hat{\lambda}+[\hat{\lambda},\hat{A}_\mu]_\ast. $$ Le tenseur de
courbure du champ de jauge sera naturellement ordinaire et abélien ou
non-commutatif et abélien de Yang--Mills :
$$ F_{\mu\nu}= \partial_\mu  A_\nu-\partial_\nu  A_\mu, $$
$$ \hat{F}_{\mu\nu}= \partial_\mu \hat{A}_\nu-\partial_\nu
\hat{A}_\mu+[\hat{A}_\mu,\hat{A}_\nu]_\ast.$$ On appelle
transformation de Seiberg--Witten l'application qui fait passer d'un
champ de jauge commutatif à un champ de jauge non-commutatif en
préservant les orbites de jauge. La  condition de pr\'eservation des orbites de jauge se lit, en rendant
explicite la dépendance du champ de jauge $\hat{A}$ vis-à-vis du
champ de jauge $A$ :

 $$ \hat{A}(A)+
 \hat{\delta}_{\hat{\lambda}}\hat{A}(A)=\hat{A}(A+\delta_\lambda A)
 .$$

Cette condition induit un flot sur l'espace des échelles de
non-commutativité :
$$\frac{\delta \hat{A}_\mu}{\delta
 \theta^{\rho\sigma}}=-\frac{1}{4}\left( \hat{A}_\rho\ast(\d_\sigma
 \hat{A}_\mu+\hat{F}_{\sigma\mu})+(\d_\sigma
 \hat{A}_\mu+\hat{F}_{\sigma\mu})\ast \hat{A}_\mu\right),$$ 
$$\frac{\delta \hat{F}_{\mu\nu}}{\delta
 \theta^{\rho\sigma}}=\frac{1}{4}\left(2\hat{F}_{\mu\rho}\ast
 \hat{F}_{\nu\sigma}+2\hat{F}_{\nu\sigma}\ast
 \hat{F}_{\mu\rho}-\hat{A}_\rho\ast(D_\sigma\hat{F}_{\mu\nu}+\d_\sigma
 \hat{F}_{\mu\nu}) - (D_\sigma\hat{F}_{\mu\nu}+\d_
 \sigma\hat{F}_{\mu\nu}) \ast\hat{A}_\rho\right),$$ avec la condition
 initiale en $\theta=0$ donnée par les quantités commutatives. On
 obtient l'équation pour $\delta\hat{F}/\delta\theta$ en développant
 le commutateur dans la définition de $\hat{F}$ au premier ordre en
 $\theta$. Il est instructif de résoudre ce problème pour un champ
 électromagnétique constant : dans ce cas, les termes 
 qui font intervenir des dérivées de $\hat{F}$, ne contribuent pas :
$$\delta \hat{F}=-\hat{F}\delta \theta \hat{F},$$
$$\hat{F}_{\theta=0}=F.$$
$$F=\hat{F}\frac{1}{1-\theta\hat{F}}.$$

 Cet argument d'invariance de jauge doit nous intriguer,
 puisque la structure même du produit de Moyal conspire contre
 l'invariance de jauge via sa non-localité. La relation
$$f(x)\ast e^{ikx}=e^{ikx}\ast f(x+\theta k)$$ implique en effet que
les transformations de jauge affecteront les différents modes de
Fourier d'un même champ de façon différente. Cet étalement des
transformations de jauge sur le spectre de Fourier devra être compensé
mode par mode afin d'obtenir des observables invariantes de jauge. Le
rôle des lignes de Wilson ouvertes dans le rétablissement
de l'invariance de jauge dans les théories non-commutatives a été mis
en évidence par Gross, Hashimoto et Itzhaki~\cite{observables}, avant
d'être exploité pour résoudre les équations de Seiberg--Witten.

\section{L'action de Dirac--Born--Infeld comme action effective de la
  théorie des cordes}

\subsection{Fonction de partition de la corde bosonique { à} l'ordre du disque}

Suivant Fradkin et Tseytlin ~\cite{FT}, calculons la fonction de
partition de cordes bosoniques ouvertes couplées { à} un champ de
jauge $U(1)$ sur leur bord.  Dans les calculs qui suivent, $x$ désigne
le mode constant de la coordonnée $X$, et $\xi$ ses fluctuations.
 Afin d'intégrer séparément par rapport aux fluctuations des
coordonnées en dimensions deux et le long du bord, nous introduisons
une nouveau param{è}tre fonctionnel $\eta$, vivant sur le bord, ainsi
qu'une contrainte liant les coordonnées { à} $\eta$. Le tenseur du
champ de jauge $F_{\mu\nu}$ est choisi constant, le couplage
$S_{\d\Sigma}$ au bord ne contient donc pas de dérivées de
$F_{\mu\nu}$. 

\begin{eqnarray}
\begin{split}
Z&=\int d^{26} x\int D\xi\int D\eta\, e ^{-
S_\Sigma[\xi]-S_{\partial\Sigma}[A,\xi|_{\partial\Sigma}]}\delta(\xi|_{\partial\Sigma}-\eta),\\&
S_\Sigma[\xi]=T\int_{\Sigma}d^2z \,\partial
\xi^\mu\bar{\partial}\xi_\mu,\\ & S_{\partial\Sigma}[A,\xi]
=\int_{\partial\Sigma}d\tau\,\dot{\xi}^\mu F_{\mu\nu}\xi^\nu.
\end{split}
\end{eqnarray}

 Représentons la contrainte { à} l'aide d'un
multiplicateur de Lagrange $\nu$ (distribution { à} support sur le
bord) et intégrons successivement par rapport { à} $\xi$ et $\nu$. Le
crochet est celui des distributions, $\star$ désigne la convolution.

\begin{eqnarray}
\begin{split}
Z&=\int d^{26} x\int D\eta \int D\nu \int D\xi
\exp\left\{-\left\langle{\xi, \frac{d^2}{d\tau^2}\xi}\right\rangle -i
\left\langle\nu, \xi\right\rangle -i\left\langle \eta,\nu
\right\rangle-S_{\d\Sigma}[\eta]\right\}\\&
=\int d^{26} x\int D\eta \int
D\nu\exp\{-\langle{\nu,G\star\nu}\rangle -i\langle \eta,\nu
\rangle-S_{\d\Sigma}[\eta]\}\\&
=\int d^{26} x\int
D\eta\exp\{-\langle{\eta,G^{-1}\star\eta}-S_{\d\Sigma}[\eta]\},\\&
\end{split}
\end{eqnarray}
o{ù} $G$ désigne la fonction de Green de l'opérateur de d'Alembert.
Diagonalisons le tenseur $F$ par blocs et considérons le bloc $(\eta
^1, \eta ^2)$, dont
la valeur propre antisymétrique associée est notée $f$. L'opérateur de
d'Alembert étant diagonal, les termes associés dans l'intégrand
s'écrivent 
$$\eta ^1 G^{-1}\eta ^1 + \eta ^2 G^{-1}\eta ^2 +if^2\dot{\eta ^1}\eta
^2,$$
de sorte que l'intégrale fonctionnelle est gaussienne en $\eta ^2$,
ce qui permet d'effectuer l'intégrale par rapport { à} la moitié
des composantes de $\eta$, en notant $f_i$ les valeurs propres
antisymétriques de $F$, pour $i$ compris entre 1 et 13.

Normalisons par rapport { à} la configuration dans laquelle le
champ électromagnétique est nul :
$$Z[F]=\int d^{26} x\int\prod_{j=1}^{13}
D\eta_j\exp\left\{-\left\langle{\eta,\left(1-f^2G\star\frac{d^2}{dt\tau^2}G\right)\star\eta}\right\rangle\right\}.$$

Or le support de la coordonnée $\eta$ est concentré sur le
bord. Donc c'est la restriction au bord de la fonction de Green $G$
qui intervient. Elle  est invariante par translation~:
$$G(e ^{i\theta}, e ^{i\theta'})=\frac{1}{2\pi}\log|e
 ^{i\frac{\theta-\theta'}{2}}-e ^{-i\frac{\theta-\theta'}{2}}
 |^2=\frac{1}{2\pi}\log(2-2\cos(\theta-\theta')).$$ Écrivons-la comme
 une série afin de la rapprocher du développement de Fourier du
 champ de coordonnées $\eta$ sur le bord~:

$$G=\log(1-e ^{-i\phi})(1-e ^{i\phi})=\sum_{m\geq 1}\frac{1}{m}(e
^{im\phi}+ e ^{-im\phi})=2\sum_{m\geq 1}\frac{1}{m}\cos\phi,$$
$$\phi~:=\theta-\theta'.$$

La convolution entre $G$ et sa dérivée seconde ne fait intervenir
que les produits dont les facteurs portent la m{ê}me fréquence, ce
qui conduit au résultat

 $$G\star\frac{d^2}{d\tau^2}G=\delta.$$
En écrivant le développement de $\eta$ en modes de Fourier,
$$\eta(\theta)=\sum_{m\geq 1}\(a_m\cos(m\phi)+b_m\sin(m\phi) \),$$
on exprime l'intégrale fonctionnelle comme
$$Z(F)\sim\int\prod_{m\geq 1}da_m db_m \(\prod_{k=1}^{13} e
^{-(1+f_k^2)(a_m^2+b_m^2) }\)\sim\prod_{k=1}^{13}\prod_{m\geq
  1}\frac{1}{1+f_k^2}=\prod_{k=1}^{13}\sqrt{1+f_k^2}.$$

Nous retrouvons donc pour l'action effective en présence d'un champ
électromagnétique uniforme dans un espace plat, l'expression
dictée par l'invariance de jauge et la covariance g\'en\'erale, proposée par Born et
Infeld\footnote{ et reconsidérée par Dirac dans un modèle de
membrane destiné à évaluer la masse du muon ; cette action sera
aussi appelée action de Dirac--Born--Infeld.} pour
l'électrodynamique non-linéaire d'un objet \'etendu :
$$\int d^{26}x\sqrt{\det(\delta_{\mu\nu}+F_{\mu\nu})}.$$

\subsection{D-branes et action effective des supercordes}
 Dans le calcul de l'action effective des supercordes, on a affaire à
une version supersymétrique de l'action classique et du couplage au
champ électromagnétique, notés $S_\Sigma[X,\psi]$ et
$S_{\partial\Sigma}[X,\psi,A]$, incluant les partenaires fermioniques
$\psi$ des coordonnées $X$ : 
$$Z=\int DX D\psi  e ^{-S_\Sigma[X,\psi]-S_{\partial\Sigma}[X,\psi,A]}.$$
 La contribution des
  fermions (hors le changement de la dimension
  de l'espace-cible, bien sûr) est un simple facteur jacobien valant 1,
  qui avait été incorrectement évalué
  dans un premier temps  à cause de la considération de conditions aux
  limites périodiques inadaptées pour les fermions. Ce fait se retrouvera plus loin dans des calculs plus pr\'ecis d'int\'egrales fonctionnelles pour les cordes : le secteur de Neveu--Schwarz fournit l'action de Born--Infeld avec ses corrections, et le secteur de Ramond fournit l'action de Wess--Zumino avec ses coorections.\\

Afin de généraliser le calcul précédent au cas des
supercordes, il convient de traduire les résultats expos\'es
jusqu'ici dans le langage des
D-branes. Pour fixer les
 notations, disons que nous cherchons l'action effective de la D9-brane
de la théorie de type IIB.\\

Tout d'abord, étant donné le r{ô}le de la T-dualité dans la
géométrie des D-branes, voyons comment une action effective { à} la
Born--Infeld en dix dimensions peut {ê}tre interprétée en physique
des particules apr{è}s une
série de T-dualités donnant lieu { à} l'action effective associée
{ à} une D0-brane. Nous savons que les champs de jauges sont remplacés par des
déplacements transverses lorsqu'une transformation de T-dualité
est effectuée dans une direction longitudinale :
$$A_\mu\longleftrightarrow \Phi ^\mu.$$
En particulier, l'action associée { à} une D0-brane est la longueur
de la ligne d'univers d'une particule relativiste, et la borne
supérieure pour la vitesse de celle-ci est la réduction
dimensionnelle du champ électrique critique ~\cite{BachasPorrati}. La
régularisation de la théorie de Maxwell donnée par le lagrangien
de Born--Infeld est donc une conséquence naturelle des symétries
de dualité des objets étendus de la th\'eorie des cordes, pourvu que l'on parvienne { à}
l'action de Born--Infeld pour les D9-branes.\\

\section{Point de vue non-commutatif sur l'action de Born--Infeld}
\subsection{Un \emph{educated guess} pour des champs constants}
Il s'agit d'abord de récrire l'action de Born--Infeld en utilisant les
paramètres de cordes ouvertes mis en évidence au moment de la
discussion du propagateur et de la limite de Seiberg--Witten. Nous
avons donc le fond de cordes fermées représenté par les deux tenseurs
adaptés aux cordes ouvertes :

$$G^{\mu\nu} = \left(\frac{1}{g+2\pi\alpha' B}g\frac{1}{g-2\pi\alpha'
B} \right)^{\mu\nu},$$

$$\Theta^{\mu\nu}=-2\pi\alpha'\left(\frac{1}{g+2\pi\alpha'
B}2\pi\alpha'B\frac{1}{g-2\pi\alpha' B}\right)^{\mu\nu}.$$
 
Le déterminant contenu dans la densité d'action de Born--Infeld se factorise de
la façon suivante pour des tenseurs de champ de jauge $F$
constants :
$$ \mathcal{L}_{DBI}=\frac{1}{G_s}\sqrt{\det(1+\theta F)}
\sqrt{\det(G+2\pi\alpha'\hat{F})}.$$ Nous avons utilisé la
transformation de Seiberg--Witten pour les champs constants et la
constante de couplage de cordes ouvertes définis par les relations
$$\hat{F}=\frac{F}{1+\theta F},$$
$$ G_s:=g_s\sqrt{\det\left(1+\frac{G\theta}{2\pi\alpha'}\right)}.$$
Bien sûr il nous faut inverser la relation donnant $\hat{F}$ en
fonction de $F$, afin de trouver une véritable action
non-commutative. L'effet de cette transformation consiste à  inverser
la contribution du pfaffien commutatif :
$$\mathcal{L}_{DBI}=\frac{1}{G_s}\frac{1}{\sqrt{\det(1-\theta\hat{F})}}\sqrt{\det(G+2\pi\alpha'\hat{F})}
.$$ Il est important de garder à  l'esprit le degré de précision de
cette expression : c'est exactement le degré de précision de l'article
de Fradkin et Tseytlin \cite{FT}, puisque nous avons  fait usage de 
la solution des équations de Seiberg--Witten à  courbure
constante. Cette hypothèse de champ constant signifie que tous les
termes non gaussiens (en les fluctuations des coordonnées) dans une
définition de l'action effective par intégrale de chemin ont été
négligés.\\ 
Cette expression nous inspire une observation formelle,
puisque le numérateur n'est autre que la densité de lagrangien de
Dirac--Born--Infeld écrite avec la métrique de cordes ouvertes, le
champ de jauge non-commutatif, la constante de couplage de corde
ouverte, et sans le champ $B$ antisymétrique.  À la vue de la
densité de lagrangien que nous venons d'écrire, nous avons un candidat
naturel pour l'action de Born--Infeld non-commutative
$$\hat{S}_{DBI}:=\frac{1}{G_s}\int
dx\sqrt{\det(G+2\pi\alpha'\hat{F})}.$$ Quelques remarques sur la
structure même de cette action permettent de mieux se convaincre de la
validité d'une telle redéfinition des champs, et de remarquer les
imperfections de cette proposition. Nous avons négligé les dérivées de
$F$. De manière effective, cette approximation équivaut à  négliger les
termes non-linéaires en $\theta$. En effet la solution en champ
constant vérifie les équations obtenues en tronquant au premier ordre
le développement en puissances $\theta$ :

$$\hat{F}_{\mu\nu}=F_{\mu\nu}-F_{\mu\rho}\theta^{\rho\sigma}F_{\sigma\nu}+\mathcal{O}(\theta^2).$$

 Cette corrélation entre l'ordre en $\theta$ et l'ordre en dérivées
 provient du rapport entre le star-produit et les crochets de Poisson
 associés à  $\theta$ dans une approximation ``semi-classique'' :

$$ [f,g]_\ast\sim_{\theta\to
0}\{f,g\}_\theta:=\theta^{\mu\nu}\partial_\mu\ f \partial_\nu g.$$ La
théorie de jauge non-commutative le long du volume d'univers d'une
D-brane est en effet reliée à  la quantification des variétés de
poisson à  la Kontsevich \cite{Kontsevich}. Cette relation a été
développé par Cattaneo et Felder \cite{CattaneoFelder} ; dans la
géométrie très simple que nous considérons dans ce mémoire, le
star-produit est déjà  explicite et nous n'aurons pas l'usage de
l'expression diagrammatique élaborée par Kontsevich. C'est au
contraire un affaiblissement du résultat (développement au premier
ordre en $\theta$) que nous sommes en train d'exposer. Cependant,
l'argument de quantification des variétés de Poisson se révèle fécond
en se qui concerne le problème de redéfinition des champs de jauge.
En effet, l'approximation semi-classique, au sens du développement au
premier ordre en $\theta$ et de l'utilisation des crochets de Poisson,
fournit déjà  des termes quadratiques dans le potentiel,  à  l'intérieur
des transformations de jauge et de la courbure :
$$\hat{F}_{\mu\nu}=\partial_\mu\hat{A}_\nu-\partial_\nu\hat{A}_\mu
+\theta^{\rho\sigma}\partial_\rho \hat{A}_\mu
\partial_\sigma\hat{A}_\nu.$$

Ce développement ordre par ordre en $\theta$, réalisant la
correspondance de Seiberg et Witten, repose au niveau semi-classique
sur le théorème de Darboux, comme l'a observé Cornalba
\cite{Cornalba,CornalbaBI}. La versatilité des expressions pour les champs de
jauge est le reflet de l'invariance d'une forme symplectique, ici la
deux-forme fermée $B$, par difféomorphisme. Le théorème de Darboux
nous dit en effet qu'il n'existe localement qu'un modèle de forme
symplectique, correspondant au tenseur antisymétrique diagonal par
blocs
$$\sum_i dx^{2i} \land dx^{2i+1}.$$   
   Le groupe de jauge au niveau des crochets de Poisson est le groupe des
   symplectomorphismes. Il est au groupe de jauge de la théorie abélienne
   non-commutative de Yang--Mills ce que le théorème de Darboux est au
   théorème de Kontsevich. Nous ne poursuivrons pas plus avant cette approche par
   quantification des crochets de Poisson, mais nous garderons à  l'esprit le
   fait que le développement en puissances de $\theta$ est compatible avec les
   arguments de Seiberg et Witten sur les orbites de jauge
   non-commutatives. Nous allons nous concentrer sur l'invariance de jauge,
   qui imposera des contraintes suffisamment fortes pour obtenir une
   solution au problème de redéfinition des champs de jauge  à  tous les ordres en $\theta$.\\

Un problème plus terre-à-terre nous vient du facteur
$\sqrt{\det(1-\theta\hat{F})}$, qu'il faudra bien interpréter, et
intégrer dans une version plus aboutie de l'action
effective. Remarquons tout de suite que la solution à  ce problème ne
se trouve pas dans le cadre des champs de jauge à  courbure
constante. Elle est forcément au-delà  de l'approximation de l'article
\cite{FT}, puisque l'action écrite est infinie pour une D-brane de
volume infini équipée d'un tenseur du champ de jauge constant. Cette
action n'est donc pas un outil valide pour l'étude de la dynamique des
D-branes. Certes, l'électrodynamique non-commutative provient du
développement à  l'ordre deux dans le champ de jauge,
$$\hat{S}_{DBI}\sim_{\hat{F}\to 0} -\frac{1}{4}\int dx
\hat{F}_{\mu\nu}\ast\hat{F}^{\mu\nu}+o(\hat{F}^2),$$ mais tous les
modes de Fourier de $\hat{F}$ doivent être pris en compte pour pouvoir
considérer de tels résultats comme des propriétés physiques
raisonnables d'une action. Nous allons voir que l'invariance de jauge
pour des champs variables donnera accès  à  la fois  à  une expression
satisfaisante pour l'action effective non-commutative, et 
 à  une expression plus précise dans le cadre commutatif.

\subsection{Champs variables et restauration de l'invariance de jauge}
Les ondes planes sur un espace non-commutatif sont intimement liées
aux opérateurs de translation, puisque l'expression la plus simple de
la non-localité est liée a une translation dépendant du mode de
Fourier :
$$ e^{ikx}\ast x^\mu=(x^\mu+\theta^{\mu\nu}k_\nu)\ast e^{ikx}.$$ Pour
un paramètre de jauge $\lambda$, une transformation de jauge de la
courbure non-commutative, même dans le cas abélien, s'écrit  à  l'aide
d'un commutateur,
$$ \delta\hat{F}_{\mu\nu}(x)=i[\lambda,\hat{F}_{\mu\nu}]_\ast.$$ Cette
relation n'est guère maniable que si nous distinguons entre les
différents modes de Fourier du paramètre de jauge,
$$ \lambda(x)=\int dk\, e^{ikx}\tilde{\lambda}(k),$$ de sorte que la
transformation de jauge de la courbure est directement liée à 
l'étalement de l'effet du mode d'impulsion $k$ entre les points $x$ et
$x+\theta k$ :
$$[\lambda(x),F_{\mu\nu}(x)]=\int dk \,\tilde{\lambda}(k)[e^{ikx},
\hat{F}_{\mu\nu}(x)]$$
$$=\int dk \,\tilde{\lambda}(k)(\hat{F}_{\mu\nu}(x+\theta
k)-\hat{F}_{\mu\nu}(x))\ast e^{ikx}.$$ Bien que nous ayons commencé
cette discussion des actions effectives par le terme de Born--Infeld,
cette observation est valable pour toute théorie effective utilisant
le produit de Moyal et des symétries de jauge : la transformation du
lagrangien (densité exprimée en fonction du point de l'espace-temps)
n'est pas imm\'ediatement invariante de jauge. Le développement de
Fourier peut cependant nous laisser l'espoir d'une densité invariante
de jauge dans l'espace de Fourier. Il suffira alors de superposer les
différents modes pour obtenir par synthèse de Fourier une théorie de
jauge non-commutative. Nous considérons donc les modes de Fourier du
lagrangien définis par la relation
$$ \mathcal{L}(k)=\int dx\, \mathcal{L}(x)\ast e^{ikx}.$$ 

Le seul mode invariant de jauge d'un lagrangien est le mode
d'impulsion nulle. Considérons une transformation de jauge finie dont
le générateur est le paramètre décrit précédemment :
$$U_\lambda(x):=\exp_\ast (i\lambda(x))=\sum_{n\geq
0}\frac{i^n}{n!}\left(\ast\lambda(x)\right)^n.$$ Pour un lagrangien
$\hat{\mathcal{L}}$ exprimé en fonction de la courbure de jauge
$\hat{F}$, cette transformation aura dans l'espace direct l'effet
 $$\hat{\mathcal{L}}(x)\to U_\lambda(x)\ast\hat{\mathcal{L}}(x)\ast
U_{-\lambda}(x),$$ de sorte que nous pouvons écrire la transformation
d'un mode de Fourier d'une manière qui pénalise toute impulsion non nulle
:
$$\hat{\mathcal{L}}(k)\to \int dx\,
U_\lambda(x)\ast\hat{\mathcal{L}}(x)\ast U_{-\lambda}(x)\ast e^{ikx}$$
$$= \int dx\, U_\lambda(x)\ast\hat{\mathcal{L}}(x)\ast e^{ikx}\ast
U_{-\lambda}(x-\theta k)$$
$$= \int dx \,U_{-\lambda}(x-\theta k)\ast
U_\lambda(x)\ast\hat{\mathcal{L}}(x)\ast e^{ikx}.$$ La dernière
équation fait usage de la cyclicité de cette trace qu'est
l'intégrale sur l'espace non-commutatif. Le recours au potentiel de
jauge (non-commutatif) apporte un remède homéopathique au manque
d'invariance de jauge, via un transport parallèle, selon la
prescription issue des résultats de Gross, Hashimoto et Itzhaki ~\cite{observables}
sur l'invariance de jauge non-commutative. En effet, un opérateur de
transport parallèle (ligne de Wilson) le long d'un segment joignant
les points $x$ et $x+\theta k$ n'est pas invariant de jauge, à cause
du caractère ouvert du segment. Les transformations de jauge
agissent aux deux extrémités, compensant les transformations du
mode $\hat{\mathcal{L}}(k)$.  La version non-commutative de l'action
de Born--Infeld s'écrit donc
$$ \hat{S}_{DBI}= \frac{1}{g_s}\int dx
\left\{\sqrt{\det(1-\theta\hat{F})}\sqrt{\det\left(g+\hat{F}\frac{1}{1-\theta\hat{F}}\right)}\ast
\exp\left( -i\int_x^{x+\theta k}\hat{A}_\mu dx^\mu \right)
\right\}\ast e^{ikx},$$ r\'esultat qui s'exprime plus simplement en
définissant la prescription d'intégration $L_\ast$, qui effectue
une moyenne sur toutes les façons d'insérer des objets le long
d'une ligne de Wilson $W_k$, tendue entre les point $x$ et $x+\theta
k$ :

$$ \hat{S}_{DBI}= \frac{1}{g_s}\int dx L_\ast \left\{
\sqrt{\det(1-\theta\hat{F})},
\sqrt{\det\left(g+B+\hat{F}\frac{1}{1-\theta\hat{F}}\right)}, W_k
\right\}\ast e^{ikx}.$$
 Nous allons écrire le terme de Chern--Simons
non-commutatif en nous inspirant de ce résultat.

\section{Couplages de Ramond--Ramond}
\subsection{Description non-commutative}
Une théorie non-commutative des champs vit donc le long du volume
d'univers de la D-brane, sur un fond de champ de magnétique constant.
Les D-branes se couplant aux champs de formes différentielles du
secteur de Ramond--Ramond, il est naturel de rechercher les
conséquences de la non-commutativité sur ces couplages. En partant du
principe que les termes de Wess--Zumino possèdent une description 
non-commutative, nous serons naturellement conduits  à une solution des
équations de Seiberg--Witten. Cette démarche est entièrement fondée
sur des arguments de symétrie : invariance de jauge et dualité entre
les descriptions commutative et non-commutative.\\

Nous allons écrire les termes de couplage aux champs de Ramond--Ramond
issus de la théorie des champs non-commutatifs~\cite{CSterms}. Le long
des D-branes instables, de tels couplages font intervenir des
tachyons, dont le rôle dans les transitions de phase vers un vide
stable de la théorie des cordes (condensation de tachyons) est l'une
des grandes idées contemporaines de la théorie des cordes. La
non-commutativité pour les scalaires tachyoniques, ainsi que les actions
effectives induites, feront l'objet du chapitre 5.\\

La recette apprise en théorie bosonique, c'est-à-dire avec l'action
effective de Born--Infeld comme fonction de partition  à l'ordre du
disque \cite{FT}, consiste  à changer de coordonnées d'intégration,
$$x^\mu\mapsto X^\mu:=x^\mu+\theta^{\mu\nu}\hat{A}_\nu,$$
$$[X^\mu, X^\nu]=\theta^{\mu\nu} -\theta^{\mu\rho}
 \hat{F}_{\rho\sigma}\theta^{\sigma\nu}=:
 \left(Q^{-1}\right)^{\mu\nu},$$ et à considérer l'ancien lagrangien
 avec les anciens champs comme un lagrangien pour les champs de jauge
 non-commutatifs ; bien sûr, tous les produits sont des $\ast$ de
 Moyal de paramètre $Q$, et la relation entre les champs commutatifs
 et non-commutatifs est non-linéaire. Afin d'obtenir une densité de
 lagrangien dépendant explicitement des champs de jauge
 non-commutatifs, tout en conservant les coordonnées $x^\mu$ et le
 paramètre de non-commutativité $\theta^{\mu\nu}$, on peut de manière
 équivalente insérer un facteur
$$\int dx \,\longrightarrow \int_{\mathbf{R}^9_\theta} dx\,\frac{\mathrm{Pf} Q}{\mathrm{Pf}
\theta},$$ 
dans la mesure d'intégration. Quant aux champs de jauge,
ils sont encore reliés par l'intermédiaire de la quantité invariante
$$B+F=Q^{-1}.$$ C'est le chemin que nous allons suivre pour écrire les
termes de Wess--Zumino. Dire qu'il existe une théorie de jauge
non-commutative de paramètre $\theta^{\mu\nu}$ couplée aux champs de
Ramond--Ramond sur la D9-brane stable de type IIB, en présence d'un
champ de fond $B$, revient  à affirmer{\footnote{Dans ce contexte,
l'intégrale est définie par dualité  à partir de l'algèbre extérieure
des coordonnées, qui n'est pas modifiée par le produit de Moyal ; la
quantité $C$ désigne la somme formelle de tous les champs de formes
différentielles de Ramond--Ramond en type IIB ; la quantité $Q^{-1}$
désigne la 2-forme associée $Q^{-1}_{\mu\nu} dx^\mu\land dx^\nu$ ;
l'exponentielle est développée formellement et l'intégration
sélectionne les termes de degré maximal.}}

$$\hat{S}_{CS}=T_9\int_{\mathbf{R}^9_\theta} \frac{{\mathrm{Pf}}
Q}{{\mathrm{Pf}} \theta}\, C \land e^{Q^{-1}}.$$ pour des champs de
jauge uniformes. La combinaison invariante $Q$, qui vaut $B+F$ dans la
description commutative, est là pour nous rappeler que le tenseur $B$
ne peut pas être éliminé sans créer un photon sur le volume d'univers.

\subsection{Solution des équations de Seiberg--Witten}

 Avec l'expérience des champs variables que nous avons acquise, nous
 pouvons avancer, suivant les travaux de Liu~\cite{Liu}, Mukhi et
 Suryanarayana \cite{Mukhi} qu'une solution des équations de
 Seiberg--Witten est donnée par le couplage au champ de formes
 différentielles $C^{(8)}$ de Ramond--Ramond :

$$F_{\mu\nu}(k)=\int dx \, L_\ast\left(\frac{\mathrm{Pf}
Q}{\mathrm{Pf}\theta},\hat{F}\frac{1}{1-\theta\hat{F}}, W_k(x)
\right)\ast e^ {ikx}.$$ Ooguri et Okawa ont démontré cette conjecture
\cite{OO} en vérifiant l'invariance de jauge, l'identité de Bianchi et
la condition initiale.\\

 Il est donc avéré que les termes de Born--Infeld
 \cite{Das--Trivedi} et de Chern--Simons \footnote{L'action de
 Born--Infeld que nous écrivons est en fait le couplage au mode
 d'impulsion $k$ du dilaton.} possèdent la description
 non-commutative
$$\hat{S}_{DBI}=\int dx\, L_\ast\left(\frac{\mathrm{Pf}
Q}{\mathrm{Pf}\theta}, \sqrt{g+Q^{-1}}, W_k(x) \right)\ast e^ {ikx},$$
$$\hat{S}_{CS}= C(-k) \wedge\int dx \, L_\ast\left(\frac{\mathrm{Pf}
Q}{\mathrm{Pf}\theta}, e^{Q^{-1}}, W_k(x) \right)\ast e^ {ikx}.$$

Il est cependant manifeste que nous n'avons pas extrait toute
l'information contenue dans cette action, en particulier à cause de la
multiplicité des termes dans les couplages de Ramond--Ramond. Le
couplage à $C^{(8)}$ ne reçoit pas de corrections faisant
intervenir les dérivées de $F$, puisque de telles corrections peuvent être
localement intégrées pour donner une forme différentielle de degré
 un, qui peut être absorbée par définition du champ de jauge. Cet
arbitraire ayant été exploité pour écrire la solution au problème de
Seiberg--Wittten, les couplages aux champs de Ramond--Ramond de degré
plus faible vont recevoir des corrections prescrites par la
non-commutativité. Ces corrections font l'objet du chapitre
suivant. Quant au couplage à $C^{(10)}$, il ne fait pas intervenir
le champ de jauge dans la description commutative et fournit donc une
identité topologique concernant la théorie de jauge
non-commutative. Nous retrouverons au chapitre 6 un exemple de cette
situation au moment de la discussion des D-branes topologiques
non-commutatives du modèle B.


\chapter{Corrections dérivatives}

\chapterprecistoc{} \chapterprecishere{Dans ce chapitre, je présente
les prédictions issues de la non-commutativité pour les
corrections dérivatives à l'action effective des champs de
jauge abéliens. Le calcul de ces corrections en théorie commutative est un
exercice qui n'avait été fait auparavant qu'à un ordre fini en
dérivées. Je reproduis, par simple utilisation des conditions de
bord, les prédictions de la théorie de jauge non-commutative pour
les termes de Chern--Simons et de Born--Infeld, dans la limite de
Seiberg--Witten. Je prolonge ensuite le calcul des couplages aux
champs de Ramond--Ramond au-delà de cette limite, en tenant compte
du terme symétrique dans le propagateur des champs scalaires. 
 J'interprète ce résultat comme la réalisation sur le bord de
la surface d'univers d'une théorie de jauge non-commutative
déformée par la métrique.
}

\section{Les star-produits modifiés, leur origine non-commutative }
Suivant Liu \cite{Liu}, nous développons en puissances du champ de
jauge des expressions du chapitre précédent pour les actions
effectives dans la description non-commutative. À chaque ordre $n$ est
associé un opérateur différentiel $\ast_n$, ou star-produit modifi\'e
d'ordre $n$. La relation entre les différents ordres est bien entendu
intimement liée à l'invariance de jauge, laquelle est responsable de
l'apparition de la ligne de Wilson. Ces produits sont en quelque sorte
des descendants du produit de Moyal, et avaient été obtenus
  pour les rangs $n=2,3$ dans les premières étapes \cite{Das--Trivedi}
du calcul de l'application de Seiberg--Witten. Ils contiennent tous
une infinité de dérivées, et sont donc susceptible de prolonger aux
grands ordres en dérivées les résultats obtenus systématiquement aux
faibles ordres \cite{Andreev--Tseytlin} sans faire appel à la notion
de théorie de jauge non-commutative. \\

Nous rappelons donc que $W_k$, ligne  de
Wilson tendue entre les points $x$ et $x+ \theta k$, permet par un transport
parallèle d'assurer l'invariance de jauge de quantités définies en théorie
de jauge non-commutative. Le segment est le chemin le plus simple que l'on
puisse fabriquer dimensionnellement à partir d'un mode de Fourier $k$ et de
l'échelle de non-commutativité $\theta$. Nous utilisons la notation suivante :

$$W_k(x)=P_\ast\exp\left(i\int_0^1 d\tau\,
  \theta^{\mu\nu}k_\nu\hat{A}_\mu(x+\tau \theta k)\right).$$

Avec un produit de champs non-commutatifs $O_i, 1\leq i\leq N,$
obtenus par transformation de Seiberg--Witten de champs constants, la
prescription de moyenne qui consiste à attacher les différentes
observables de la manière la plus simple possible pour adapter les
observables à la non-localité, est désignée par $L_\ast$, et définie
par la relation donnant le mode d'impulsion $k$ du produit
d'observables :

$$Q(k)=\int dx \left( \prod_{i=1}^N \int_0^1 d\tau_i \right)P_\ast\left(
  W(x),\prod_{i=1}^N  O_i (x+\tau_i\theta k)\right).$$
Le transport parallèle
le long du segment de Wilson induit un développement en puissances du champ
de jauge $\hat{A}$ :
$$Q(k)=\sum_{p\geq 0} Q_p(k,\hat{A}),$$ où $Q_p(k,\hat{A})$ reçoit
une contribution d'ordre $p$ en $\hat{A}$ du développement de la
ligne de Wilson. Considérons d'abord le terme d'ordre 0. En écrivant
les observables $O_i$ dans l'espace des impulsions, et en intégrant sur
l'espace des positions, Liu a obtenu \cite{Liu} la formule plus
symétrique en les arguments,
$$
Q_0(k)=\left( \prod_{i=1}^N \int\frac{dk_i}{(2\pi)^d}
  \delta(k-\sum_{i=1}^N k_i)\right)\prod_{i=1}^N O_i(k_i)J_N(k_1,\dots,k_N).$$
sur laquelle on lit le noyau d'intégration qui correspond dans l'espace de
Fourier à l'opérateur $\ast_n$ :
$$ J_N(k_1,\dots,k_N):=\left( \prod_{i=1}^N \int_0^1 d\tau_i
\right)\exp\left(
-\frac{i}{2}\sum_{i<j}k_{\mu,i}\theta^{\mu\nu}k_{\nu,j}(2\tau_{ij}-\epsilon(\tau_{ij}))\right),$$
qui se lit dans l'espace des positions comme un opérateur différentiel agissant sur $N$ arguments :
$$\ast_N :=\left( \prod_{i=1}^N \int_0^1 d\tau_i
\right)\exp\left(
-\frac{i}{2}\sum_{i<j}\d_{\mu,i}\theta^{\mu\nu}\d_{\nu,j}(2\tau_{ij}-\epsilon(\tau_{ij}))\right).$$

En particulier, on a dans l'espace des positions :
$$\ast_2=\frac{\sin\left(\frac{1}{2}\d_\mu\theta^{\mu\nu}\d'_\nu\right)}{\frac{1}{2}\d_\mu\theta^{\mu\nu}\d'_\nu}$$

Les corrections que nous allons calculer dans ce chapitre sont
codées par ces opérateurs, et certaines de leurs d\'eformations,
notées $\tilde{\ast}_n$, qui sont apparues dans \cite{MSbeyond} au
rang $n=2$, et que j'ai généralisées aux ordres supérieurs
dans \cite{Grangebeyond}.

\section[Corrections dérivatives aux actions effectives]{Corrections dérivatives aux actions effectives : le point de vue commutatif}

Plusieurs voies ont été explorées pour calculer les
couplages des modes légers de la théorie des cordes.  Il y a la
méthode de la matrice de diffusion, qui consiste à évaluer des
amplitudes de diffusion avec les modes en question pour états
asymptotiques. L'action effective est alors une fonctionnelle qui
reproduit les amplitudes. Il y a la méthode dite de la fonction
$\beta$ de renormalisation : elle consiste à évaluer le flot de
renormalisation des couplages de cordes comme modèle sigma
non-linéaire ; l'équation $\beta=0$, qui commande l'invariance
conforme, est alors interprétée comme équation du mouvement pour les
différents champs, dérivant d'une certaine fonctionnelle. Notre
approche sera davantage liée à la méthode dite de la fonction de
partition, via la technique des états de bord, plus adaptée au point
de vue sur les D-branes comme états étendus. Les couplages effectifs
du secteur de jauge sont des éléments de  matrice d'un opérateur qui
allume le champ de jauge le long du volume d'univers de la D-brane.\\

 Les relations algébriques~\cite{CLNY} vérifiées par un état
 de bord ne sont autres que les conditions de bord habituelles,
 considérées non pas comme des équations aux dérivées
 partielles, mais comme des opérateurs annihilant un état
 $|B\rangle$. Une condition de Neumann s'écrira
$$P_\mu |B\rangle =0,$$
et une condition de Dirichlet 
$$
\partial_\tau X^\mu |B\rangle=0,$$
alors qu'en présence d'un champ électromagnétique de fond, l'état de bord \'etant noté $|B(F)\rangle$, on a la
relation $$(P_\mu+F_{\mu\nu}\partial_\tau X^\nu)|B(F)\rangle=0,$$
qui est en
fait une définition de l'état de bord $|B(F)\rangle$, ou plutôt de sa partie
bosonique. Cette équation peut être résolue en fonction  de
l'état de bord $|B\rangle$ avec conditions de Neumann dans toutes les directions et sans
champ de jauge
$$
|B(F)\rangle=\exp\left(-\frac{i}{2\pi\alpha'}\int d\tau\partial_\tau
  X^\mu A_\mu(X(\tau))\right)|B\rangle,$$
dont la partie supersymétrique
s'obtient en substituant des super-champs et des super-dérivées aux objets
usuels.
$$
|B(F)\rangle=\exp\left(-\frac{i}{2\pi\alpha'}\int d\tau d\theta D\phi^\mu A_\mu(\phi)\right)|B\rangle,$$
$$\phi^\mu:= X^\mu +\theta\psi^\mu, $$
$$ D:=\d_\theta -\theta \d_\tau.$$

Pour expliciter les deux termes de l'action effective, il faut calculer les
deux intégrales de chemin à l'ordre du disque correspondant aux quantités

$$\langle 0| \exp\left(-\frac{i}{2\pi\alpha'}\int d\tau\partial_\tau \phi^\mu
  A_\mu(\phi)\right)|B\rangle_{R}= \int C\land e^F +\Delta S_{CS},$$

$$\langle C |\exp\left(-\frac{i}{2\pi\alpha'}\int d\tau\partial_\tau
  \phi^\mu A_\mu(\phi)\right)|B\rangle_{NS}= \int dx\,\sqrt{\det(g+F)}+\Delta S_{DBI}.$$
  Les indices R et NS indiquent le secteur de Ramond (avec conditions
  de bord périodiques pour les fermions, qui autorisent un mode de
  fréquence nulle) et le secteur de Neveu--Schwarz (avec conditions
  anti-périodiques, qui interdisent les modes de fréquence
  nulle). Le développement de
  Taylor du champ de jauge est un développement en puissances de
  l'opérateur de dérivation agissant sur ce champ, et en puissances des
  fluctuations des coordonnées, lesquelles seront intégrées par
  contractions de Wick. Dans le secteur de Neveu--Schwarz, ce
  développement ne comporte que la première somme de contributions :

\begin{multline}
\label{developpement}
\int d\sigma d\theta \,D\phi^\mu A_\mu(\phi) =-\int d\sigma
d\theta\,\sum_{k\geq
0}\frac{1}{(k+1)!}\frac{k+1}{k+2}D\tilde{\phi}^\nu
\tilde{\phi}^{\mu}\tilde{\phi}^{\mu_1}\dots\tilde{\phi}^{\mu_{k}}\d_{\mu_1}\dots\d_{\mu_{k}}F_{\mu\nu}(x)+\\
-\int
d\sigma(\tilde{\psi}^\mu\psi_0^\nu+{\psi}_0^\mu\psi_0^\nu)\sum_{k\geq
0}\frac{1}{k!}\tilde{X}^{\mu_1}\dots\tilde{X}^{\mu_k}\d_{\mu_1}\dots\d_{\mu_k}F_{\mu\nu}(x)\pt
\end{multline}

Quant aux corrections dépendant des dérivées du champ de jauge,
elles tiennent essentiellement à la structure de la théorie des cordes,
alors que l'action de Born--Infeld est dictée par l'invariance de
jauge et la covariance générale. Nous utiliserons ce
développement de Taylor tout au long du chapitre pour \'evaluer les corrections
 en théorie {\emph{commutative}} des cordes.

\section{Prédictions non-commutatives}
À présent que la redéfinition des champs de jauge a été explicitée
conformément aux équations de Seiberg--Witten, il est naturel de
demander que l'action effective tout entière, et pas seulement le
couplage à $C^{(8)}$, admette une expression non-commutative. Cette
expression est remarquablement concise et géométrique, grâce à la
prescription d'intégration de long de la ligne de Wilson
ouverte. Cependant, un développement en dérivées est induit dans le
langage commutatif, qui a beaucoup moins de structure apparente, et
prend beaucoup plus de place. Ce développement constitue une solution
simple à un problème compliqué, celui de l'évaluation des corrections
dérivatives aux grands ordres en dérivées. C'est ce qu'ont expliqué
Das, Mukhi et Suryanarayana \cite{DMS}.\\

Les corrections à l'action de Dirac--Born--Infeld viennent du couplage
au dilaton (étudié dans l'espace des impulsions sous la forme de ses
modes $\tilde{D}(k)$ dans le secteur de Neveu--Schwarz), tandis que
les couplages aux modes de Fourier des champs $C^{(2p)}$, pour $p\leq
3$, constituent les termes dits de Chern--Simons. La limite de
Seiberg--Witten fait tendre vers 0 toutes les corrections dans
lesquelles les dérivées sont accompagnées d'un facteur
$\alpha'$. Quant aux termes dans lesquels les dérivées sont
accompagnées de tenseurs $\theta$, il se trouvent en nombre infini
dans les actions non-commutatives, mais ce ne sont pas des corrections
dans la description non-commutative ! Ils deviennent des corrections lorsque
l'expression est traduite dans le langage commutatif. C'est le sens
des équations
$$S_{CS}+\Delta S_{CS}=\hat{S}_{CS},$$ 
$$S_{DBI}+\Delta S_{DBI}=\hat{S}_{DBI}.$$
 Explicitement, l'analogue de (l'inverse de) $B+F$, soit 

 $$Q^{\mu\nu}=\theta^{\mu\nu}-\theta^{\mu\rho}\hat{F}_{\rho\sigma}\theta^{\sigma\nu},$$
 contient la contribution du tenseur $\hat{F}$, de sorte que
$$
S_{DBI}+\Delta S_{DBI}=\frac{\tilde{D}(-k)}{g_s}\int d^{10}x\, L_\ast\left(
  \frac{\mathrm{Pf} Q}{\mathrm{Pf}\theta},\sqrt{\det(g+2\pi\alpha'
    Q^{-1})}, W_k(x)\right)\ast e^{ikx},$$
$$
S_{CS}+\Delta S_{CS}=\tilde{C}(-k)\land \int L_\ast \left(
  \frac{\mathrm{Pf} Q}{\mathrm{Pf} \theta},e^{2\pi\alpha'
    Q^{-1}},W_k(x)\right)\ast e^{ikx}.$$

Le star-produit modifié $\ast_{2p}$ intervient dans l'expression du couplage
proportionnel à $F^p$, dans la limite de Seiberg--Witten (not\'ee SW) suivant la formule corrigeant le couplage au
caractère de Chern, dans l'espace des positions :
$$(S_{CS}+\Delta S_{CS})|_{SW}= \int C\land \sum_{p\geq 2}\frac{1}{p!}\ast_p
[F^p].$$ Quant au terme de Dirac--Born--Infeld, il a été développé
à l'ordre quadratique en $\hat{F}$ dans \cite{DMS}, ce qui fournit le
développement suivant, toujours dans l'espace des positions :
$$ (S_{DBI}+\Delta S_{DBI})|_{SW}=\frac{1}{g_s}\int
d^{10}x\sqrt{\det(g+2\pi\alpha' B)} \left(
1+\frac{1}{4}\theta^{\mu\nu}\theta^{\rho\sigma}\ast_2[F_{\nu\rho},F_{\sigma\mu}]-\frac{1}{8}\theta^{\mu\nu}\theta^{\rho\sigma}\ast_2[F_{\mu\nu},F_{\rho\sigma}]\right).$$
J'ai établi cette dernière relation dans dans \cite{corrections}, ce
qui constitue le premier calcul direct de corrections à tous les
ordres en dérivées à l'action de Born--Infeld. Étant donné le rôle
prépondérant joué par le secteur de Ramond dans la suite de cette
thèse, je ne passerai en revue que le calcul de $\Delta S_{CS}$,
renvoyant au premier article figurant en annexe pour celui de $\Delta
S_{DBI}$. La différence essentielle tient à l'absence des modes de
fréquence nulle pour les fermions, ce qui impose d'effectuer des
calculs avec le propagateur complet, incluant les partenaires
supersymétriques des coordonnées. Dans le texte de cette thèse, nous
n'aurons besoin que du propagateur des scalaires. De plus, seule
l'action $S_{CS}$ a été étudiée au-delà de la limite de
Seiberg--Witten.

\section{Calcul en théorie (commutative) des cordes}

 Nous avons évoqué, via la dualité non-commutative, un argument de
symétrie, qui s'avère fécond pour la déduction de
termes correctifs au secteur de jauge dans les actions effectives. Il
est cependant naturel de poursuivre l'évaluation de l'intégrale de
chemin, afin de déceler l'existence de ces termes dans l'approche
(commutative) de Fradkin et Tseytlin. Écrivons la théorie de jauge
effective comme le résultat de l'intégrale de chemin à l'ordre du
disque.\\

Mukhi a retrouvé l'expression de $\ast_2$ par un calcul direct \`a l'aide d'états
de bord \cite{commutative} pour le terme se couplant à $C^{(6)}$. J'ai
montré par récurrence dans \cite{corrections} comment les
star-produits modifiés d'ordre supérieur apparaissent dans ce
formalisme. J'ai en effet calculé dans l'article \cite{corrections} les termes de
$2p$-forme d'ordre $p$ en $F$ dans l'action de Chern--Simons, de
manière à corriger le couplage au caractère de Chern.
L'organisation du calcul dans le secteur de Ramond repose sur le fait
suivant : les relations d'anticommutation entre les modes zéro des
fermions reproduisent l'algèbre extérieure des formes différentielles
dans l'espace-cible plat.

 $$ \{\psi_0^\mu, \psi_0^\nu\}= \delta ^{\mu\nu},$$
 $$
 \psi^\mu \longleftrightarrow dx^\mu.$$
 C'est cette même identification
  qui est utile pour calculer des invariants topologiques tels que
 l'indice de Witten à partir d'une théorie supersymétrique
 \cite{SUSYMorseW}. Nous devons donc extraire $2p$ modes de fr\'equence nulle $\psi_0$
de fermions et $p$ puissances du champ de jauge, de la quantit\'e suivante
 dans le secteur de Ramond, qui est égale à l'action de
Chern--Simons complète, corrections incluses :
$$\langle C | \exp \left(-\frac{i}{2\pi\alpha'}\int d\tau d\theta
D\phi^\mu A_\mu(\phi) \right) |B\rangle_R.$$ Nous
effectuerons la substitution
$$\langle C |\frac{1}{2}\psi_0^\mu\psi_0^\nu
F_{\mu\nu}|B\rangle_R\longrightarrow -i\alpha ' F$$ à la fin du
calcul.\\

Il y a donc, dans le développement de l'exponentielle en puissances de $F$,
des termes quadratiques, auxquels nous allons d'abord nous intéresser, afin
d'illustrer l'approche et d'initialiser la récurrence qui va suivre :
 $$\frac{1}{2} \left(\frac{i}{2\pi\alpha'}\right)^2\int_0^{2\pi}
 d\sigma_1\int_0^{2\pi} d\sigma_2$$
 $$ \langle C
 |\left(\frac{1}{2}\psi_0^\mu\psi_0^\nu\right)\left(\frac{1}{2}\psi_0^\alpha\psi_0^\beta\right)\sum_{n\geq
 0}\sum_{p\geq 0}\frac{1}{n!}\frac{1}{p!}\tilde{X}^{\lambda_1}\dots
 \tilde{X}^{\lambda_n}\tilde{X}^{\rho_1}\dots
 \tilde{X}^{\rho_p}\partial_{\lambda_1}\dots\partial_{\lambda_n}
 F_{\mu\nu}
 \partial_{\rho_1}\dots\partial_{\rho_p}F_{\alpha\beta}|B\rangle_R.$$
 
 Ces termes quadratiques donnent bien l'expression de $\ast_2$ par sélection des termes
 réguliers avec $n=p$ dans la double somme \cite{commutative}. En
 effet, le propagateur des scalaires est proportionnel à
 $\theta^{\mu\nu}$ dans la limite de Seiberg--Witten,
$$\langle X^\mu(\sigma_1) X^\nu(\sigma_2)\rangle \mapsto
 \frac{\theta^{\mu\nu}}{2\pi\alpha'}\lim_{\epsilon\rightarrow
 0}\left(\frac{1-e^{-\epsilon+i\sigma_{12}}}{1-e^{-\epsilon+i\sigma_{12}}}\right)
 =i(\sigma_{12}-\pi),$$ et les contractions donnent un facteur
 combinatoire de $n!$ qui compte les apppariements, alors qe
 l'intégrale par rapport aux abscisses $\sigma_1$ et $\sigma_2$
 sélectionne les termes avec un nombre pair de dérivées agissant
 sur chaque champ $F$. Le couplage de degré quatre comme forme
 différentielle, et quadratique dans les champs de jauge, se lit
 donc :

\begin{align}
\label{star2}
(S_{CS}+\Delta S_{CS})|_{(2)} &=\sum_{p\geq
0}\frac{(-1)^p(\alpha')^{2p}}{(2p)!}\int_0^{2\pi}\frac{d\sigma}{2\pi}(\sigma_\pi)
^{2p}\prod_{k=1}^{2p}\theta^{\mu_k\nu_k} \d_{\mu_i}\d'_{\nu_i}
F(x)\wedge F(x')|_{x'=x} \\\nonumber{}
&= \sum_{p\geq
0}\frac{(-1)^p}{2^{2p}(sp+1)!}\theta^{\mu_1\nu_1}\dots\theta^{\mu_{2p}\nu_{2p}}\d_{\mu_1}\dots\d_{\mu_{2p}}
F\land \d_{\nu_1}\dots\d_{\nu_{2p}} F \\\nonumber{}
 &=\frac{\sin\left(\frac{1}{2}\d_\mu\theta^{\mu\nu}\d'_\nu\right)}{\frac{1}{2}\d_\mu\theta^{\mu\nu}\d'_\nu} F\land F|_{x'=x} \\\nonumber{}
&= F\ast_2\land F.
\end{align}

 Il s'agit à présent de
 montrer par récurrence comment les opérateurs différentiels de rang
 supérieur peuvent se déduire des termes du développements contenant des
 puissances supérieures de $F$. Le principe de la preuve est une
 représentation graphique des contractions de Wick, dont le facteur de
 symétrie doit venir compenser les factorielles intervenant au
 dénominateur via le développement de la fonction exponentielle. À
 l'ordre $K$ dans le champ de jauge, pour chaque terme contenant des
 puissances des fluctuations de scalaires, nous avons $K$ nuages de
 points bien séparés ; à chaque nuage est associé l'un des tenseurs
 $F$, et le nuage compte autant de points qu'il y a d'opérateurs
 $\tilde{X}^\lambda\partial_\lambda$ agissant sur le tenseur $F$ en
 question. Une contraction de Wick est symbolisée par un trait reliant
 deux points. Nos termes réguliers sont ceux qui sont symbolisés par
 un graphe dans lequel tous les points sont connectés par paires (un
 point ne peut être l'extrémité que d'un seul trait), et où les traits
 relient toujours des nuages différents. Il est instructif de refaire
 explicitement une étape $K=3$, non pas pour initier la récurrence
 mais pour gagner un peu de pratique et fixer les notations.

Il nous reste à compter les graphes permis, afin de connaître  le facteur  de symétrie intervenant dans l'évaluation de la triple somme

$$\(S_{CS}+\Delta S_{CS}\)|_{(4)}=\frac{1}{3!}\sum_{n\geq
0}\sum_{p\geq 0}\sum_{q\geq
0}\(\frac{i}{\l}\)^3\int_0^{2\pi}d\sigma_1\int_0^{2\pi}d\sigma_2\int_0^{2\pi}d\sigma_3$$
$$\langle C|\(\frac{1}{2}\psi^\mu_0\psi^\nu_0\)\(\frac{1}{2}\psi^\alpha_0\psi^\beta_0\)\(\frac{1}{2}\psi^\kappa_0\psi^\tau_0\)×$$
$$\frac{1}{n!}\tilde{X}^{\lambda_1}(\sigma_1)\dots\tilde{X}^{\lambda_n}(\sigma_1)\frac{1}{p!}\tilde{X}^{\rho_1}(\sigma_2)\dots\tilde{X}^{\rho_p}(\sigma_2)\frac{1}{q!}\tilde{X}^{\phi_1}(\sigma_3)\dots\tilde{X}^{\phi_q}(\sigma_3)×$$
$$\d_{\lambda_1}\dots\d_{\lambda_n}
F_{\mu\nu}(x)\d_{\rho_1}\dots\d_{\rho_p}
F_{\alpha\beta}(x)\d_{\phi_1}\dots\d_{\phi_q}
F_{\kappa\tau}(x)|B\rangle_R\pt$$

Nous désignerons les nuages par les lettres grecques provenant des groupes
d'indices. Nous avons donc affaire à trois nuages, notés $\{\lambda\}$,
$\{\rho\}$ et $\{\phi\}$.  Avec des notations adaptées à la formule
ci-dessus, il y a $A+C$ indices dans $\{\lambda\}$, $A+B$ indices dans
$\{\rho\}$, et $B+C$ indices dans $\{\phi\}$ ; nous choisissons A indices du
nuage $\{\lambda\}$ à contracter avec des points du nuage $\{\rho\}$. Nous
effectuons les contractions correspondantes à l'aide du propagateur
$D^{\lambda\rho}(\sigma_{12})$. Il y a $C_{A+C}^A ×(A+B)!/B!$ manières
d'effectuer ces contractions. Il reste $B$ indices dans le nuage $\rho$, à
contracter avec $B$ indices du nuage $\phi$ à l'aide du propagateur 
$D^{\rho\phi}(\sigma_{23})$. Il y a $(B+C)!/C!$ manières d'effectuer ces
contractions. Il ne reste alors que $C$ points libres dans $\{\rho\}$ et
$\{\phi\}$, et les contractions à l'aide du propagateur
$D^{\phi\lambda}(\sigma_{13})$ apportent un dernier facteur de symétrie égal
à $C!$. Nous trouvons donc au numérateur le produit de trois factorielles
qui venaient du développement de l'exponentielle :
$$\frac{(A+C)!}{A!C!}×\frac{(A+B)!}{B!}×\frac{(B+C)!}{C!}× C!.$$
Le
dénominateur, à l'intérieur de la triple somme, est donc réorganisé de la
façon suivante :

$$\(S_{CS}+\Delta S_{CS}\)|_{(4)}=\frac{1}{3!}\sum_{A\geq 0}\sum_{B\geq 0}\sum_{C\geq 0}\(\frac{i}{\l}\)^3\int_0^{2\pi}d\sigma_1\int_0^{2\pi}d\sigma_2\int_0^{2\pi}d\sigma_3$$
$$\langle
C|\(\frac{1}{2}\psi^\mu_0\psi^\nu_0\)\(\frac{1}{2}\psi^\alpha_0\psi^\beta_0\)\(\frac{1}{2}\psi^\kappa_0\psi^\tau_0\)×$$
$$\frac{1}{(C+A)!}\(\tilde{X}^{\lambda_1}\dots\tilde{X}^{\lambda_{C+A}}\)(\sigma_1)\frac{1}{(A+B)!}\(\tilde{X}^{\rho_1}\dots\tilde{X}^{\rho_{A+B}}\)(\sigma_2)\frac{1}{(B+C)!}\(\tilde{X}^{\phi_1}\dots\tilde{X}^{\phi_{B+C}}\)(\sigma_3)×$$
$$\d_{\lambda_1}\dots\d_{\lambda_{(C+A)}} F_{\mu\nu}(x)\d_{\rho_1}\dots\d_{\rho_{(A+B)}} F_{\alpha\beta}(x)\d_{\phi_1}\dots\d_{\phi_{(B+C)}} F_{\kappa\tau}(x)|B\rangle_R\pt$$

Aux ordres supérieurs, il nous faut considérer tous les triangles de
nuages tels que celui que nous venons d'examiner. À l'ordre $K$ dans
les champs de jauge, pour les corrections au
couplage\footnote{Remarquons que la récurrence fonctionne même pour
$K$ supérieur à 5, sans forme de Ramond--Ramond
correspondante. Ce fait sera utilisé au prochain chapitre.}
$C^{(10-2K)}\land F^K$, nous avons $K$ nuages de points numérotés de 1
à $K$. Il y a $P_K:= K! /(2! (K-2) !)$ propagateurs de types différents
$D^{\mu\nu}(\sigma_{ij})$ selon le choix des indices $i$ et $j$ entre
1 et $K$. Ces types sont numérotés de 1 à $P_K$, et nous disons qu'il
y a $N_i$ propagateurs du type numéro $i$, notés $H_i$, dans une contraction.
L'expression du couplage est réorganisée par un facteur de symétrie
constitué de factorielles de nombres de propagateurs, au dénominateur,
donnant
$$\frac{1}{K!}\sum_{N_1\geq 0}\sum_{N_2\geq 0}\dots\sum_{N_{P_K}\geq
  0}\left(\prod_{i=1}^{P_K}\frac{1}{N_i
    !}\right)\left(\prod_{i=1}^{P_K}(H_i)^{N_i}\right)×$$
$$ C\land(\partial\dots \partial F)^K.$$ Les dérivées apparaissent
avec les indices correspondant aux propagateurs, de façon
automatique, ce qui nous dispense d'\'ecrire ces indices.\\


Pour passer de l'ordre $K$ à l'ordre $K+1$, il faut ajouter un nuage
de points et compter les nouveaux graphes permis. Ils contiennent
comme sous-graphes des graphes qui étaient permis à l'ordre $K$. Dans
chacun de ces sous-graphes, appelons $S_i$ le nombre de liens
connectés au sommet numéro $i$, pour $1\leq i\leq K$. Le sommet
numéro $i$ est connecté par $N_{K+1}$ propagateurs au nouveau
sommet (numéro $K+1$), et par $S_i$ propagateurs à des sommets $i$  du
sous-graphe ($S_i$ est le nombre de points du nuage au sommet $i$
reliés à d'autre points du sous-graphe). Disons qu'il y a $P$
types de propagateurs (numérotés) dans le sous-graphe, et pour
tout entier $J\leq P$, appelons $N_J$ le nombre de propagateur du
type $J$. Le facteur de symétrie d'un graphe à $K+1$ sommets est
le produit du facteur de symétrie du sous-graphe par le nombre de
façons de relier le sommet numéro $K+1$ au sous-graphe.
D'après l'hypothèse de récurrence au rang $K$, nous pouvons écrire

$$\sum_{N_{1}}\dots\sum_{N_{P+K}}\left(\(\prod_{j=1}^K\frac{1}{(S_j+N_{P+j})!}\)\times\frac{1}{(N_{P+1}+\dots+N_{P+K})!} \right)\times$$
$$\(C^{N_{P+1}}_{N_{P+1}+\dots +N_{P+K}}C^{N_{P+2}}_{N_{P+2}+\dots+N_{P+K}}\dots C^{N_{P+K-1}}_{N_{P+K-1}+N_{P+K}} \)\times
\prod_{j=1}^K \frac{(S_j+N_{P+j})!}{S_j!}\times$$
$$ \(\frac{\prod_{j=1}^K S_j!}{\prod_{I=1}^P N_I!}\)\pt$$ 
Le premier facteur provient du développement de
l'exponentielle. Le deuxième est le nombre de choix d'un point du
nuage du sommet $K+1$, et de contractions associées. Le facteur
$\prod_{j=1}^K S_j!$ était compensé au rang $K$ à cause du
développement de l'exponentielle, mais doit être pris en compte
ici.

$$\frac{1}{(N_{P+1}+\dots+N_{P+K})!} C^{N_{P+1}}_{N_{P+1}+\dots
+N_{P+K}}C^{N_{P+2}}_{N_{P+2}+\dots+N_{P+K}}\dots
C^{N_{P+K-1}}_{N_{P+K-1}+N_{P+K}}\frac{1}{\prod_{i=1}^P N_i!}=$$
$$\frac{1}{\prod_{i=1}^P N_i!}\frac{(N_{P+K}+N_{P+K-1})!}{N_{P+K}!N_{P+K-1}!}\frac{(N_{P+K}+N_{P+K-1}+N_{P+K-2})!}{(N_{P+K}+N_{P+K-1})!N_{P+K-2}!}\times$$
$$\dots\frac{(N_{p+K}+N_{p+K-1}+\dots+N_{p+1})!}{\dots
N_{p+1}!}\times$$
$$\frac{1}{(N_{P+1}+\dots+N_{P+K})!}=$$

$$\frac{1}{\prod_{i=1}^{P_K+K}N_i!},$$
ce qui termine la démonstration, puisque $P_{K+1}=P_K+K$.\\

\section{Au-delà de la limite de Seiberg--Witten}

\subsection{Contribution du terme symétrique}
 
La limite de Seiberg--Witten, dans laquelle nous avons travaillé
jusqu'à présent, ne laisse subsister que les corrections dérivatives
qui accompagnent chaque paire de dérivées par un tenseur
$\theta$. Or le paramètre $\alpha'$, qui tend vers 0 dans la limite de
Seiberg--Witten, a la même dimension que l'échelle de
non-commutativité $\theta$. Toute une classe de corrections a donc été
effacée par la considération de la limite de
Seiberg--Witten. L'origine des ces corrections sous-dominantes est
claire au vu du propagateur des champs scalaires : c'est le terme
symétrique, avec la métrique de corde ouverte, qui contient le facteur
$\alpha'$. Or, du point de vue de la théorie des champs
non-commutatifs, il est important de n'avoir qu'un terme
antisymétrique puisque c'est ce terme qui donne lieu au commutateur
des coordonnées et mesure la non-commutativité ; l'interprétation du
terme symétrique du propagateur, et des corrections sous-dominantes
induites, n'est donc pas fournie par un argument de théorie quantique
des champs tel que celui donné dans \cite{SW}. Les corrections
dérivatives ne seront donc pas calcul\'ees par la vertu d'une symétrie
de dualité. Notre approche directe en théorie {\emph{commutative}} des
cordes, pour lourde qu'elle soit, constitue un recours naturel. Il
s'avère, en y regardant de plus près, que nous avons fait l'essentiel
du travail nécessaire dans l'évaluation par récurrence des facteurs de
symétrie. Si nous utilisons le même propagateur régularisé, nous
aurons encore usage des résultats de cette récurrence.\\

 Nous nous intéressons donc au propagateur
$$\langle X^\mu(\sigma_1) X^\nu(\sigma_2)\rangle=:
D^{\mu\nu}(\sigma)=\theta^{\mu\nu} \log\left( \frac{1- e^{-\epsilon
+i\sigma}}{1- e^{-\epsilon -i\sigma}} \right)+\alpha' G^{ab}\log|1-
e^{-\epsilon +i\sigma}|^2,$$ dont l'invariance par translation a été
prise en compte en définissant
$$ \sigma:= \sigma_1-\sigma_2.$$ Pour la partie régulière du couplage
à $C^{(6)}$, nous pouvons écrire le terme d'ordre $2n$ en dérivées, en
faisant entrer tous les opérateurs de dérivées dans un seul opérateur
différentiel agissant sur le produit $F\land F$ :
 
$$ \frac{{\alpha '}^n}{n !}\int_0^{2\pi} d\sigma \left(  \frac {\theta^{\mu\nu}}{2\pi\alpha '}  \log\left( \frac{1- e^{-\epsilon +i\sigma}}{1- e^{-\epsilon -i\sigma}}  \right)+G^{\mu\nu}\partial_\mu\partial'_\nu\log|1- e^{-\epsilon +i\sigma}|^2 \right) ^n F(x)\land F(x')|_{x'=x}.$$

Nous avons à effectuer la somme sur l'ordre $n$, que nous faisons passer dans l'intégrand, et qui donne l'exponentielle d'un opérateur différentiel :
$$\int_0^{2\pi} \sum_{n\geq 0} \frac{{\alpha '}^n}{n !} d\sigma \left(
\frac {\theta^{\mu\nu}}{2\pi\alpha '} \log\left( \frac{1- e^{-\epsilon
+i\sigma}}{1- e^{-\epsilon -i\sigma}}
\right)+G^{\mu\nu}\partial_\mu\partial'_\nu\log|1- e^{-\epsilon
+i\sigma}|^2 \right) ^n$$
$$=\int_0^{1} d\tau|2\sin(\pi\tau) |^{2t}\exp((ia\pi)(2\tau-1)).$$ Cet
opérateur est donc construit à partir de deux opérateurs différentiels
d'ordre 2, en provenance directe des parties antisymétrique et
symétrique du propagateur. La partie symétrique restitue les
dimensions anormales qui avaient été évitées par la considération de
la limite d'échelle :
$$ a := \partial_\mu \theta^{\mu\nu} \partial_\nu,$$
$$ t :=\alpha'\partial_\mu G^{\mu\nu} \partial_\nu.$$

 L'opérateur codant les corrections dérivatives au couplage $F\land
 F$, noté $\tilde{\ast}_2$ possède la bonne limite $\ast_2$ parce que
 les paramètres de cordes ouvertes restent fixés quand $\alpha'$ tend
 vers 0. L'opérateur $t$ disparaît donc simplement, et la forme
 précédemment identifiée pour $\ast_2$ est récupérée grâce à une
 identité de fonctions Gamma :

$$\tilde{\ast}_2:= \int_0^{1} d\tau|2\sin(\pi\tau) |^{2t}\exp((ia\pi)(2\tau-1))=\frac{\Gamma(1+2t)}{\Gamma(1-a+t)\Gamma(1-a+t)}.$$
$$ \mathrm{lim}_{\alpha' \to 0}\left(\tilde{\ast}_2\right)=
\frac{1}{\Gamma(1-a)\Gamma(1+a)}=\frac{\sin(a\pi)}{a\pi}.$$ De plus,
ce résultat reproduit l'amplitude de diffusion évaluée par Liu et
Michelson dans \cite{startrekIII}. Seul l'accord au premier ordre en
$G$ avait été vérifié par Mukhi et Suryanarayana dans \cite{MSbeyond},
qui n'avaient pas déduit l'expression entière de $\tilde{\ast}_2$ par
la méthode de la fonction de partition, mais avaient postulé
l'existence d'un tel opérateur compatible avec les résultats des
amplitudes de diffusion. Les conséquences sur la description
non-commutative qu'ils ont tirées de ce raisonnement sont correctes,
et je peux préciser leurs attentes sur l'invariance de jauge
non-commutative, grâce aux corrections aux plus grands ordres en $F$,
que j'ai également étudiés. C'est ici que le raisonnement par
récurrence exposé plus haut intervient, pour donner les opérateurs
différentiels suivants :

 \begin{eqnarray*}  
\tilde{\ast}_3 &=& \sum_{A,B,C\geq
0}\frac{1}{A~!}\frac{1}{B~!}\frac{1}{C~!}
\int_0^{2\pi}\frac{d\sigma_1}{2\pi}\int_0^{2\pi}\frac{d\sigma_2}{2\pi}\int_0^{2\pi}\frac{d\sigma_3}{2\pi}\;
Q_{12} ^A Q_{23}^B Q_{31}^C\\
&=&\int_0^{1}d\tau_1\int_0^{1}d\tau_2\int_0^{1}d\tau_3\exp\{ia_{12}\pi(2\tau_{12}-\epsilon(\tau_{12}))
+2t_{12}\log|2\sin(\pi\tau_{12})|\\ & &
\;\;\;\;\;\;\;\;\;\;\;\;\;\;\;\;\;\;\;\;\;
\;\;\;\;\;\;\;\;\;\;\;\;\;\;\;\;\;
+ia_{23}\pi(2\tau_{23}-\epsilon(\tau_{23}))+2t_{23}\log|2\sin(\pi\tau_{23})|\\
& &\;\;\;\;\;\;\;\;\;\;\;\;\;\;\;\;\;\;\;\;\;
\;\;\;\;\;\;\;\;\;\;\;\;\;\;\;\;\; +
ia_{31}\pi(2\tau_{31}-\epsilon(\tau_{31}))+2t_{31}\log|2\sin(\pi\tau_{31})\},
\end{eqnarray*}
et plus généralement

$$\tilde{\ast}_p=\int_0^1 d\tau_1\dots\int_0^1 d\tau_p \prod_{1\leq
i<j\leq p}\exp\left\{i\pi
a_{ij}(2\tau_{ij}-\epsilon(\tau_{ij}))\right\}.$$

\subsection{Interprétation non-commutative}

Au cours de ce chapitre, nous avons décrit des situations
correspondant à trois cases du tableau suivant, correspondant aux 
réf\'rences indiquées :

\begin{table}[htbp]
  \centering
  \begin{tabular}{|c|c|c|}
    \hline
    {} & description commutative & description non-commutative \\\hline
{limite de Seiberg--Witten} & \cite{FT,Andreev--Tseytlin} & \cite{Liu,Mukhi} \\ \hline 
{au-delà} & \cite{MSbeyond, Grangebeyond} & ? \\ \hline
  \end{tabular}
  \caption{Versatilité et limite de Seiberg--Witten}
  \label{tab:e-uno}
\end{table} 
Il nous manque encore une interprétation commutative de la déformation
des produits $\ast_n$ au-delà de la limite de Seiberg--Witten. Il est
tentant de penser, comme dans \cite{MSbeyond}, que la théorie de jauge
non-commutative reste un outil valide\footnote{C'est en particulier
encourageant pour les développements moins formels dans lesquels
l'échelle de non-commutativité n'est plus invariante par
translation.}. L'argument topologique d'invariance des orbites de
jauge vis-à-vis du choix de description reste valide, et nous héritons
d'une déformation de l'application de Seiberg--Witten (couplage à
$C^{(8)}$), après avoir calculé des corrections dérivatives (couplage
à $C^{(6)}$). Nous parcourons en quelque sorte en sens inverse le
chemin qui avait été suivi pour extraire les prédictions
non-commutatives dans la limite de Seiberg--Witten.\\

J'ai confirmé dans \cite{Grangebeyond} cette déformation de la
 théorie de jauge non-commutative anticipée dans \cite{MSbeyond}
 à partir de l'expression de $\tilde{\ast}_2$.  Les calculs de
 produits d\'eform\'es au-delà de la limite de Seiberg--Witten nous
 invitent à redéfinir le tenseur du champ de jauge, ainsi que les
 transformations de jauge, afin de les rendre cohérentes avec une
 prescription d'intégration le long d'une ligne de Wilson
 déformée. Les expressions naturelles sont
$$\hat{F}_{\mu\nu}=\d_\mu\hat{A}_\nu-\d_\nu\hat{A}_\mu-i[\hat{A}_\mu,\hat{A}_\nu]_{\tilde{\ast}},$$
$$\delta\hat{A}_\mu=\d_\mu\hat{\lambda}-i[\hat{A}_\mu,\hat{\lambda}]_{\tilde{\ast}},$$
 où $\tilde{\ast}$ est le star-produit déformé, à l'aide duquel une
 ligne de Wilson peut être écrite, de manière à donner les
 opérateurs déformés $\tilde{\ast}_p$ dans un développement en
 puissances du champ de jauge.  Conformément aux conjectures de
 ~\cite{MSbeyond}, ces expressions assurent que la relation de
 récurrence entre les différents opérateurs différentiels est
 équivalente à l'invariance de jauge. Nous nous souvenons en effet que
 la relation de récurrence écrite dans ~\cite{Liu} entre les $\ast_n$
 d'ordres $n$ distincts était dans la limite de Seiberg--Witten
 équivalente à la symétrie de jauge. Nous allons redéfinir les
 opérateurs modifiés à partir d'observables se transformant suivant la
 loi
$$O_i\mapsto -i[O_i,\hat{\lambda}]_{\tilde{\ast}},$$
et intervenant dans les couplages de Ramond--Ramond.\\

Écrivons les couplages de Ramond--Ramond pour un mode de Fourier
$k$. Leur forme implique des insertions d'objets $O_i$ du type
$(\theta -\theta\hat{F}\theta)^{\mu\nu}$ se transformant par symétrie
de jauge suivant la loi

$$O_i\mapsto -i[O_i,\hat{\lambda}]_{\tilde{\ast}},$$
ce qui invite à reproduire avec nos produits déformés le raisonnement de
\cite{Liu} vérifiant l'invariance de jauge à partir de l'expression des
opérateurs $\ast_n$ : 
$$Q(k)=\sum_{m\geq 0} Q_m(k),$$

$$\,\,\,\,\,\,\,Q_m(k)=\frac{1}{m~!}(\theta\d)^{\mu_1}\dots(\theta\d)^{\mu_m}\langle
O_1,\dots,O_p,\hat{A}_{\mu_1},\dots,\hat{A}_{\mu_m}
\rangle_{\tilde{\ast}_{p+m}}.$$
\noindent{Une transformation} de jauge agit en effet sur les termes de
cette dernière expression suivant la loi

\begin{eqnarray*}
\delta
Q_m&=&-\frac{i}{m~!}(\theta\d)^{\mu_1}\dots(\theta\d)^{\mu_m}\sum_{i=1}^p\langle
O_1,\dots,[\hat{\lambda},O_i]_{\tilde{\ast}},\dots,
O_p,\hat{A}_{\mu_1},\dots,\hat{A}_{\mu_m} \rangle_{\tilde{\ast}_{p+m}
}\\
&&-\frac{i}{m~!}(\theta\d)^{\mu_1}\dots(\theta\d)^{\mu_m}\sum_{i=1}^m\langle
O_1,\dots,
O_p,\hat{A}_{\mu_1},\dots,[\hat{\lambda},\hat{A}_{\mu_i}]_{\tilde{\ast}},\dots,\hat{A}_{\mu_m}
\rangle_{\tilde{\ast}_{p+m} } \\
&&+\frac{1}{(m-1)~!}(\theta\d)^{\mu_1}\dots(\theta\d)^{\mu_{m}}\langle
O_1,\dots,O_p,\hat{A}_{\mu_1},\dots,\hat{A}_{\mu_{m-1}},\d_{\mu_{m}}\hat{\lambda}
\rangle_{\tilde{\ast}_{p+m}},
\end{eqnarray*} 
\noindent{de} sorte que la variation des courbures de jauge dans $Q_m$
est compensée par la variation du potentiel dans $Q_{m+1}$, selon le
schéma connu de \cite{Liu}. Nous avons montré la cohérence interne
entre la déformation de la prescription $L_\ast$, celle de la courbure
de jauge et celle des transformations de jauge. Le chapitre suivant
aborde la question naturelle du couplage au champ de tachyon, pour
lequel les résultats du présent chapitre facilitent le calcul
 des intégrale de chemin.


\chapter{Tachyons, actions effectives et transitions de phase}

\chapterprecistoc{}

 \chapterprecishere{Dans ce chapitre, nous considérons l'action
effective pour le champ de tachyon des cordes ouvertes. Nous
retrouvons les produits déformés rencontrés dans le secteur de
jauge. Ils sont compatible avec l'universalité du potentiel des
tachyons. Le couplage des tachyons au champ de jauge, avec les
corrections dérivatives que nous calculons, nous invite à rapprocher
les résultats de ce chapitre et ceux du précédent par
l'intermédiaire des superconnexions.}

 \section{Motivation}

\subsection{Condensation de tachyons}

Sen a conjecturé l'existence d'une transition de phase entre un vide
instable présentant cordes ouvertes, D-branes et scalaire tachyonique,
et un vide stable de cordes fermées.  Nous avons mentionné plus haut
l'annihilation de D-branes, afin de motiver l'introduction de la
K-théorie pour décrire les charges de Ramond--Ramond. Nous sommes
également habitués à considérer une D-brane comme un état lié de
D-branes de dimension plus petite. Les solitons tachyoniques sur le
volume d'univers font également l'objet d'une conjecture, qui les
assimile à des D-branes de dimension plus petite.\\

Ces conjectures donnent lieu à une intense activité en théorie des cordes,
et ont pour enjeu rien de moins que la formulation non-perturbative de la
théorie des cordes, qui fait l'objet de la seconde quantification ou
{\emph{théorie des champs de cordes}}. Une notion de non-commutativité a été
associée à la théorie des champs de cordes par Witten en 1986 dans l'article
fondateur \cite{SFTNC}. Elle tient aux deux ordres cycliques distincts qui
peuvent être assignés aux trois cordes ouvertes se rencontrant en un vertex.


\subsection{Tachyons et bords}
Nous avons utilisé l'invariance de jauge au chapitre précédent comme
guide dans l'écriture des couplages de Ramond--Ramond dans la
description non-commutative. Le problème de la transformation de
Seiberg--Witten est issu du caractère topologique des orbites de
jauge. Or le spectre des cordes ouvertes contient un scalaire, le
tachyon $T$, dont la dynamique tient à la structure non-perturbative
de la théorie des cordes, problème universel qui fait l'objet de la
conjecture de Sen~\cite{Senreview}.\\

 L'objet de ce chapitre est d'abord l'extension des techniques
 précédentes au secteur tachyonique, ce qui impose de considérer
 l'intégrale de chemin
$$Z[T]=\int DX \,e ^{-S_\Sigma -\int_{\partial\Sigma} T(X(\tau))},$$
avec le même fond de cordes fermées que précédemment, et en
particulier avec un champ antisymétrique constant.  Nous
retrouverons les star-produits et leur déformation par la métrique,
déformation qui pourra être entièrement contenue dans une modification
de la loi de produit, sans remettre en cause la structure de l'action
effective. Formellement, le caractère scalaire du champ de tachyons,
par opposition au caractère gradué des champs de jauge, autorise des
puissances arbitrairement élevées du champ à contribuer à l'action
effective. Les élégantes structures identifiées lors du calcul des
corrections dérivatives dans le secteur de jauge, donnent lieu à une
extrapolation à des degrés arbitrairement élevés, qui ont inspiré à
 Wyllard~\cite{Wyllard} une identité de fonctions Gamma d'Euler, qui
apparaît dans le contexte de l'évaluation des potentiels effectifs
de tachyons. Mes calculs montrent qu'il ne s'agit pas d'un accident,
et que le potentiel quadratique (exactement soluble):
$$ T(X)=a+u_{\mu\nu} X^\mu X^\nu,$$ 
où $a$ et $u_{\mu\nu}$ sont un scalaire et un tenseur symétrique constants, correspond \`a l'identit\'e sugg\'er\'ee par Wyllard.\\

\section{Non-commutativité et condensation de tachyons}
Les travaux présentés dans ce chapitre ont été motivés par des
articles de Cornalba \cite{tachyonCornalba} et Okuyama
\cite{Okuyama}. L'article \cite{tachyonCornalba} teste la
conjecture de Sen sur des calculs de tension de produits de
désintégration dans la limite de Seiberg--Witten, et nous présenterons
une conjecture de Sen à l'aide de cet exemple.  L'article
\cite{Okuyama} calcule le potentiel de tachyons, et il est
naturel après les calculs du secteur de jauge, de se demander comment
les produits modifiés peuvent s'insérer dans ce résultat.\\

La conjecture de Sen est d'essence non-perturbative, puisqu'elle relie
différents points critiques de la théorie des champs de cordes. La valeur
du potentiel de tachyons doit refléter les tensions des D-branes de
différentes dimensions impliquées dans la condensation de tachyons. La
méthode de Cornalba repose sur la fonction $\beta$ des couplages insérés sur le
bord de la surface d'univers, à l'ordre du disque. Shatashvili a argumenté
\cite{problemsShatashvili} en faveur d'une expression de l'action effective pour les cordes
ouvertes tenant compte de la renormalisation des couplages sur le bord. Pour
une intégrale de chemin de la forme
$$Z=\int DX\, e^{-\int_\Sigma\left(dX \land\ast dX
+B\right)-\int_{\partial\Sigma} d\tau T[X(\tau)]},$$
la valeur du champ de
tachyon en chaque point du bord est un couplage, et l'1action effective
s'écrit
$$S[T]=\left (1-\int dx \,\beta[T(x)]\frac{\delta}{\delta
    T(x)}\right)Z[T].$$
    Dans la limite d'un fort champ antisymétrique, les dimensions anormales
  disparaissent, comme dans le développement en produit d'opérateurs de
  deux ondes planes que nous avons écrit au moment de la discussion de la
  limite de Seiberg--Witten. La fonction $\beta$ du champ de tachyon est donc
  simplement linéaire dans le champ :
  
$$\beta[T(x)]= -T(x).$$ Ainsi l'action effective s'écrit, en fonction de
  l'opération de trace sur l'espace de Hilbert évoquée au chapitre
  précédent :
  $$ a:= \frac{1}{\sqrt{2\theta}}(x^1+ix^2),
  \;\;\;\;\;\;\;\;\;\;\;\;\;a^\dagger:= \frac{1}{\sqrt{2\theta}}(x^1-ix^2),$$
$$S[T]=2\pi T_{25}\, {\mathrm{Tr}}[(1+T) e^{-T}].$$
  
Pour la donnée du champ antisymétrique $B$, un seul bloc de taille
deux d'une matrice antisymétrique est considéré, et les configurations
de tachyon insérées sur le bord ne dépendent que des coordonnées
\footnote{Dans toutes les actions effectives, nous avons omis un
facteur égal au volume de l'espace transverse aux directions 1 et 2.}
associées aux directions 1 et 2. On s'attend donc à obtenir une
D23-brane comme produit de la condensation de tachyons. Le problème
est donc réduit à montrer que le minimum de $S$ coïncide avec la
tension d'une D23-brane. Pour le modèle exactement soluble
$$T(x):= a+ u B\left((x^1)^2+(x^2)^2\right),$$
que nous reprendrons plus
loin, la trace peut être évaluée,
$$
S=\left( 1-a\frac{\partial}{\partial a} -u\frac{\partial}{\partial a }
\right)\left( 2\pi T_{25} Tr(e^{-a-u(2a^\dagger a+1)})\right),$$
par une somme de série géométrique.
Un point
stationnaire $a^\ast(u)$ par rapport au terme constant $a$ peut ensuite être écrit
en fonction de $u$. La limite de la valeur
$$S( a^\ast(u), u)=2\pi T_{25} \frac{1}{1-e ^{-2u}}\exp\left( \frac{2u}{e^{2u}-1}\right)$$
lorsque $u$ tend vers l'infini, est bien $T_{23}$, ce
qui confirme sur un fond de tenseur antisymétrique, et par des méthodes de
théorie non-commutative, un résultat de \cite{KMM1}.

\section{Un \emph{toy model} : les cordes $p$-adiques}
\subsection{Cordes $p$-adiques}
  Décrivons tout d'abord la théorie des cordes $p$-adiques, et
 expliquons en quoi elle constitue un \emph{toy model} de la théorie des
 cordes. Pour une introduction plus complète, le rapport de Brekke
 et Freund \cite{Brekke--Freund} fait autorité. Soit un nombre
 premier $p$. On lui associe le corps des nombres $p$-adiques
 $\mathbf{Q}_p$ en complétant le corps des nombres rationnels pour la
 norme $p$-adique, définie par $|x|_p=p^{-n}$, où $p^n$ est la plus
 grande puissance de $p$ divisant le nombre $x$.\\

 L'utilisation d'un corps de nombres non-archimédien n'est pas qu'une
 fantaisie : elle s'avère féconde pour au moins deux raisons. D'abord,
 il existe remarquablement peu de façons de faire entrer les
 nombres $p$-adiques dans la théorie des cordes bosoniques.  De plus,
 cette introduction aboutit à une solution exacte, le calcul de la
 fonction de partition des tachyons de la corde $p$-adique. Concernant
 la multiplicité des modèles $p$-adiques, on peut imaginer un
 choix entre les nombres complexes et les nombres $p$-adiques pour les
 coordonnées de la surface d'univers, les coordonnées de
 l'espace-temps, l'action de la surface d'univers et les amplitudes de
 diffusion, soit un surplus de quinze nouvelles théories pour toute
 théorie des cordes. Ce serait un véritable boîte de Pandore. Or il
 s'avère que le seul endroit dans lequel les nombres $p$-adiques
 peuvent venir prendre la place des nombres complexes de manière
 cohérente\footnote{c'est-à-dire donnant des amplitudes
 méromorphes, invariantes de Moebius.}, est le bord de la surface
 d'univers. Il n'y a donc qu'un seul modèle $p$-adique possible, et il est tout
 à fait adapté à notre problème concernant l'action effective des
 champs scalaires sur les D-branes, puisque la corde $p$-adique
 possède un scalaire dans son spectre, et des bords sur sa surface
 d'univers.\\

Jusqu'à quel point les résultats habituels peuvent-ils être récrits
via de simples changements de notations ? Les notions de mesure,
intégration, changements de variables, transformation de Fourier,
trouvent des analogues sur le corps des nombres $p$-adiques, cependant
la notion de dérivée ne se généralise pas, à cause de l'absence
d'ordre total qui interdit de comparer les valeurs d'une fonction ``à
deux instants $p$-adiques successifs'', à moins de considérer un modèle discret de
l'espace sur lequel nous travaillons. C'est ce qu'a fait Zabrodin
\cite{Zabrodin}, en réalisant la surface d'univers à l'ordre du
disque, munie d'un bord $p$-adique, comme un arbre infini, dans lequel
chaque noeud réunit $p+1$ branches. Le bord de la surface d'univers
n'est autre que l'ensemble (projectif) des feuilles de cet arbre, le
développement $p$-adique de chaque feuille étant dicté
algorithmiquement par le chemin qui la relie à la racine de
l'arbre. Les reparamétrages de la surface d'univers engendrent dans
cette description des changements de racine pour l'arbre, et des changements de
coordonnées $p$-adiques sur le bord. Un tel arbre est connu
sous le nom d'arbre de Bruhat--Tits.\\


 L'action de la corde bosonique
comportant des dérivées, elle doit être traduite dans le langage
discret adapté à la structure d'arbre. Avec une numérotation des
sommets et la notation $j(i)$ désignant un sommet numéro $j$ voisin du
sommet numéro $i$, nous écrivons sans surprise
$$\int_\Sigma d^2\sigma \partial_\alpha X^\mu \partial^\alpha X_\mu
\to
\sum_i\sum_{j(i)}\left(X^\mu(j(i))-X^\mu(i))(X_\mu(j(i))-X_\mu(i)\right).$$
L'intégrale de chemin par rapport aux degrés de liberté portés par
l'intérieur du disque peut être effectuée, ce qui fournit une action
non-locale sur le bord $p$-adique. Les amplitudes de diffusion de
tachyons s'obtiennent soit en calculant des intégrales de chemin avec
des conditions de bord de Neumann et l'action discrète
bidimensionnelle ci-dessus, soit par une méthode du col à partir de
l'action non-locale sur le bord. Les deux méthodes
\cite{Frampton,FO,BFW} donnent au tachyon $\phi$ une dynamique dictée
par le lagrangien effectif suivant :
$$\mathcal{L}=-\frac{1}{2}\phi
p^{-\frac{1}{2}\square}\phi+\frac{1}{p+1}\phi^{p+1}.$$

\subsection{Coupler les cordes $p$-adiques à  un fond de tenseur antisymétrique}

L'action de la surface d'univers qui conduit au lagrangien effectif
exact exposé plus haut ne comprend qu'un terme, couplant les
coordonnées discrètes sur l'arbre au tenseur symétrique qu'est la
métrique de l'espace-temps. La question de l'inclusion du champ de
fond antisymétrique est naturelle ; se fondant sur l'expérience
acquise dans l'étude des solitons non-commutatifs dans le cadre
archimédien, Ghoshal a conjecturé \cite{Ghoshal} que l'influence d'un fond de tenseur
antisymétrique constant était convenablement décrite par une
déformation du produit des champs, sur le modèle du produit de Moyal
longuement étudié plus haut :
$$\mathcal{L}=-\frac{1}{2}\phi\ast
p^{-\frac{1}{2}\square}\phi+\frac{1}{p+1}(\ast\phi)^{p+1}.$$ Si cette
action est un honnête point de départ pour l'étude d'une classe de
solitons $p$-adiques non-commutatifs en théorie des champs, il n'est
pas certain {\emph{a priori}} qu'elle provienne de la théorie des
cordes $p$-adiques via une action de surface d'univers complétant le
couplage de la métrique à l'arbre de Bruhat--Tits. J'ai proposé dans
\cite{padicGrange} une telle action de surface d'univers, et obtenu le
star-produit attendu, en adaptant la démarche de Seiberg et Witten à
un bord dépourvu de relation d'ordre.\\

Il est naturel de se demander comment coupler le champ antisymétrique
au bord, car on peut interpréter ce champ comme un champ magnétique de fond,
ce qui est indifférent au niveau de la théorie des cordes. Notre
démarche sera donc à rapprocher de l'action non-locale obtenue en
intégrant les degrés de liberté liés aux sommets intérieurs de
l'arbre. Les ingrédients spécifiques de la corde $p$-adique sont les
extensions quadratiques qui se subsituent au demi-plan complexe
supérieur habituel, avec certaines de leurs fonctions spéciales.\\

  Comme nous l'avons signalé plus haut, en l'absence d'une composante
  antisymétrique dans les champs de fond, les amplitudes de diffusion
  à $N$ particules, notées $A_N$, induites par le schéma discret du
  laplacien sur l'arbre de Bruhat--Tits,

$$A_N(k_1,\dots, k_N)=\int DX\exp(-S[X]+i\sum_{i=1}^N k_{i,\mu}
 X^\mu(u_i))=\prod_{1\leq i < j \leq N} |u_i-
 u_j|_p^{k_{i,\mu}g^{\mu\nu}k_{j,\nu}},$$ sont reproduites pour tout
 $N$ par une action non-locale unidimensionnelle sur le bord
 $p$-adique. Cette action a été écrite par Zhang \cite{Zhang} :
$$S_{\mathrm{sym}}[X]=\int_{\mathbf{Q}_p}du \,(g_{\mu\nu}
X^\mu (-u) |u|_p X^\nu(u)).$$ La structure de cette action, à une
transformation de Fourier près, nous rappelle fortement l'intégrale
obtenue plus haut par le théorème de Stokes dans le cas archimédien en
couplant le tenseur antisymétrique. C'est un signe encourageant, qui
nous fait penser qu'il doit bien exister une antisymétrisation de
l'action de Zhang possédant encore certaines des propriétés du terme
archimédien à l'origine du produit de Moyal. Cependant, pour des
raisons d'antisymétrie précisément, une antisymétrisation minimale
donne un résultat identiquement nul d'après les règles de changement
de variables dans l'intégrale sur les nombres $p$-adiques :
$$\int_{{\mathbf{Q}_p}}du\, B_{\mu\nu} X^\mu (-u) |u|_p
X^\nu(u)=0.$$ C'est ennuyeux, mais il semble que la recette consistant
à définir l'analogue de la dérivée par des polynômes dans l'espace de
Fourier, pour intégrer ensuite et trouver une action valide d'après le
théorème de Plancherel, échoue pour une structure d'indices
antisymétrisée une seule fois, comme si l'absence d'ordre sur le bord
conspirait pour effectuer une sorte d'opération de moyenne interdisant
le couplage à $B_{\mu\nu}$. Il faut donc jouer une seconde fois, ce
qui est possible. En effet, nous pouvons antisymétriser en les points
d'insertion des tachyons sur le bord. Autrement dit, c'est d'une
fonction signe que nous avons besoin. Une telle fonction existe dans
le contexte $p$-adique, mais par sur le corps de scalaires $p$-adiques
lui-même (qui n'est pas ordonné). Elle existe sur une extension
quadratique de ${\mathbf{Q}}_p$, ce qui nous rappelle le caractère
bidimensionnel de la surface d'univers. Le disque archimédien est en effet
conformément équivalent au demi-plan complexe, extension quadratique
de la droite réelle qu'est le bord.\\

Considérons une extension quadratique du corps des nombres $p$-adiques
correspondant à un nombre $\tau$ non carré. La fonction signe
$\epsilon_\tau$ correspondante est définie sur l'extension
$\mathbf{Q}_p$ comme suit (définitions et propriétés des fonctions
signe sont prises dans le cours d'arithmétique de Serre \cite{Serre}),
à l'aide de la notion de conjugaison dans l'extension quadratique :
elle prend la valeur $+1$ sur les nombres qui peuvent s'écrire comme
le produit de deux nombres conjugués dans $\mathbf{Q}_p(\sqrt{\tau})$,
et la valeur $-1$ partout ailleurs. Elle a les propriétés
multiplicatives habituelles. La propriété qu'il nous faut pour
restaurer l'alternance de signe attachée à la relation d'ordre s'écrit
$$\epsilon_\tau(x-y)=-\epsilon_\tau(y-x),$$ c'est-à-dire que le signe
attribué au nombre $-1$ dans $\mathbf{Q}_p(\sqrt{\tau})$ doit être
négatif. Or ce n'est pas le cas pour toutes les extensions du corps
des nombres $p$-adiques. Nous ferons dorénavant l'hypothèse que
l'extension est choisie de manière à vérifier cette propriété, et nous
supprimons l'indice $\tau$. Le caractère {\emph{ad hoc}} de ce choix
devrait être éclairci par une prescription bidimensionnelle pour le
couplage au champ de fond antisymétrique. Une telle prescription
serait l'analogue du travail de Zabrodin sur l'arbre de Bruhat--Tits,
alors que nous n'abordons que le travail de Zhang sur les couplages
aux feuilles. Toujours est-il que la seule propriété du
paramètre $\tau$ que nous utilisons est
$$\epsilon_\tau(-1)=-1.$$ 
L'action unidimensionnelle sur le
bord $p$-adique est finalement
$$S[X]:=\int_{{\mathbf{Q}}_p} du\left( g_{\mu\nu} X^\mu (-u)
|u|_p X^\nu(u)+B_{\mu\nu} X^\mu (-u) |u|_p\epsilon_\tau(u)
X^\nu(u)\right).$$ Cette proposition a également été émise
également par Ghoshal et Kawano\cite{Ghoshal--Kawano}, qui ont
développé davantage les questions de symétrie.\\

Notre prescription suffit à calculer les modifications induites par
 le terme antisymétrique sur les amplitudes $A_N$.  Nous
 introduisons les notations habituelles $G$ et $\Theta$ pour les
 parties symétrique et antisymétrique de l'inverse de $\delta+B$:
  $$G^{\mu\nu}:=
\(\frac{1}{\delta+B}\)^{\mu\alpha}g_{\alpha\beta}\(\frac{1}{\delta-B}\)^{\beta\nu},$$
 $$ \Theta
 ^{\mu\nu}:=-\(\frac{1}{\delta+B}\)^{\mu\alpha}B_{\alpha\beta}
 \(\frac{1}{\delta-B}\)^{\beta\nu}.$$ Le propagateur est
 régularisé comme dans~\cite{Zhang} par un paramètre $s$. Par
 bonheur nous avons un arsenal de fonctions Gamma $p$-adiques
 modifiées $\hat{\Gamma}_p$ incluant notre fonction signe :
$$\Delta_{reg} ^{\mu\nu}(x-y):=\int_{{\mathbf Q}_p} du \,e
^{i(x-y)u}\(G
^{\mu\nu}\frac{1}{|u|^{1+s}}+\Theta^{\mu\nu}\frac{\epsilon(u)}{|u|^{1+s}}\).$$

$$\hat{\Gamma}_p(s):=\int_{{\mathbf{Q}}_p}du\, e
 ^{i2\pi[u]}\epsilon_\tau(u)|u|^{s-1}.$$ 
$$\Delta_{reg} ^{\mu\nu}(x-y)= |x-y|^{-s}\Big{(}G^{\mu\nu}\Gamma_p(s)
+ \Theta^{\mu\nu}\epsilon(x-y)\hat{\Gamma}_p(s)\Big{)}.$$
 La
 partie finie dans un développement près de $s=0$ est extraite via la formule

$$\hat{\Gamma}_p(s) =\sqrt{\epsilon_\tau(-1)} p^{s-1/2}.$$ Le
propagateur que nous utiliserons pour l'évaluation des amplitudes
dans une approximation de col s'écrit enfin  
$$\Delta^{\mu\nu}(x-y)=G^{\mu\nu}\log |x-y|
+\frac{i}{\sqrt{p}}\Theta^{\mu\nu}\epsilon_\tau(x-y).$$ 

Les techniques habituelles donnent lieu aux phases qui sont la
signature dans l'espace de Fourier des produits de Moyal :

$$A_N(k_1,\dots,k_N)=\prod_{1\leq i < j\leq N} e ^{\frac{i}{2\sqrt{p}}
  k_{i,\mu} \Theta^{\mu\nu} k_{j,\nu}\epsilon(u_i-u_j)} |u_i-u_j|^{
  k_{i,\mu} G^{\mu\nu} k_{j,\nu}}.$$ Ceci légitime l'étude des
  solitons $p$-adiques non-commutatifs par optimisation de l'action
  effective écrite avec le produit
$$\ast:=\exp\left(
\frac{i}{2\sqrt{p}}\d'_\mu\Theta^{\mu\nu}\d_\mu\right)|_{x'=x}.$$

\section{Dictionnaire pour l'évaluation de l'action effective pour les scalaires }

 En lieu et place de l'intégrale de chemin correspondant au secteur
 de jauge, considérons l'intégrale de chemin correspondant à
 l'action effective du champ de tachyons. Le nouveau terme d'action de la
 surface d'univers n'est autre que l'intégrale du champ scalaire le
 long du bord:

 $$S_{\d\Sigma}= \int_{\d\Sigma} T(X(\tau))d\tau.$$ 
Il est naturel de
se demander comment coupler le tachyon à la théorie
non-commutative lorsque le fond de cordes fermées contient un champ
magnétique constant. Néanmoins, notre expérience des calculs
d'action effective nous fournit à bon marché des résultats
concernant l'intégrale de chemin
$$Z[T]=\int DX \,e^{-S_\Sigma[X]-S_{\d\Sigma}[X]},$$ sans qu'il soit besoin
d'invoquer la dualité géométrique exposée dans le cadre de la
théorie de jauge. Nous reviendrons plus loin
à la question du couplage entre les tachyons et la théorie de
jauge.\\

Il se trouve que le calcul de cette intégrale de chemins est un
exercice proche de celui consistant à faire émerger les star-produits
de la théorie des cordes commutative (nous travaillons avec le même
fond de cordes fermées, et des conditions de Neumann dans toutes les
directions). Bien sûr nous n'avons pas, dans cette théorie bosonique,
ces modes fermioniques de fréquence nulle qui nous ont servi dans le
chapitre précédent à construire le degré des champs de formes
différentielles de Ramond--Ramond. Néanmoins le développement en série
de Taylor du champs de tachyon induit des puissances arbitrairement
élevées des fluctuations des coordonnées, dont la contraction à l'aide
du propagateur des scalaires peut être menée à bien en utilisant la
combinatoire du chapitre précédent. En séparant les coordonnées
$X^\mu$ en une partie $x^\mu$ de fréquence nulle et une partie
fluctuante $\xi^\mu(\sigma)$, nous avons en effet
$$T[X(\sigma)]=T(x)+\sum_{n\geq 1}\frac{1}{n
!}\xi^{\mu_1}(\sigma)\dots\xi^{\mu_n}(\sigma)\partial_{\mu_1}
\dots\partial_{\mu_n} T(x).$$

En développant l'intégrand de $Z[T]$  à un certain ordre en $T$, on fait
apparaître un polynôme en $\xi^\mu\partial_\mu$, dont la contribution
à l'intégrale de chemin vient des contractions de Wick entre
fluctuations du champ de coordonnées en des points du bord deux à deux
distincts. Ces contributions commencent donc à l'ordre deux en $T$,
lorsque deux points du bord, $\sigma_1$ et $\sigma_2$, supportent un
tachyon :
$$\int d\sigma_1 \int d\sigma_2\sum_{p\geq
1}\left(\frac{1}{k!}\right)^2\xi^{\mu_1}(\sigma_1)\dots\xi^{\mu_p}(\sigma_1)\xi^{\nu_1}(\sigma_2)\dots\xi^{\nu_p}(\sigma_2)\partial_{\mu_1}\dots\partial_{\mu_p}
T(x)\partial_{\nu_1}\dots\partial_{\nu_p}T(x).$$ Grâce à l'invariance
par translation le long du bord du disque, l'intégrand ne dépend que
de la différence $\sigma_1-\sigma_2$, et une seule intégrale doit être
calculée. Comme d'habitude, un facteur combinatoire $k!$ vient du
nombre de contractions, et nous utilisons le propagateur régularisé
$$ D^{\mu\nu}(\sigma)=
\left(\frac{\theta^{\mu\nu}}{2\pi\alpha'}\log\left(\frac{1-e^{-\epsilon+i\sigma}}{1-e^{-\epsilon-i\sigma}}\right)+G^{\mu\nu}\log|1-e^{-\epsilon+i\sigma}|^2\right).$$
Nous savons que ce propagateur intervient sous la forme antisymétrique
de son premier terme pour restituer les prédictions de la théorie de
jauge non-commutative. Cependant, en calculant les corrections
sous-dominantes au-delà de la limite de Seiberg--Witten, nous avons
remarqué que la combinatoire des contractions de Wick était la même,
que l'on tienne compte ou non du second terme. L'article de Cornalba
\cite{tachyonCornalba} faisait l'hypothèse d'un fort champ magnétique,
ce qui conduisait à négliger les dimensions anormales des champs de
tachyons ; cette approximation est à notre calcul ce que la limite de
Seiberg--Witten est à la théorie de jauge non-commutative déformée par
le produit $\tilde{\ast}$. Dans le contexte des termes de Wess--Zumino
en théorie de jauge non-commutative le long d'une D-brane, le degré
maximal du couplage était bien sûr la dimension du volume
d'univers. Nous avions cependant remarqué que la récurrence liant les
opérateurs $\tilde{\ast}_n$ les uns aux autres ne s'arrêtait pas pour
un ordre $n$ fini. Le présent contexte va rendre justice à cette
remarque. Mais commençons par écrire le résultat de la discussion \cite{grangepotential} à
l'ordre deux en $T$, qui fait intervenir le star-produit binaire
déformé :

$$Z[T]=\int dx \,\sqrt{\det(1+B)}\left(1+T(x)+\frac{1}{2 !} T\tilde{\ast}_2 T(x)+ \dots \right),$$
 $$ \tilde{\ast}_2=\frac{\Gamma(1+2\partial
G\partial')}{\Gamma(1-\partial \theta\partial'+\partial
G\partial')\Gamma(1+\partial \theta\partial'+\partial G\partial')}.$$
Le terme d'ordre zéro est le terme de Born--Infeld habituel ; le terme
linéaire en $T$ ne reçoit pas de corrections, car celles-ci ne
pourraient venir que de contractions entre scalaires insérés au même
point du bord. Aux ordres supérieurs, la récurrence du chapitre
précédent nous autorise à écrire
$$Z[T]=\int dx\, \sqrt{\det(1+B)} \left( 1+T(x)+\sum_{k\geq
2}\frac{1}{k !}\tilde{\ast}_k [T\tilde]^k \right),$$ ce qui peut se
symboliser par
$$Z[T]=\int dx\, \sqrt{\det(1+B)} \exp
\left(-\tilde{\ast}T(x)\right).$$ La limite dans laquelle les
dimensions anormales sont négligées donne sans surprise les
star-produits $\ast_k$, ce qui veut dire que les
tachyons sont transportés parallèlement le long d'une ligne de
Wilson. Nous reviendrons sur la contribution des champs de jauge au
moment de rapprocher les résultats sur les tachyons du secteur de
jauge.\\

Un modèle exactement soluble, celui d'un tachyon quadratique, nous
 fournit une illustration de la validité de ce résultat. Nous allons
 rapprocher la situation présente du calcul par Wyllard des termes à
 $2n$ dérivées agissant sur une $2n$-forme. Établissons un
 dictionnaire formel entre les deux calculs, ce qui nous permettra
 d'évaluer le potentiel de tachyons et d'observer que l'influence du
 fond de cordes fermées considéré peut être entièrement décrite comme
 une déformation de l'algèbre des champs scalaires tachyoniques, sans
 que la forme \cite{Gerasimov--Shatashvili,KMM2} du potentiel
$$U(T)=(1+T)e^{-T}$$
soit bouleversée.\\

 Nous devons donc établir tous les changements induits dans le calcul
 de $Z[T]$ par la substitution
$$ \int d\tau d\theta \,D\phi^\mu A_\mu(\phi)\longrightarrow \int
d\tau\, T[X(\tau)],$$ en remplaçant les tenseurs du champ de jauge
$F_{\mu\nu}$ par des champs de tachyon $T$, et en annulant la
contribution des termes qui ne contiennent pas deux modes $\psi_0$ de
fermions pour chaque champ $F_{\mu\nu}$. En effet ces termes contiennent des
fluctuations de champs fermioniques, qui n'ont pas d'analogue dans les
contractions de Wick pour l'action effective des tachyons de la corde
bosonique.\\

 Nous avons affaire à un modèle exactement soluble, et de nombreuses
 simplifications sont induites par le caractère quadratique du champ
 $T$. En effet l'action des dérivées s'arrête très vite. Avec la
 substitution que nous avons prescrite,
$$ F_{\mu\nu}\longrightarrow T,$$ les termes de la théorie de
jauge qui engendrent les termes contenant deux dérivées agissant sur
chaque champ de tachyon, sont précisément les termes
d'ordre $n$ en $F$ et portant $2n$ dérivées, ceux-là même que Wyllard
a évalués. Il nous suffit donc de traduire les résultats de Wyllard
dans notre langage.\\

 Les expressions à traduire ont une forme élégante, en fonction d'une
 deux-forme de courbure, correspondant formellement au tenseur
 non symétrique $h_{\mu\nu}=\delta_{\mu\nu}+B_{\mu\nu}$. Cette
 deux-forme, comme tout tenseur de Riemann, contient deux termes avec
 deux dérivées chacun :
$$\mathrm{S}_{\rho_1\rho_2}=\partial_{\rho_1}\d_{\rho_2}F+2\iota_h\partial_{\rho_1}
 F\land \partial_{\rho_2} F.$$ Le premier se réduit au tenseur $u$ des
 fréquences de l'oscillateur harmonique $$T(X):=a+u_{\mu\nu}X^\mu
 X^\nu,$$ alors que le second est issu de fluctuations de fermions et
 n'engendre donc pas de correction dérivative à l'action effective des
 tachyons. Ainsi notre dictionnaire traduit le tenseur de
 Riemann en tenseur des fréquences :
$$ \mathrm{S}_{\rho_1\rho_2}\longrightarrow u_{\rho_1\rho_2}.$$
 Wyllard définit les
tenseurs $W_{2n}$ pour $n\geq 2$, construits à partir de la courbure $\mathrm{S}$,
qui apportent les corrections suivantes au terme de Wess--Zumino sur la
D9-brane :
$$ S_{WZ}=\int C\land e^F\land \left( 1+\sum_{n\geq 2 }W_{2n}
\right),$$ sans aucune limitation d'origine dimensionnelle sur
l'indice $n$. Les termes calculés explicitement par Wyllard aux ordres
les plus bas donnent lieu à une forme exponentielle :
$$W_4=\frac{\zeta(2)}{2}\mathrm{tr}[(hu)^2] ,$$
$$W_6= \frac{\zeta(3)}{3}\mathrm{tr}[(hu)^3],$$
$$W_8=\frac{\zeta(4)}{4}\mathrm{tr}[(hu)^4]+\frac{1}{2}\left(\frac{\zeta(2)}{2}\right)^2(\mathrm[(hu)^2])^2,$$
$$ 1+\sum_{n\geq 2 }W_{2n} = \exp\left(\sum_{n\geq
2}\frac{\zeta(n)}{n}\mathrm{tr}[(hu)^n]\right).$$

Ainsi, les corrections exprimées plus haut engendrent un facteur exponentiel
$$ \exp\( \sum_{n\geq 2}\frac{\zeta(n)}{n} \tr[(hu)^n]\).$$ Puisque le
tenseur $u_{\mu\nu}$ est symétrique, l'opération de trace peut
être effectuée en utilisant le tenseur $h$ ou son transposé,
contrairement à la situation que nous avons rencontr\'ee en théorie de
jauge. Nous retrouvons donc le résultat de~\cite{Okuyama}:
$$\tr[((G \inv+\theta)u)^n]=\tr[(hu)^n]=\tr[(u ^t h^t)^n]= \tr[(u
(G\inv-\theta))^n].$$ Nous utilisons l'identité que Wyllard avait
écrite comme une remarque formelle :
$$-e ^{-\gamma z}\Gamma(1-z)=\exp\(\sum_{n\geq 2}\frac{\zeta(n)}{n}
z^n \),$$
$$Z=e ^{-a}\exp\(\frac{1}{2}\sum_{n\geq 2}\tr[((G
\inv+\theta)u)^n]\)\exp\(\frac{1}{2}\sum_{n\geq 2}\tr[((G
\inv-\theta)u)^n]\)$$
$$=\e{-a}e ^{-\gamma\tr{G\inv u }}\sqrt{\det\Gamma(1-(G
  \inv+\theta)u)}\sqrt{\det\Gamma(1-(G \inv-\theta)u)}.$$ La limite dans
  laquelle l'influence du champ de fond antisymétrique est
  négligée, correspond à la situation étudiée dans l'article
  ~\cite{CFGNO}, où la méthode de la fonction $\beta$ a été
  appliquée jusqu'à l'ordre trois dans le champ de tachyon.\\

 En ce qui concerne l'action effective, l'intervention de la métrique
 donne lieu à un terme cinétique~\cite{GS},
ce qui correspond, dans le modèle exactement soluble du tachyon
quadratique, à l'équation
$$S(a,u)=\(1+{\mathrm{tr}}(G^{-1}u)-a\frac{\d}{\d
a}-{\mathrm{tr}}\(u\frac{\d}{\d u}\)\)Z(a,u).$$ Quant au potentiel, il
garde la forme prédite par Gerasimov et Shatashvili, à la
modification près de l'algèbre des fonctions
$$V(T)=(1+T)\tilde{\ast} e^{-\tilde{\ast}T}.$$ Dans leurs travaux sur
les tachyons non-commutatifs, Dasgupta, Mukhi et Rajesh~\cite{DMR},
ont expliqué que l'universalité du potentiel de tachyon
contraignait les modes d'impulsion nulle. En accord avec leur
remarque, l'influence de nos opérateurs modifiés $\tilde{\ast}$
disparaît pour les modes de Fourier de fréquence nulle.

 \section{Tachyons non-commutatifs et superconnexion}
  Nous allons citer quelques arguments
 dus à Witten \cite{Wnctachyons} en faveur d'une description des
  D-branes en K-théorie non-commutative. L'introduction de la notion de
 superconnexion est la principale nouveauté de cette section ; elle
 sera utilisée dans la suite pour écrire une action effective
 associant champs de jauge et tachyons pour une paire
 ${\mathrm{D}}p$-brane- anti-$\bar{{\mathrm{D}}p}$brane. \\

La superconnexion code dans un même objet les connexions sur les fibrés
portés par chacune des deux branes, numérotées 1 et 2, et le tachyon,
considéré comme un morphisme entre les deux fibrés.

$$\mathcal{A}=\begin{pmatrix}
d+ A^1 && T \\
\bar{T} && d +A^2\\
\end{pmatrix}
 $$
 
 Le tenseur du champ de jauge sur cette paire de fibrés (ou super-fibré, via
 la graduation $\mathbf{Z}_2$induite par la numérotation des deux branes), à
 l'aide duquel Quillen définit un caractère de Chern, est simplement le
 carré de la superconnexion, qui contient naturellement les deux tenseurs
 $F^1$ et $F^2$ et les dérivées covariantes du tachyon :

$$\mathcal{F}=\begin{pmatrix}
 F^1-T\bar{T} && T \\
\bar{T} &&   F^2-T\bar{T}\\
\end{pmatrix}
 $$

  La théorie perturbative des cordes couple les tachyons aux photons
 par une dérivée covariante, comme il a été démontré dans les travaux
 de Garousi~\cite{Garousi} sur les amplitudes de diffusion. Au niveau
 géométrique, la perte de structure évoquée plus haut (absence de la
 notion de degré et de structure fibrée pour les tachyons), n'est en
 fait que superficielle, un artefact de notre calcul, qui
 momentanément ne considère que des champs scalaires. Une situation
 physique à la mesure de la conjecture de Sen (et du contexte des
 D-branes topologiques dans le modèle B, que nous aborderons dans les
 chapitres suivants), impose de considérer les tachyons comme des
 objets reliant deux fibrés (une paire brane/anti-brane).\\

 L'objet mathématique adapté à cette
 situation est la superconnexion introduite par
  Quillen~\cite{Quillen}. La proximité entre cet objet et le
 caractère de Chern a incité Alishahiha, Ita et Oz~\cite{AIO} à
 récrire les actions effectives à l'aide de la courbure
 associée. Le détail des corrections dérivatives n'a cependant
 pas été calculé dans cet article, ce qui a limité la
 précision des comparaisons avec les prédictions la théorie des
 champs de corde avec bord~\cite{KMM1,KMM2,GS}. J'essaierai de donner
 une version non-commutative de cette prescription de manière à
 assurer la compatibilité des deux approches.

Les séries de corrections calculées au chapitre précédent valent par
l'élégance de leur interprétation non-commutative. La question de leur
application renvoie à celle de leur compatibilité. Ayant obtenu les
corrections à l'action effective des tachyons avec très peu de calculs
supplémentaires, nous sommes naturellement convaincus que le tachyon
et le champ de jauge doivent admettre une description conjointe,
dans laquelle les deux champs seraient inclus, et multipliés à l'aide
des produits $\ast_n$ ou $\tilde{\ast}_n$ que nous avons construits.\\

 Nous devrions pouvoir reconstruire un analogue non-commutatif,
naturel du point de vue de la K-théorie, de cette action effective, et
compatible avec les termes de Chern--Simons \cite{CSterms} calculés
par Mukhi et Suryanarayana. Les corrections dérivatives habilleraient
naturellement la proposition de \cite{AIO} pour composer le résultat
d'une prescription d'intégration $L_\ast$ le long d'une ligne de
Wilson. C'est cette dernière qui nous pose problème : deux sortes
de champs de jauge sont impliqués par la paire de fibrés
attachée à la situation que nous étudions. Il faudrait une
notion de ligne de Wilson graduée traitant les deux transports
parallèles de manière symétrique.


\chapter[Modèle topologique B]{Modèle topologique B, de la non-commutativité à la géométrie complexe généralisée}
\chapterprecistoc{} \chapterprecishere{ Les fibrés holomorphes stables
correspondant aux D-branes du modèle topologique B en espace-cible plat sont rendus
non-commutatifs en présence d'un fond de tenseur
antisymétrique. Nous allons utiliser les techniques précédemment
développées pour montrer la stabilité de ces D-branes non-commutatives du mod\`ele B.}

\section{D-branes : surface d'univers et espace-temps}

 Dans ce qui suit, je vais introduire un doublement de certains
 degrés de liberté (liés à la structure complexe), doublement
 qui sera formalisé au chapitre suivant dans le cadre de la
 géométrie complexe généralisée à la
 Hitchin--Gualtieri~\cite{HitchinGCY,Gualtieri}. Nous considérons
 les D-branes comme des objets support\'es par une sous-variété de
 l'espace-cible. Les champs de jauge du spectre de cordes ouvertes
 font de cette sous-variété la base d'un fibré en cercles (nous
 ne considérons que des champs de jauge abéliens).\\

 Les champs de fond issus du spectre de cordes fermées ont un effet
 sur la dynamique des D-branes, comme nous l'avons observé en
 exhibant les couplages sous la forme des star-produits, à l'aide
 desquels s'écrivent les actions effectives. Quant à la géométrie de
 ces fibrés, nous allons l'étudier avec le même fond de cordes
 fermées, dans le modèle B des cordes topologiques.\\

 Les travaux de Mariño, Minasian, Moore et Strominger~\cite{MMMS}, qui
 donnent les équations hermitiennes de Yang--Mills comme conditions de
 supersymétrie sur une D$p$-brane,
\begin{equation}\label{holomorphicity}  \hat{F}^{(0,2)}=0,\end{equation}
\begin{equation}\label{stability}\hat{F}\land \omega^{\frac{p-1}{2}}=0.\end{equation}
 illustrent le point de vue d'espace-temps sur les objets
 étendus. Ces équations sont en effet obtenues par un principe de
 symétrie appliqué à une action effective.\\

 En revanche, Kapustin et Li ont adopté dans~\cite{Kapustin--Li} le
 point de vue de la surface d'univers, dans lequel les D-branes sont
 définies par un ensemble de conditions de bord. L'article de Kapustin
 sur la non-commutativité dans les théories de cordes
 topologiques~\cite{NCKapustin} contient la conjecture suivante :
 {\emph{les branes du modèle B sont équipées d'un fibré holomorphe, cette 
 condition d'holomorphicit\'e portant sur le tenseur du champ de jauge non-commutatif lié
 au tenseur ordinaire par la transformation de Seiberg--Witten}}. J'ai
 établi cette conjecture, assortie d'une version non-commutative de
 la condition de stabilité~(\ref{stability}) dans
 l'article~\cite{NCholomorphic}.

\section{Conditions de bord et structures complexes}
Nous allons donc nous efforcer de produire une version des conditions
d'holomorphicité et de stabilité adaptée à la théorie de jauge
non-commutative émergeant de conditions de recollement de courants. De
plus, et c'est là une application des résultats précédents concernant
les star-produits, les non-linéarités de la transformation de
Seiberg--Witten, et le grand nombre de dérivées intervenant dans
l'expression des produits, nous fourniront un moyen de tester la
condition d'holomorphicité dans le cas de champs non-uniformes.\\

   Les théories superconformes nous ont habitués à travailler avec
   deux copies de l'algèbre superconforme, provenant des secteurs
   droit et gauche des modes des champs de coordonnées sur la surface
   d'univers. Nous considérons dans toute la suite un tore plat
   $X\simeq T^6$ comme espace-cible. La métrique est notée $G$, et un
   champ de fond antisymétrique uniforme est présent. Quant à la
   structure complexe, nous ne la restreignons pas pour l'instant, et
   gardons les deux structures complexes $I_+$ et $I_-$ héritées des
   deux copies de l'algèbre superconforme. L'usage que nous allons en
   faire est le suivant : en imposant des condition de bord aux
   courants, nous allons définir une structure complexe sur la somme
   directe du fibré tangent et du fibré cotangent. Les contraintes sur
   cette structures seront ensuite assez fortes pour prescrire une
   structure complexe au sens usuel, en fonction des données $I_+$,
   $I_-$, $G$ et $B$, donnant ainsi lieu à la notion d'holomorphicité
   dont nous avons besoin.\\

 Une simple écriture des structures complexes sous forme de tableau
 diagonal nous fournit un seul objet sur un espace de dimension double
 :

$$ 
\begin{pmatrix}
I_+ & 0\\
 0 & I_-
\end{pmatrix}.\\
$$
Le champ antisymétrique $B_{\mu\nu}$ interviendra en géométrie complexe
généralisée dans les isométries de la somme du tangent et du
cotangent ; dans notre cas, un changement de coordonnées sur les
fermions
$$ \psi^i=\frac{1}{2}(\psi^i_+ + \psi^i_-),$$
$$ \psi^i=\frac{1}{2}G_{ij}(\psi^j_+ - \psi^j_-),$$
induit sur la matrice ci-dessus une transformation qui place les
 indices en position convenable, de sorte que la nouvelle matrice
 s'interprète dans cette base comme une structure complexe
 $\mathcal{I}$ sur $X\oplus X^\ast$,

 $$
\mathcal{I}=\begin{pmatrix}\tilde{I} +(\delta P)B & -\delta P \\
\delta\omega+B(\delta P)B+B\tilde{I} +\tilde{I}^tB & -\tilde{I}^t-B\delta P
\end{pmatrix},\\
$$

 L'hypothèse de \cite{NCKapustin} consiste à considérer le cas d'une
 matrice triangulaire supérieure. Cette condition sera retrouvée de
 manière naturelle au cours du raisonnement qui va suivre. Les
 éléments de matrices sont notés de façon adaptée aux écarts entre
 les versions droite et gauche des différentes structures ($\omega_+$
 et $\omega_-$ désignent les formes de Kähler droite et gauche
 induites, $\delta\omega$ mesure leur différence, $\delta P$ la
 différence entre leurs formes de Poisson inverses, alors que
 $\tilde{I}$ désigne la structure complexe moyenne, qui n'est pas
 celle en fonction de laquelle nous aurons un fibré holomorphe) :

$$\delta\omega=\frac{1}{2}(\omega_+ -\omega_-) ,$$
$$\delta P=\frac{1}{2}(\omega_+^{-1}-\omega_-^{-1}) ,$$    
$$\tilde{I}=\frac{1}{2}(I_+ -I_-) .$$
 
 Nous allons faire l'hypothèse d'une courbure localisée : le tenseur
 du champ de jauge $F$ est supposé nul hors d'une certaine
 région. Cette hypothèse a également été utilisée par les auteurs de
 \cite{MMMS} afin d'évaluer une constante, qui leur a servi à proposer
 les équations de Yang--Mills hermitiennes
 non-commutatives{\footnote{Ils ont cependant utilisé la
 transformation de Seiberg--Witten des champs constants, ce qui rend
 l'apport de la théorie de jauge non-commutative nécessaire pour
 compléter les résultats obtenus du point de vue d'espace-temps.}}. De même
 nous allons travailler loin de la courbure non-triviale pour
 identifier une structure complexe unique, puis incorporer les champs
 de jauge en nous plaçant dans la région où la courbure se trouve
 concentrée. Ce sera chose faite en substituant la combinaison $B+F$
 au champ $B$, par l'argument habituel d'invariance de jauge. Loin de
 la courbure, donc, les conditions de bord pour les fermions
 s'écrivent simplement en fonction du champ $B$ et des sections des
 fibrés tangent et cotangent utilisées pour transformer la structure
 complexe :
$$\rho_i=-B_{ij}\psi^j.$$ En substituant ces équations dans la
condition de recollement entre les R-courants, et en incorporant les
champs de jauge de la manière décrite plus haut, nous obtenons une
contrainte portant sur le tenseur du champ de jauge (ordinaire) $F$, qui
n'est pas une contrainte d'holomorphicité comme celle que nous avons exposée
au chapitre 1. En effet, elle contient un second membre :
\begin{equation}\label{default}FI + I^t F= -F\delta P F,\end{equation}
avec l'écart de Poisson et la structure complexe suivants 
$$\delta P= I\theta + \theta I ^t$$
$$I:=\tilde{I}+(\delta P)B,$$
$$ \delta P=I\theta+ \theta I^t,$$ où $\theta$ désigne bien sûr
l'inverse du tenseur $B$.  Deux remarques formelles motivent la
transformation de Seiberg--Witten que nous allons appliquer :\\
1. L'annulation de l'échelle de non-commutativité $\theta$ entraîne
l'annulation du second membre, et l'holomorphicité du fibré
ordinaire. Le second membre est donc susceptible d'être corrélé à la
prise en compte de la non-commutativité ;\\
 2. La non-linéarité du
second membre rappelle la non-linéarité de la transformation de
Seiberg--Witten.\\

Kapustin a justifié ces deux remarques pour des champs constants, mais
ce test, à ce stade de notre raisonnement, n'est que formel. Il
souffre du même manque de cohérence interne que la proposition
non-commutative à la MMMS : une étape du raisonnement suppose une
concentration de la courbure, et partant des champs variables, or la
proposition utilise la transformation entre champs constants. Nous
allons voir que les star-produits modifiés apportent une solution
à ce problème.

\section{Condition de stabilité}
La condition de stabilité (supersymétrie) donnée par le
recollement des opérateurs de flot spectral, une fois imposée aux D-branes
topologiques, fait d'elles des D-branes supersymétriques de la théorie
des cordes. En d'autres termes, les conditions de bord sur les
R-courants et le flot spectral dans le modèle B doivent conduire aux
contraintes d'holomorphicité et de Yang--Mills hermitiennes à la MMMS,
c'est-à-dire à des solitons supersymétriques d'espace-temps.\\

En supposant que la non-commutativité permette effectivement
d'absorber par ses non-linéarités le défaut quadratique
d'holomorphicité, nous pouvons d'ores et déjà appliquer l'une des
leçons apprises en théorie de jauge non-commutative. Les raisonnements que
nous avons faits plus haut nous ont convaincus que les couplages de
Ramond--Ramond, quel que soit leur degré, possédaient une expression
non-commutative. Cette expression  appara\^it lorsque nous rendons invariante de jauge
l'expression écrite pour des champs constants. Nous avons fait des
calculs concernant les couplages aux formes de Ramond--Ramond de degré
inférieur à la dimension du volume d'univers. Nous avons mentionné
l'existence d'une identité topologique correspondant au couplage à la
forme de degré maximal $C^{(10)}$. Elle s'écrit
$$\delta(k)=\int
dx\;L_\ast\(\sqrt{\det\(1-\theta\hat{F}\)},W_k(x)\)\ast e ^{ikx}.$$ 

 Nous reconnaissons ici une situation analogue : une certaine quantité
 définie dans le contexte de D-branes portant des champs de jauge, la
 phase $e^{i\alpha}$, possède une signification indépendante du champ
 de jauge, en l'occurrence elle décrit une supercharge conservée. Si
 la théorie de jauge possède une version non-commutative, alors cette
 phase doit être insensible à la transformation de Seiberg--Witten. Il
 suffit donc d'appliquer le principe qui nous a réussi dans le
 contexte des couplages de Ramond--Ramond, en écrivant une expression
 valable pour des champs constants (c'est ce qui a été fait dans la
 proposition MMMS), puis en promenant cette expression le long d'un
 segment de Wilson de longueur proportionnelle à l'impulsion de chaque
 mode. Le relation (\ref{stability}) découle donc comme une identité
 topologique de la condition d'holomorphicité de la transformée de
 Seiberg--Witten. Il nous reste à établir cette condition
 d'holomorphicité.

\section{Non-commutativité et holomorphicité}

Puisque le tenseur du champ de jauge apparaît à des puissances
différentes dans les deux membre de notre condition d'holomorphicité
éventuelle, il est naturel de traiter la transformation de
Seiberg--Witten ordre par ordre en le champ de jauge. Le second membre
devrait être compensé par les premières non-linéarités, tandis que les
non-linéarités d'ordre supérieur se compenseraient ordre par ordre,
sans donner de contrainte nouvelle sur le fibré.\\

 La relation (\ref{default}) exprimant le défaut d'holomorphicité est
 adaptée au sens dans lequel se présente la solution des équations de
 Seiberg--Witten exposée au chapitre précédent. En effet, en
 substituant

\begin{equation} \label{map}F_{ij}(k)=\int dx\, L_\ast\left(\sqrt{\det(1-\theta
    \hat{F})},\,\hat{F}_{ik}\(\frac{1}{1-\theta\hat{F}}\)^k_j,\,
    W_k(x)\right) \ast e ^{ikx}\end{equation}
dans la transformée de Fourier de la condition (\ref{default}), nous
obtenons une équation concernant le champ de jauge non-commutatif. Il
nous reste à tenir compte des non-linéarités afin de décider si une
telle équation peut se lire comme une honnête condition
d'holomorphicité sur $\hat{F}$, au sens de la structure complexe $I$
incluant le champ antisymétrique. Nous devons donc développer en
puissances du champ de jauge non-commutatif pour traiter séparément
les non-linéarités de degrés différents. Le membre de gauche de
l'équation de défaut (\ref{default}) ne contient qu'une observable
dans la prescription d'intégration $L_\ast$.  Il est donc permis de
le récrire comme
$$FI+I^tF=\int dx\, L_\ast\left(\sqrt{\det(1-\theta
    \hat{F})},\left(\hat{F}\frac{1}{1-\theta\hat{F}}I+ I
    ^t\hat{F}\frac{1}{1-\theta\hat{F}}\right),W_k(x)\right) \ast e
    ^{ikx},$$ puisque les champs de jauge effectuant le transport
    parallèle ne portent pas d'indices (la structure complexe $I$ ne
    voit pas la prescription $L_\ast$). Les calculs commencent avec le
    membre de droite, puisque les deux tenseurs $F$ doivent être
    attachés à la ligne de Wilson indépendamment l'un de l'autre.
    Le mode de Fourier d' impulsion $k_\mu$ du membre de droite prend la forme d'un
    produit de convolution entre deux segments de Wilson, dont les
    longueurs ont pour somme $\theta^{\mu\nu}k_\nu$, chacune portant
    l'un des champs attaché à son extrémité :


 Cette image est équivalente à la combinaison de deux
 segments en un seul, de longueur $\theta^{\mu\nu}k_\nu$, avec
 insertion d'une seconde observable à un point arbitraire de
 l'intérieur. Le mode $k_\mu$ dans l'espace de Fourier donne lieu à la
 relation
$$\int dx\, L_\ast\left(\sqrt{\det(1-\theta
    \hat{F})},\hat{F}\frac{1}{1-\theta\hat{F}},\left(I\theta\hat{F}\frac{1}{1-\theta\hat{F}}+ \theta I ^t\hat{F}\frac{1}{1-\theta\hat{F}}\right),W_k(x)\right) \ast e
    ^{ikx}.$$

Grâce à la technologie développée au chapitre précédent, nous allons
pouvoir écrire des conditions de cohérence entre le membre de droite
et le membre de gauche de ~(\ref{default}) ordre par ordre dans le
champ de jauge. les dérivées contenues dans les opérateurs $\ast_n$
sont là pour prolonger la validit\'e du raisonnement au cas des champs non
uniformes.\\

le développement de (\ref{default}) au premier ordre en $\hat{F}$
n'est autre que la condition d'holomorphicité (\ref{holomorphicity})
que nous cherchons à établir, puisque le défaut d'holomorphicité ne
débute que par un terme quadratique :
 
$$\hat{F}I+I ^t\hat{F}=o(\hat{F}).$$
 À l'ordre suivant, le membre de droite contient un seul terme, parce que
 chacun des deux indices de forme différentielle doit être apporté par un
 champ $\hat{F}$, sans contribution possible du pfaffien ou du
 dénominateur. Dans le membre de gauche, ce terme peut être comparé au
 terme quadratique qui apparaît dans le développement de l'application de
 Seiberg--Witten dans l'espace des positions : 
   
$$F_{ij}=\Fhat_{ij} +\theta ^{mn}
\langle\Fhat_{im},\Fhat_{nj}\rangle_{\ast_2} -\frac{1}{2}\theta ^{mn}
\langle\Fhat_{nm},\Fhat_{ij}\rangle_{\ast_2} +\theta ^{mn}
\partial_n\langle \hat{A}_m, \Fhat_{ij}\rangle_{\ast_2} +
O(\Fhat^3).$$ Nous pouvons lire la condition à vérifier par le
champ de jauge pour assurer la cohérence du développement :
$$0= -\frac{1}{2} \theta ^{mn}\langle\Fhat_{nm},(\Fhat I + I
^t\Fhat)_{ij}\rangle_{\ast_2} +\theta ^{mn} \partial_n\langle
\hat{A}_m,(\Fhat I + I ^t\Fhat)_{ij}\rangle_{\ast_2}.$$ Nous observons
que cette condition est vérifiée dès que la condition
d'holomorphicité non-commutative (\ref{holomorphicity}) l'est. Nous
avons donc dépassé le cas des champs constants, grâce aux dérivées
contenues dans $\ast_2$.\\

Un développement plus poussé en puissances du champ de jauge engendre
des contributions plus compliquées, mais grâce à la transparence de la
structure complexe vis-à-vis de la prescription d'ordre $L_\ast$, la
structure que nous avons identifiée aux petits ordres n'est pas
bouleversée. Dans le membre de gauche, la condition d'holomorphicité
$\hat{F}^{(0,2)}=0$ assure que l'un des arguments sur lesquels
agissent les différents opérateurs différentiels est nul, celui
précisément dont les deux indices de formes différentielles sont
portés par le même champ $\hat{F}$. Les champs de jauge provenant du
développement de la ligne de Wilson ayant déjà tous leurs indices
contactés, ils ne changent rien à l'affaire.

Nous allons argumenter en faveur d'une compensation entre les défauts aux
ordres plus élevés dans le champ de jauge. Le pfaffien et la ligne de
Wilson, avec leurs contributions aux indices déjà contractés, sont les
mêmes de part et d' autre. En revanche, le développement des dénominateurs
contenus dans l'application de Seiberg--Witten~(\ref{map}) libère des
indices. Il engendre des termes dans lesquels deux indices sont portés par
des champs de jauge différents. En passant d'un ordre au suivant dans le
développement d'un dénominateur, nous pouvons jouer le même jeu que
précédemment, et ce qui ne peut pas être comparé entre les deux membres sera
annulé par la condition d'holomorphicité.

Le pfaffien et la ligne de Wilson, remis en place des deux côtés, complètent
la cohérence entre les deux différents développements de la version
non-commutative de ~(\ref{default}), à des termes près qui s'annulent sur un
fibré holomorphe non-commutatif suivant ~(\ref{holomorphicity}).\\

\section{Vers la géométrie complexe généralisée}
Nous avons donc appliqué le formalisme des produits de Moyal
modifiés pour montrer la robustesse de la non-commutativité
vis-à-vis du twist topologique du modèle B. La
structure complexe a été fixée à partir de conditions de bord
sur un espace dédoublé de degrés de liberté, obtenu en
considérant sur un pied d'égalité des champs de vecteurs et de
formes différentielles.\\

 La somme du fibré tangent et du fibré conormal à une sous-variété
 supportant une D-brane est un objet naturel, à cause des champs de
 jauge portés par le volume d'univers, et des scalaires
 transverses. De plus, les symétries $O(d,d)$ qui apparaissent dans
 les situations indépendantes de $d$ coordonnées
 \cite{Meissner--Veneziano} ont été étudiées par Hassan dans
 \cite{OddHassan} en relation avec la T-dualité et le problème de
 l'inclusion du champ antisymétrique. Au chapitre suivant, nous allons
 aborder la symétrie miroir en présence de D-branes, ce qui nous
 forcera à introduire la notion de spineur
 pur~\cite{Cartan,Chevalley,HitchinGCY,Gualtieri}, somme de formes
 différentielles sur laquelle des sommes de vecteurs et de formes
 différentielles agissent comme une algèbre de Clifford.


\chapter{Symétrie miroir en présence de D-branes}

\chapterprecistoc{} 

\chapterprecishere{Dans ce chapitre, nous reprenons les conditions de
  bord définissant les D-branes topologiques, sans nous limiter aux
  D-branes lagrangiennes dans le modèle A. Nous utilisons d'abord des
  techniques de réduction symplectique. Nous partons des relations de
  symétrie miroir obtenues par Leung dans le modèle B et par Thomas
  dans le modèle A, et nous les complétons par des contributions du
  champ de jauge dictées par l'action des champs de vecteurs
  hamiltoniens.  La comparaison entre les deux mod\`eles topologiques
  fait émerger une notion de spineur pur modifié par les champs de
  jauge, compatible avec la symétrie miroir sur une variété de
  Calabi--Yau fibrée en tores $T^3$.}

\section{Argument de Strominger--Yau--Zaslow : fibration en tores $T^3$}
\subsection{Espaces de modules et T-dualité}

Suivons l'argumentation de Strominger, Yau et Zaslow (SYZ) exposé dans leur
article \cite{SYZ}, ce qui nous fera entrer dans le domaine de la
symétrie miroir en présence de D-branes. Dans cette section, les D-branes du
modèle A seront considérées comme lagrangiennes et équipées d'une
connexion plate. La question des A-branes non-lagrangiennes ne sera
abordée que dans la section sur la modification des spineurs purs par
les champs de jauge. Nous considérons une variété de Calabi--Yau de
dimension complexe trois, notée $X$, et nous supposerons qu'elle
possède une variété miroir $\tilde{X}$ qui vérifie encore la condition
de Calabi--Yau. La variété $X$ est supposée simplement connexe,
hypothèse qui est utilisée pour affirmer que la D-brane occupant tout
l'espace n'a pas de modules : le seul fibré dont elle peut être la
base est un fibré trivial. À l'opposé, nous pouvons considérer une
D0-brane ; elle n'a pas non plus de module associé au fibré qu'elle
porte, puisque celui-ci est automatiquement trivial faute de
changement de cartes sur une base de dimension zéro. Ses modules
proviennent de sa position sur la variété $X$. Comme la D0-brane peut
se trouver en un point quelconque, son espace de modules est l'espace
$X$ tout entier.\\

Les cycles miroirs sur $\tilde{X}$ des deux cycles discutés précédemment doivent
posséder les mêmes espaces de modules, mais leurs dimensions ne sont
pas 0 et 6. Ainsi, soit $L$ le cycle miroir du cycle de dimension six
$X$. Il doit posséder un espace de modules de dimension six, et cet
espace n'est autre que le miroir de $\tilde{X}$. Nous apprenons donc
que la variété miroir doit pouvoir être réalisée comme espace de
modules d'une D-brane. Quant au cycle $L$, c'est un tore lagrangien. Comme on peut voir l'espace de modules $\mathcal{M}(L)$ de la
D-brane $L$, comme un espace fibré au-dessus de son espace
de modules de variété spéciale lagrangienne
$\mathcal{M}_{\mathrm{sLag}}(L)\simeq X$, la fibre étant simplement la
donnée de la fibration $U(1)$, c'est-à-dire un tore dont la dimension
réelle est le premier nombre de Betti de $L$, soit 3. La variété de
Calabi--Yau $X$ est donc fibrée en tores $T^3$.\\

De plus, un calcul  utilisant l'action de Dirac--Born--Infeld, effectué dans un
régime suffisamment classique, donne à voir la symétrie miroir comme
T-dualité le long des trois cercles de chaque fibre.  Tout ce
raisonnement est effectué dans une limite de grand volume ou de grande
structure complexe, dans laquelle les corrections provenant des
instantons de disques peuvent être négligées. 

\section{Spineurs purs}
\subsection{Géométrie complexe généralisée}
L'introduction par Hitchin de la géométrie complexe
généralisée a fourni un cadre unifié aux aspects symplectiques
et complexes de la géométrie des D-branes topologiques.\\

 Considérons l'action suivante de la somme du tangent $T$ et du
 cotangent $T^\ast$ d'une variété, sur une somme formelle $h$ de
 formes différentielles : $$(v,\phi).h=\iota_v h +\phi\land h.$$
 C'est l'action la plus simple que l'on puisse écrire en utilisant
 les règles algébriques de multiplication intérieure et
 extérieure.\\

Un fait conduit à considérer les sommes de formes
 différentielles comme des spineurs : il est possible, toujours par
 simple utilisation de ces mêmes relations algébriques, d'écrire
 une algèbre de Clifford sur la somme du tangent et du cotangent,
$$ \gamma^\mu:=dx^\mu,$$
$$ \gamma_\nu=\frac{\partial}{\partial x^\nu},$$
$$\{\gamma^\mu,\gamma_\nu\}=\delta^\mu_\nu.$$
 Le tenseur
$\delta^\mu_\nu$ apparaissant dans le membre de droite n'est pas une
métrique sur un espace plat mais un simple tenseur unité sans
interprétation en géométrie riemannienne.\\ 

Munis de l'action de $T\oplus T^\ast$, considérons les vides de
l'algèbre de Clifford, ou \emph{spineurs purs}, c'est-à-dire les formes différentielles
annihilées par la moitié des générateurs de l'algèbre.
 Si l'espace-cible considéré est une variété de Calabi--Yau de
 dimension complexe trois, deux spineurs purs de rangs différents
 peuvent être exhibés : \\
- l'un provenant de la géométrie
 complexe, qui n'est autre que la trois-forme holomorphe $\Omega$,
  annihilée par les vecteurs de l'espace de dimension réelle
 trois $T^{(1,0)}\oplus T^{\ast (0,1)}$, par manque d'indices
 holomorphes,\\
 - l'autre provenant de la géométrie de Kähler, et
 donné par l'exponentielle de la forme de Kähler $e ^\omega$. Il
 est en effet clair que la forme différentielle 1 est annihilée
 par tout l'espace tangent ; d'autre part on montre que le produit de
 ce spineur pur par l'exponentielle de la deux-forme $\omega$ est
 encore pur, son annihilateur étant l'ensemble des sommes
 s'écrivant $v+\iota_v\omega,\; v\in T$.\\

\subsection{Échange de spineurs purs}
La  symétrie miroir échange les deux
spineurs purs que nous avons identifiés en géométrie complexe généralisée
sur toute variété de Calabi--Yau possédant une fibration en tores $T^3$ :
$$
\Omega\longleftrightarrow e^{i\omega}.$$
Nous reviendrons sur cette
relation en même temps que nous établirons sa version modifiée par les
champs de jauge.  La condition de Calabi--Yau a été affaiblie en une
condition de structure $SU(3)$ dans \cite{FMT} en présence d'un champ
antisymétrique $B_{\mu\nu}$.

\section{Inclusion de D-branes}
Revenons aux notions concernant les D-branes topologiques introduites au
début de ce mémoire.  Pour introduire un secteur de cordes ouvertes, il faut
préciser un ensemble de conditions de bord pour les équations du mouvement.
Ces conditions de bord donnent les points o{ù} peuvent s'appuyer les
extrémités des cordes ouvertes. Elles donnent l'équation des D-branes.
De même que ces conditions de bord brisent une partie de la symétrie de
Poincaré en introduisant un défaut topologique dans l'espace-temps,
elles brisent une partie de la supersymétrie. Dans le cas présent, seule
une certaine combinaison des deux générateurs de supersymétrie
correspondra à une supersymétrie non brisée. Le choix de cette
combinaison
$$Q=\cos\alpha Q_1 + \sin\alpha Q_2$$
 est donné par une certaine phase complexe $e^{i\alpha}$.
On retrouve cette phase dans la condition de bord sur les opérateurs de flot
spectral $S$, puisque ces opérateurs assurent la supersymétrie
d'espace-temps. Sur la D-brane, ces conditions s'écrivent
$$S_+=e^{i\alpha} S_-.$$
Le
 paramètre  $\alpha$ doit être constant sur toute la D-brane, de
 fa{ç}on à bien définir une supersymétrie non brisée
 associée au défaut topologique dans sa totalité. L'existence de
 ce paramètre constant sur la sous-variété lagrangienne supportant
 la D-brane dans le modèle A, confère à cette sous-variété
 la qualité de sous-variété {\emph{spéciale}} lagrangienne.\\

 Dans le modèle B, la dimension 3 n'est plus privilégiée, mais les
 dimensions paires le sont, puisque les indices holomorphes et
 anti-holomorphes ne sont pas mélangés par les conditions de bord.
 L'interprétation de la phase $\alpha$ n'est donc plus la même que dans
 le modèle A, puisque les branes ne se trouvent plus sur des
 sous-variétés lagrangiennes. Néanmoins, la prescription de
 proportionnalité entre les courants de flot spectral est toujours valable
 comme condition de stabilité, et peut être utilisée pour obtenir les
 D-branes stables du modèle B des équations d'instantons par une approche
 de surface d'univers.

\section{Des A-branes spéciales lagrangiennes aux B-branes}

Puisque nous avons accepté l'idée (ou fait l'hypothèse) d'une
fibration en tores $T^3$ de la variété de Calabi--Yau $X$ de dimension complexe
 trois, nous pouvons effectuer une
transformation de T-dualité le long de chacun des trois cercles
  de chaque fibre, et obtenir ainsi la variété miroir
notée $\hat{X}$. Nous travaillons formellement, avec les
ingrédients complexes et kählériens que nous avons à notre
disposition.\\

 Nous allons d'abord passer en revue la démarche de Leung, Yau et
 Zaslow~\cite{LYZ}, qui relie par ces trois T-dualités une A-brane
 spéciale lagrangienne à une B-brane vérifiant la condition de
 stabilité déduite par Mariño, Minasian, Moore et
 Strominger~\cite{MMMS} via une approche d'espace-temps. J'adopte ici
 non pas ce point de vue d'espace-temps, mais le point de vue
 géométrique de la symétrie miroir en présence de D-branes,
 dans le but de présenter les résultats obtenus en collaboration
 avec Minasian~\cite{GM}, qui complètent ceux de \cite{LYZ}, en
 autorisant des configurations plus générales pour les champs de
 jauge dans le modèle A, et par la même occasion des
 géométries non-lagrangiennes pour les A-branes. Nous mettrons
 donc en évidence les restrictions de l'approche de~\cite{LYZ} afin
 de motiver la généralisation qui suivra.\\
 
 L'objet donné au départ est une A-brane spéciale lagrangienne (avec
 une connexion plate).  Nous avons vu plus haut, lors de la discussion des
 conditions de bord pour les cordes topologiques, que les A-branes
 lagrangiennes équipées d'une connexion plate apparaissaient comme des
 solutions évidentes aux conditions de bord, du point de vue de la
 surface d'univers. Il est donc intéressant en soi de partir de telles
 branes comme données dans le modèle A, et de retrouver les équations
 de~\cite{MMMS}.\\

 D'après les travaux de Calabi, une variété
 de Calabi--Yau possède un potentiel de Kähler vérifiant les
 équations de Monge--Ampère, c'est-à-dire qu'il existe une
 fonction scalaire $\phi$ telle que la métrique et la forme de
 Kähler s'écrivent :
$$g=\sum_{i,j} \phi_{ij}(dx^i dx^j+dy^i dy^j),$$
$$\omega=\sum_{i,j}\phi_{ij} \,dz^i\land d\bar{z}^j.$$ Nous avons muni
la base de la fibration de coordonnées locales $x^i (1\leq i\leq 3)$, et la
fibre de coordonnées $y^i (1\leq i\leq 3)$. Les quantités notées
$\phi_{ij}$ sont les dérivées secondes de la fonction $\phi$, qui
par hypothèse\footnote{Comme dans l'argument de
Strominger--Yau--Zaslow, une brane lagrangienne ne coupe la fibre
qu'en un point.} ne dépend que des coordonnées sur la base,
$$\phi_{ij}:=\frac{\partial ^2\phi}{\partial x^i\partial y^j}.$$

Pour obtenir une sous-variété miroir équipée d'un fibré en cercles muni d'une connexion,
nous modifions la connexion plate en lui ajoutant un terme écrit à l'aide
des coordonnées duales $\tilde{y}$ du revêtement universel, mais qui sera
bien défini sur le tore $T^3$ dual, une fois inséré dans l'argument d'une
exponentielle :
$$ d\longrightarrow d+A:= d+\tilde{y_i} dy^i,$$
$$ e^{F'}:=\int _{T^3_y} dy\, e^{F+dA}.$$ La dernière ligne exprime le
caractère de Chern du modèle B comme une transformée de Fourier
de celui du modèle A (en l'occurrence $F=0$). Nous reviendrons sur
les caractères de Chern plus complexes qui sont autorisés dans le
membre de droite par les conditions de bord.\\

 Tout d'abord nous devons transformer la condition de stabilité par
 T-dualité le long des fibres. Partant de la condition spéciale de
 stabilité :
 \begin{equation}\label{stab} 
 \mathrm{Im} (e^{-i\theta}dz^1\land dz^2\land dz^3)=0,
\end{equation}
 écrite dans les
 coordonnées complexes
$$z^i:= x^i+ iy^i,$$
nous recherchons une formule dans les coordonnées complexes T-duales
$$\tilde{z}_i=x_i+i\tilde{y}_i.$$ La sous-variété lagrangienne $L$
considérée étant transverse par rapport à chaque fibre, la restriction
des coordonnées $y$ à cette sous-variété fournit une section $y(x)$ de
la fibration en tores $T^3$. Nous allons exprimer la condition
lagrangienne dans les coordonnées de la base et substituer la relation
obtenue dans la condition spéciale, avec pour la forme symplectique la restriction suivante :
$$\omega|_L=\frac{i}{2}\sum_{i,j}\phi_{i,j}\left(dx^i+i\frac{\partial
    y^i}{\partial x^l} dx^l\right)\land \left(dx^j+i\frac{\partial
    y^j}{\partial x^l} dx^l\right).$$
Ceci implique une condition de
fermeture sur une 1-forme :
$$
d(\phi_{ij} y^j dx^i)=0,$$
qui s'intègre localement d'après le lemme de
Poincaré. Nous d\'eduisons donc l'existence locale d'une fonction scalaire $f$
telle que les différentielles des coordonnées complexes peuvent, une fois
restreintes à la sous-variété lagrangienne $L$, s'écrire en fonction de la
hessienne de $f$:
$$\phi_{ij} y^j=\frac{\partial f}{\partial x ^i},$$
$$
dz^i|_L=(\delta^i_k+i\phi^{il}\mathrm{Hess}_{lk}f)dx^k.$$
Nous obtenons
la restriction à $L$ de la trois-forme holomorphe, comme un
multiple de la forme volume dans les coordonnées de la base:
$$dz^1\land dz^2\land dz^3|_L=\det(I_3+ ig ^{-1} \mathrm{Hess} f)dx^1\land
dx^2\land dx^3.$$
En d'autres termes, avant toute transformation de T-dualité,
nous approchons via (\ref{}stab) de la forme des équations à la MMMS :
$$\cos\theta\,\mathrm{Re}(g+\mathrm{Hess} f)=\sin\theta\,
\mathrm{Im}(g+\mathrm{Hess} f),$$
au point de pouvoir anticiper le fait que
la fonction $f$ aura à voir avec la connexion sur le cycle holomorphe obtenu
par symétrie miroir.\\

Dans le modèle B, toute la courbure provient de la connexion de Poincaré,
puisque nous sommes partis d'une connexion plate. L'annulation de la
composante de type $(2,0)$ de cette courbure s'écrit
  $$0=\frac{\partial y^i}{\partial x ^k}-\frac{\partial y^k}{\partial x
    ^i},$$
  ce qui garantit l'existence locale d'une fonction scalaire, notée
  $f$, telle que
  $$y_i=\frac{\partial f}{\partial x^i},$$
  ce qui identifie bien la
  hessienne de la primitive précédente au tenseur du champ de jauge sur la
  B-brane:
$$\cos\theta\,\mathrm{Re}\left(\exp (\tilde{\omega}+F)|_{\mathrm{top}}\right)=\sin\theta\,
\mathrm{Im}\left(\exp (\tilde{\omega}+F)|_{\mathrm{top}}\right).$$

\subsection{Des B-branes holomorphes aux A-branes non-lagrangiennes sur le tore}Nous allons montrer, sur l'exemple de la fibration triviale en tores $T^3$ qu'est le tore $T^6$, que les données précédentes sur le modèle A ne contiennent pas toutes les configurations accessibles par symétrie miroir à partir d'un fibré holomorphe. cet exemple a été exhibé dans~\cite{vanEnckevort} et est instructif comme critique de l'approche précédente, et aussi comme exemple du traitement de $F$ comme une distribution, dont le support définit la position d'une D-brane. Sur une variété de Calabi--Yau plus générale, nous ferions ce calcul dans une carte locale sur la base. Nous décomposons le tenseur du champ de jauge associé à un fibré holomorphe suivant ses composantes dans la base adaptée au produit $T^3\times T^3$ (fibration triviale) :

\begin{align}
F &=F_{\mu\nu}(dx^\mu\wedge dx^\nu -dy^\mu\wedge dy^\nu) +
iF_{\mu\nu}( dx^\mu\wedge dy^\nu+ dy^\mu\wedge dx^\nu)\nonumber\\ &=
F_{\mu\nu}(dx^\mu\wedge dx^\nu -dy^\mu\wedge dy^\nu)+
iF_{\mu\nu}(dx^\mu\wedge dy^\nu -dy^\mu\wedge dx^\nu) \nonumber \\ &=
F^{(A)}_{\mu\nu}(dx^\mu\wedge dx^\nu -dy^\mu\wedge dy^\nu)+
iF^{(S)}_{\mu\nu}(dx^\mu\wedge dy^\nu).\nonumber
\end{align}
Les blocs diagonaux ont un rang pair, puisque ce sont des matrices
antisymétriques. Le cas lagrangien correspond au rang zéro, alors
que le rang deux donne par transformation de Fourier--Mukai une
A-brane non-lagrangienne équipé d'une fibré à courbure non
nulle. Nous introduisons de nouvelles notations pour les blocs:
 
 $$F = {\mathcal{A}}_{\mu\nu}(dx^\mu\wedge dx^\nu+dy^\mu\wedge
dy^\nu)+{\mathcal{S}}_{\mu\nu}dx^\mu\wedge dy^\nu,$$ 
La transformation décrite dans \cite{LYZ} opère en sens inverse :
\begin{align}
e^{F'}&=\int_{T_{y}^3} e^{dy_\mu d\tilde{y}^\mu} e^F\nonumber\\
& =  e^{{\mathcal{A}}_{\mu\nu}(dx^\mu\wedge dx^\nu)}\int_{T_{y}^3} e^{dy_\mu
d\tilde{y}^\mu} e^{{\mathcal{A}}_{\mu\nu}dy^\mu\wedge
dy^\nu+{\mathcal{S}}_{\mu\nu}dx^\mu\wedge dy^\nu}.\nonumber
\end{align}
Le cas ${\mathcal{A}}=0$ donne bien
$$\delta(\tilde{y}_\mu-{\mathcal{S}}_{\mu\nu} x^\nu),$$ qui n'est
autre qu'une D-brane lagrangienne équipée d'une connexion
plate. En revanche, si ${\mathcal{A}}$ a pour rang deux, alors le
calcul fait intervenir l'inverse d'une sous-matrice inversible de
rang deux, notée ${\mathcal{A}}^{-1}$. Nous écrivons le résultat
à l'aide de l'opération $V^\perp\lrcorner(\cdot )$ définie dans
~\cite{FMT}, et liée à la dualité de Hodge
\[
V^\perp \lrcorner(e^{\alpha_1}\ldots e^{\alpha_k}) = \frac1{(3-k)!}
\epsilon^{\alpha_1\ldots \alpha_3} e_{\alpha_{k+1}} \ldots
e_{\alpha_3} \ , k=1\ldots 3\ .
\]
Le dual de Hodge de la deux-forme ${\mathcal{A}}_{\mu\nu} dy^\mu
dy^\nu$ est une un-forme le long de la fibre, à laquelle
l'opération $V^\perp$ associe un vecteur $(V^\perp\lrcorner
{\mathcal{A}})$, qui définit la direction normale au support de
${\mathcal{A}}$. La codimension un de la D-brane résultante se lit
sur l'argument du facteur $\delta$ dans le résultat suivant :

$$e^{F'}=\delta\Big{(}{(\tilde{y}_\mu-{\mathcal{S}}_{\mu\nu}
 x^\nu)(V^\perp\lrcorner
 {\mathcal{A}})^\mu}\Big{)}\exp\Big{(}{\mathcal{A}}_{\mu\nu}(dx^\mu\wedge
 dx^\nu)+ \frac{1}{2}{(\mathcal{S}}dx)_\mu( {\mathcal{A}}^
 {-1})^{\mu\nu} ({\mathcal{S}}dx)_\nu\Big{)}.$$
   Cette D-brane n'est donc pas lagrangienne, et ne relève pas du traitement donné dans \cite{LYZ}.

\section{Réduction symplectique et appariements à la Atiyah--Bott}
\subsection{Notion d'application moment}

Cette notion est apparue en analyse globale dans les travaux de
Souriau. La terminologie provient du contexte de la mécanique
classique. Les applications moment sont l'équivalent des charges de
Noether associées aux symétries, lorsqu'une structure symplectique est
présente. L'exemple de l'homogénéité de l'espace en mécanique
classique donne une première illustration. Soit une situation où l'on dispose d'une forme symplectique $\omega$
et d'une symétrie de g\'en\'erateur $X$. Un champ de
vecteurs appelé $X^\sharp$ est induit par ce g\'en\'erateur sur la variété
symplectique. Si le flot de ce champ de vecteurs dérive d'un hamiltonien
$\mu^X$, ce hamiltonien est appelé application moment associée à
la symétrie générée par $X$. Autrement dit :
$$d\mu^X=\iota_{X^\sharp}\omega.$$

Considérons la mécanique classique avec la forme symplectique
usuelle sur l'espace des phases $\mathbf{R^6}$:
$$\omega:= dp_i\land dq^i.$$ L'action des translations sur l'espace
de phases est hamiltonienne. En effet, soit $X$ un générateur de
la translation dans la direction $i$ ; le champ de vecteurs induit est
$X^\sharp=\partial/\partial q^i$. Il suffit de calculer la contraction
entre ce champ de vecteurs et la forme symplectique~\cite{lectures}:
$$\iota_{X^\sharp}\omega=dp_i.$$ On lit l'application moment associée au
générateur des translations dans la direction $i$, qui n'est autre
que l'impulsion $p_i$.\\

La généralisation de cette procédure dans le contexte des
théories de jauge est l'objet de travaux d'Atiyah et
Bott~\cite{AB}. Nous allons l'appliquer aux symétries des deux
modèles de cordes topologiques en présence d'une D-brane.

\subsection{Structure symplectique sur la théorie de jauge du modèle B}
Cette structure, exhibée par Leung~\cite{LeungStructures}, provient
de l'intégration naturelle d'une forme différentielle de degré
pair, fabriquée à partir du caractère de Chern en présence
d'une structure de Kähler:

$$W(a,b):= \int_{Y^{2m}} a\land b^\ast\land\exp(\omega+F).$$
Montrons qu'il existe une application moment $\mu^\psi$ associée à
l'ajout d'une forme exacte $d\psi$ au potentiel de jauge :
$$\mu^\psi(b):=\int_Y \psi\exp(\omega +F(b)).$$ Il suffit de
d\'eriver $\mu^\psi$, dans une direction notée $h$, pour lire
la relation souhaitée :
\begin{align}
d\mu^\psi(b).h &=\frac{d}{dt}{\Large{|}}_{t=0}\int_Y \psi\exp(\omega + F(b+th))\\
&= \int_Y \psi \,dh\land\exp(\omega+ F(b))\\
&=\iota_{d\psi} W.h.
\end{align}
L'annulation de la limite de faible champ $F$ est la condition de
stabilité dont les équations de MMMS ont donné des
déformations. Les équations de MMMS se retrouvent donc dans
l'approche symplectique des champs de jauge sur les B-branes.

\subsection{Le modèle A}
Nous avons exposé au début de ce mémoire les conditions vérifiées par
la géométrie et la théorie de jauge sur les A-branes. Nous avons
mentionné la possibilité de l'existence de certaines A-branes
coisotropes, suivant l'observation de Kapustin et Orlov
\cite{KapOr}. Sans être lagrangiennes, ces A-branes sont encore
définies par une condition de géométrie symplectique, ce qui induit
une invariance sous l'action des champs de vecteurs hamiltoniens.  Or
cette symétrie implique le tenseur du champ de jauge $F$, via un terme
qui ne peut pas être considéré comme une simple transformation de
jauge $U(1)$. Nous allons identifier ce terme comme une modification
de l'appariement à la Atiyah--Bott.\\

 En effet, pour un champ de vecteurs hamiltonien,
 $$V_h:=\omega^{-1} dh,$$
 dont le flot préserve la géométrie des A-branes,
 et notamment le noyau de $\omega|_Y$, nous pouvons calculer la dérivée de
 Lie de la connexion
$$\mathcal{L}_{V_h}A=\iota_{V_h}dA+d(\iota_{V_h}A).$$
 Le premier terme s'annule avec la courbure sur les A-branes spéciales
lagrangiennes abéliennes\footnote{Nous essaierons dans la section suivante,
  en faisant allusion aux commutateurs de scalaires transverses à la Myers, propres au
  cas non-abélien, de donner une preuve de notre formule miroir qui dépende
  moins de l'hypothèse d'une symétrie de jauge $U(1)$.}. Le second terme
est une forme exacte. Le flot d'un champ de vecteurs hamiltonien n'est donc
rien d'autre qu'une transformation de jauge abélienne. En prenant en
considération la possibilité de A-branes coisotropes (de dimension 5), nous
devons par la même occasion prendre en compte le premier terme, qui lie
l'action des champs de vecteurs hamiltoniens à une courbure de jauge
non nulle.\\

Commençons par rappeler le raisonnement de Thomas \cite{Thomas} à l'origine
de l'application moment pour l'action des champs de vecteurs hamiltoniens
sur les A-branes spéciales lagrangiennes. Soit $L$ une sous-variété
lagrangienne. Pour une densité scalaire $h$ et un champ de vecteurs
$V_u=\omega^{-1}du$, nous pouvons écrire la dérivée de la quantité
$$\int_L h\,\Omega $$
par le flot de $V_u$ noté $\Phi_t$, comme un appariement entre deux
formes de degré un, via une intégration par parties
$$V_u \int_L h\,\Omega =\int_L
h\frac{d}{dt}|_{t=0}(\Phi_t^\ast\Omega)=\int_L hd\left(
  \iota_{V_h}\Omega\right)=\int_L dh\land (\iota_{V_u}\Omega)$$
$$=: \rho (dh,du) $$
Nous lisons donc l'appariement 
$$\rho(a,b)=\int_L a\land (\iota_{\omega^{-1}b}\Omega).$$
Ainsi, après une
fixation de jauge pour la phase de la condition spéciale lagrangienne, nous
lisons l'application moment
$$\mu= \mathrm{Im}\,\Omega|_L.$$
  
Modifions les quantités ci-dessus en les adaptant au cas d'une A-brane
non-lagrangienne supportée par une sous-variété $Y$ de dimension cinq.
L'inclusion du tenseur $F$ dans le flot du champ de vecteurs hamiltonien
ajoute un terme, 
$$
\int_Y \left(d(\iota_{V_u}\Omega)\land F +\Omega\land d(\iota_{V_u}F)
\right)= \int_Y hd\iota_{V_u}\left(\Omega\land F\right),$$
de sorte que cette
somme correspond à la dérivée d'une application moment, associée à un nouvel
appariement entre formes différentielles de degré un, contenant le tenseur
du champ de jauge. Nous avons en effet
$$ d\mu'^{dh}\cdot du=\iota_{dh} \rho'\cdot du,$$
moyennant les définitions
$$
\rho'(a,b):=\int_L a\land \left(\iota_{\omega^{-1}b}(\Omega\land
  F)\right).$$
Il s'agit du terme d'ordre un dans le développement en
puissances de $F$ de $\Omega\land e^F$, où la courbure de jauge $F$ est
considérée comme une distribution, d'où une opération automatique de
pull-back sur la D-brane. Cette quantité doit être l'image miroir de
l'application moment évaluée pour les B-branes. Nous avons eu accès à cette
conjecture par réduction symplectique, méthode alternative à la fois à
l'approche d'espace-temps et à l'approche de surface d'univers.

\section{T-dualité et symétrie-miroir}
Nous arrivons donc à la formule d'échange par symétrie miroir de deux
quantités qui se réduisent aux spineurs purs mentionn\'es plus haut si nous oublions les
champs de jauge :
\begin{equation}\label{miroir}\Omega\land e^F\leftrightarrow e^{i\omega}\land e^F.\end{equation}
 Passons à une preuve directe de cette formule miroir. La T-dualité, identifiée par
Strominger, Yau et Zaslow avec la symétrie-miroir, permet de
transformer les tenseurs reliés par notre proposition.\\

Il est important de considérer les deux quantités \'ecrites ci-dessus comme étant définies
dans tout l'espace-cible, et pas seulement le long des D-branes. La
restriction au volume d'univers des D-branes est automatique si les
caractères de Chern sont consid\'er\'es comme des distributions supportées par
les D-branes. De plus, les formules intrinsèques sur l'espace-cible
sont à rapprocher des couplages de Ramond--Ramond obtenus par Hassan
et Minasian \cite{HM}. L'inclusion des scalaires transverses à
la Myers pourrait être un moyen d'obtenir des résultats pour des
champs de jauge de rang plus élevé.\\

\subsection{T-dualité pour les cordes fermées}

Les conventions sont les suivantes : nous désignerons par $(x^1, x^2, x^3)$
les coordonnées locales sur la base, et par $(y^1, y^2, y^3)$ les
coordonnées sur la fibre ; les différentes quantités étudiées seront
localement des fonctions des coordonnées $x^i$. La métrique et le tenseur
antisymétrique s'adaptent à la fibration via les notations
suivantes, avec des indices latins du milieu de l'alphabet pour les
  directions de base, et des indices grecs du début de l'alphabet pour les
  directions de fibre :
$$G_{\mu\nu}dx^\mu dx^\nu= g_{ij} dx^i dx^j + h_{\alpha\beta } e^\alpha
e^\beta.$$
$$B= \frac{1}{2} B_{ij}dx^i\land dx^j+ B_\alpha\land\left( dy^\alpha
  +\frac{1}{2} \lambda^\alpha\right)+ \frac{1}{2} B_{\alpha\beta }
e^\alpha\land e^\beta,$$
$$B_\alpha=B_{\alpha i} dx^i,$$
$$\lambda^\alpha=\lambda^\alpha_i dx^i.$$
$$ e^\alpha= dy^\alpha+\lambda^\alpha.$$

Dans le cadre des variétés à structure $SU(3)$, nous avons, et c'est là une
caractérisation des variétés à structure $SU(3)$, deux formes, de degrés
deux et trois, notées $\omega$ et $\Omega$, vérifiant
$$\omega\land\Omega=0,  $$
$$
\Omega\land\bar{\Omega}= \frac{(2\omega)^3}{3 !}.$$
Dans le cas des
variétés de Calabi--Yau, elles ne sont autres que que la forme de Kähler et
la trois-forme holomorphe. Elles s'expriment en fonction du vielbein
holomorphe $E^A$, qui contient la donnée d'une structure presque complexe
$V_\alpha^A$ via la relation :
$$E^A= i e_i^A dx^i+ V_\alpha^A e^\alpha.$$
Les formes différentielles en
question sont construites sur le modèle de $dz\land d\bar{z} $ et
$dz^1\land dz^2\land dz^3$ comme
$$\Omega=E^1\land E^2\land E^3,$$
$$
\omega=\frac{i}{2}\delta_{AB} E^A\land E^{\bar{B}}=-V_{i\alpha}
dx^i\land e^\alpha.$$
La T-dualité ne concernant que les directions de 
fibre, les composantes $g_{ij}$ et $B_{ij}$ sont invariantes. Quant aux
composantes avec un ou deux indices de fibre, elles  se transforment selon
$$ h_{\alpha\beta}\longleftrightarrow h^{\alpha\beta},$$
   $$ B_{\alpha\beta}\longleftrightarrow B^{\alpha\beta},$$
$$ B_{\alpha}\longleftrightarrow \lambda^{\alpha}.$$

Le vielbein de la métrique T-duale $ \hat{h}^{\alpha\beta}$ s'obtient à
partir de la structure presque complexe, du bloc $h_{\alpha\beta}$ de la
métrique et du tenseur antisymétrique via
$$\hat{V}^{a\alpha}=\left( \frac{1}{h+B}\right)^{\alpha\beta}V_\beta^a=
V_\beta^a \left( \frac{1}{h-B}\right)^{\beta\alpha}.$$

Les transformations de T-dualité sont alors complétées par 
$$ V_\alpha^a\longleftrightarrow \hat{V}^{a\alpha},$$
$${V}^{a\alpha} \longleftrightarrow \hat{V}_\alpha^a.$$ Nous faisons
l'hypothèse d'un tenseur antisymétrique $B$ ne comportant que des
composantes munies d'un indice de base et d'un indice de
fibre\footnote{les composantes munies de deux indices de base sont
inertes sous la T-dualité et peuvent être incorporées {\emph{in fine}}
sans frais.}, ce qui simplifie beaucoup le calcul du vielbein dual. La
  T-dualité se réduit alors à un échange de $B_\alpha$ et $\lambda^\alpha$,
ainsi qu'à des manipulations d'indices.\\

En l'absence du tenseur antisymétrique, nous avons déjà un échange de
spineurs purs par T-dualité. Il s'exprime à l'aide d'un dual de Hodge
généralisé, défini à l'aide de nos conventions de T-dualité, comme une
simple succession de contractions et de manipulations d'indices. Nous
récrivons d'abord la trois-forme $\Omega$ comme une exponentielle de la
structure complexe $V_i^\alpha e_\alpha dx^i$. Il suffit de développer
$E^1\land E^2 \land E^3$ en substituant l'expression du vielbein holomorphe
comme une somme de deux termes adaptés à la décomposition entre directions
de base et direction de fibre. L'opération $V^\perp \lrcorner(.)$ est
ensuite appliquée afin de préparer l'expression à l'application de la
T-dualité (il s'agit d'une dualité de Hodge sur la fibre qui associe à une
forme de degré $k$ un vecteur à $3-k$ indices).  L'action de la T-dualité,
par simple application des règles de transformation des indices, transforme
l'argument de l'exponentielle en deux-forme $\omega$ :
$$V_i^\alpha e_\alpha dx^i\longleftrightarrow V_{i\alpha} e^\alpha dx^i=\omega $$

$$ V^\perp \lrcorner \Omega\longleftrightarrow \frac{i}{3 !} e^{i\omega}.$$ 

Il nous reste à modifier cette formule par la contribution du tenseur
antisymétrique, ce qui prépare l'inclusion des D-branes et du tenseur du
champ de jauge.
$$\frac{i}{3! } T(e ^{i\omega})=V^\perp\lrcorner (e^B \Omega )
e^{-B_\alpha \lambda^\alpha},$$
$$\Omega\longleftrightarrow \frac{i}{3! }V^\perp\lrcorner (e^B e^{i\omega}
)e^{+B_\alpha \lambda^\alpha}.$$
Ces relations se récrivent en fonction de
la courbure $\mathcal{P}$ sur le fibré de Poincaré :
 
$$
\frac{i}{3! }e^{i\omega+ B}\longleftrightarrow \int_{T^3}e^{\mathcal{P}}
e^B\, \Omega.$$

\subsection{T-dualité pour les cordes ouvertes}
Nous allons inclure les D-branes et les champs de jauge dans cette
transformation de T-dualité. Il est instructif de remarquer que les
spineurs purs et les champs de Ramond--Ramond se transforment comme
des spineurs de l'algèbre de Clifford $Cl(d,d)$. Cette analogie invite
à exploiter les résultats de l'article de Hassan et Minasian \cite{HM}
sur les couplages de Ramond--Ramond et la multiplication de
Clifford. Nous allons considérer le tenseur $F$ du
champ de jauge comme un objet défini dans l'espace ambiant et non sur
une sous-variété.\\
 
En particulier, le tenseur $F$ est considéré comme une distribution,
et la restriction au support du champ de jauge dans toute expression
où apparaît $F$ est sous-entendue. De plus, le champ de jauge et les
scalaires transverses sont codés dans un même vecteur en dimension
six. Ainsi, les composantes pourvues de deux indices de directions
longitudinales de la D-brane sont les composantes habituelles du
tenseur du champ de jauge. Les composantes pourvues de deux indices de
directions tranverses sont des commutateurs de scalaires transverse,
qui s'annulent dans le cas abélien que nous étudions\footnote{Le fait
que ces objets soient naturellement prévus par le langage que nous
utilisons laisse penser que notre formalisme pourrait être adapté à
l'étude de cas non-abéliens.}. Quant aux termes portant un indice
longitudinal et un indice transverse, ce sont des dérivées covariantes
de scalaires transverses.\\
 
Les composantes à indices mixtes (base et fibre) du tenseur $F$ sont
  transparentes pour l'opération $V^\perp\lrcorner(.)$, comme
l'étaient les composantes du tenseur $B$ considéré dans le cas des cordes
fermées. Les composantes à deux indices le long de la base sont insensibles
à la T-dualité. En sous-entendant la présence de ces composantes, que nous
pouvons restaurer facilement, nous renotons les composantes restantes en
oubliant l'indice $i$ de la coordonnée de base :
 $$F:=(F_\alpha, f^\alpha),$$
 où $F_\alpha$ rassemble les composantes du
 tenseur du champ de jauge avec un indice le long de la base et un indice le long
 de la fibre, et où $f^\alpha$ rassemble les dérivées de scalaires
 transverses étendus dans la direction de la fibre.  Avec la contrainte de
 transversalité
$$F_\alpha f^\alpha=0,$$
il n' y a aucune solidarité entre la notion de
transversalité par rapport à la D-brane et celle de séparation entre
directions de base et directions de fibre sur l'espace tangent.\\

Par les règles habituelles d'échange entre directions longitudinales et
directions transverse, nous avons sous l'action de la T-dualité :

 $$F_\alpha\longleftrightarrow f^\alpha.$$
 La combinaison $F_\alpha
 V^\alpha +V_\alpha f^\alpha$ est invariante, ce qui permet d'insérer toutes
 les composantes de $Q:=e^F$ dans la formule d'échange des spineurs purs par
 symétrie miroir en théorie des cordes fermées :

$$\frac{i}{3!}(Q\land e^{i\omega}) e^B \longleftrightarrow \int_{T^3}
e^{\mathcal P} e^B(Q\land \Omega).$$

Une sous-variété généralisée semble s'identifier naturellement à une
combinaison du fibré tangent et du fibré conormal
\cite{HM,Ben-Bassat}. Nous avons donc confirmé la formule issue de la
réduction symplectique sous les hypothèses de \cite{SYZ}, tout en
prolongeant l'étude de la symétrie miroir en présence de D-branes
selon le programme de Vafa \cite{withbundles} au-delà du cas des
A-branes spéciales lagrangiennes envisagé par Leung, Yau et Zaslow
\cite{LYZ}.\\
  
  La formule miroir (\ref{miroir}) que nous venons d'établir a bien
  sûr un caractère plus intrinsèque que celle que nous avait inspiré
  l'échange des applications moments. Seuls deux termes impliquant le
  champ de jauge (et correspondant aux D-branes de dimensions trois et
  cinq) sont permis par le développement du caractère de Chern, pour
  des raisons dimensionnelles. Néanmoins, le résultat exprimé en
  fonction de $e^F$ se prête à une interprétation topologique en
  K-théorie.  En ce qui concerne les contributions gravitationnelles,
  que nous avons ignorées, une description en K-théorie des D-branes
  qui s'échangent par T-dualité devrait naturellement impliquer une
  contribution du genre elliptique de l'espace-cible, selon
$$Q= \sqrt{\hat{A}(X)}\, e^F.$$

 À ce stade de la discussion, nous avons donc donné un traitement
 plus symétrique de la T-dualité en présence de D-branes, dans
 une limite de grand volume où les corrections induites par les
 instantons de cordes ouvertes peuvent être négligées. Prendre en
 compte de telles corrections imposerait de renoncer à l'hypothèse
 d'une fibration à la SYZ, ou tout au moins de la rendre plus
 précise en considérant par exemple des fibres
 singulières. Notre discussion est plus symétrique dans la mesure
 ou elle ne privilégie pas les fibrés holomorphes qui proviennent
 de A-branes lagrangiennes par T-dualité.\\ 

 De plus, le traitement par la géométrie complexe généralisée
 donne l'espoir d'une interpolation entre les modèles A et B, dans
 un cadre géométrique suffisamment ample pour incorporer tous les
 aspects sur lesquels se localisent les twists topologiques. Les A-branes
 coisotropes ont plutôt servi comme outils de cohérence interne,
 via leurs symétries, que d'exemples. Il est crucial que
 l'intervention des champs de jauge dans le modèle A puisse
 mélanger le tenseur du champ de jauge à la trois-forme holomorphe
 dans l'application-moment. Remarquons que l'hypothèse d'une
 variété de Calabi--Yau simplement connexe a été utilisée
 dans l'argument SYZ afin de pouvoir ignorer les modules du fibré
 porté par la D6-brane. Il n'y a donc pas de cycle en homologie pour
 enrouler une A-brane de dimension cinq dans la géométrie que nous
 avons considérée. Néanmoins, un développement plus ample de
 la K-théorie dans le cadre de la géométrie complexe
 généralisée pourrait permettre de réaliser des A-branes
 non-lagrangiennes stables.


\chapter{Conclusions et problèmes ouverts}

\section{Actions effectives de cordes ouvertes}
\subsection{Théories de jauge non-commutatives}
 La théorie de jauge non-commutative sur le volume d'univers des
D-branes a été identifiée comme une limite d'échelle de la
théorie des cordes, dans une situation où le fond de cordes
fermées contient un fort champ antisymétrique $B_{\mu\nu}$.  La
dualité entre les descriptions commutative et non-commutative du
secteur de jauge des cordes ouvertes, proposée par Seiberg et
Witten, peut être explicitée dans le cas abélien, grâce aux
contraintes très fortes imposées par l'invariance de jauge. Le
haut degré de non-linéarité de cette transformation a de
profondes conséquences sur la structure des actions effectives. En
particulier, la non-localité de la théorie non-commutative induit
des séries de corrections dérivatives qui couplent de champ de
cordes fermées $B_{\mu\nu}$ aux d\'erv\'ees du champ de jauge (champ
de cordes ouvertes).\\

 J'ai retrouvé ces corrections dans l'article \cite{corrections} via
un calcul direct d'intégrales de chemin en théorie des cordes,
dans le langage commutatif ordinaire. Ce travail contient le premier
calcul direct à tous les ordres en dérivées du champ de
jauge, de corrections au secteur de jauge de l'action de
Born--Infeld. Ce résultat constitue d'une part une vérification de
la validité de la description non-commutative, et d'autre part une
approche systématique permettant d'examiner les corrections aux
actions effectives {\emph{au-delà}} de la limite d'échelle. J'ai
réalisé ce programme dans \cite{Grangebeyond}, et j'ai
interprété les résultats comme une déformation de la
description non-commutative. Ces corrections nous poussent à poser la
question du statut des couplages de Ramond--Ramond au-delà de la
K-théorie. En effet, c'est le caractère de Chern, sans corrections,
qui est naturel en K-théorie. Ses modifications dans le couplage avec
champs de jauge variables sont équipées d'une structure mathématique
intimement liée à l'invariance de jauge. En effet, que ce soit dans la
limite de Seiberg--Witten, avec les produits $\ast_n$, ou au-delà avec les produits déformés
$\tilde{\ast}_n$, les rangs $n$ sont liés les uns aux autres de
manière à assurer l'invariance de jauge.\\

La versatilité des actions effectives et les redéfinitions
non-linéaires des champs ne nous permettent cependant pas de dire que
nous sommes allés au-delà de la K-théorie. Nous nous sommes plutôt
approchés de la notion d'universalité dans les potentiels de
tachyons. Disons que les redéfinitions des champs, telles que
l'application de Seiberg--Witten, fournissent des moyens d'aborder des
champs variables de manière adaptée à des configurations dans
lesquelles la courbure de jauge est localisée.\\

  Des développements plus anciens sur les orbifolds asymétriques, dont
  nous avons donné un aperçu au moment du calcul du propagateur, ont
  associé le champ de fond antisymétrique à des rotations
  asymétriques. Les travaux récents de Pantev et Sharpe sur les
  symétries inefficaces jaugées et la symétrie miroir
  \cite{Pantev--Sharpe1,Pantev--Sharpe2}, pourraient donner lieu à des
  prolongements de cette approche, grâce notamment à l'asymétrie de la
  T-dualité et à son caractère inefficace sur le secteur gauche.\\

Certains développements \cite{Bars,butterfly} ont fait usage du
produit de Moyal en dimension infinie pour écrire la théorie des
champs de cordes comme une théorie des champs non-commutatifs. Des
anomalies d'associativité, discutées dans les années 80 par
Horowitz et Strominger \cite{Horowitz--Strominger}, et liées à la
brisure de l'invariance par translation lors du recollement de deux
cordes ouvertes en leur milieu, ont refait une apparition à cette
occasion. Je pense qu'elles pourraient faire l'objet d'une étude
plus systématique à l'aide d'équations de descente
\cite{Stora,ZuminoHouches}.

\subsection{Potentiel effectif pour les tachyons} 

 Certaines intégrales de chemin pour les cordes ouvertes, dont
 l'action classique contient un terme couplant un bord au champ de
 tachyon, codent la dynamique
 du tachyon sur un fond donné de cordes fermées. Dans le cas d'un
 champ $B_{\mu\nu}$ infini ou nul, le calcul a été poussé
 jusqu'à la vérification de la conjecture de Sen sur la
 condensation des tachyons. La limite du champ infini est en rapport
 avec la non-commutativité évoquée plus haut.\\

  Les techniques que j'ai développées dans ce contexte ont permis
 d'interpoler entre les deux régimes précédemment étudiés,
 via un dictionnaire formel \cite{grangepotential}. Les déformations
 rencontrées dans le secteur de jauge se révèlent pertinentes
 pour les tachyons. Une interprétation non-commutative a également
 été conjecturée récemment pour le modèle exactement soluble
 des cordes $p$-adiques. J'ai établi cette conjecture en montrant
 dans \cite{padicGrange} comment coupler un champ magnétique à un
 bord $p$-adique tout en conservant la validité de l'approche
 non-commutative dans l'\'etude des solitons scalaires.\\

  Les résultats décrits dans cette thèse, notamment lors du détour
  par les cordes $p$-adiques et du passage aux cordes
  topologiques, confirment la robustesse de la non-commutativité.
  Pour aller plus loin, je mentionnerai les deux pistes de la
  non-commutativité dépendant du temps, étudiée par Dolan et Nappi
  \cite{timeNappi}, et du relèvement en supergravité des théories des
  champs non-commutatifs amorcé par Robbins et Sethi
  \cite{liftSethi}.\\

\section{D-branes topologiques et symétrie miroir}
 
 Le problème de la stabilité des membranes de Dirichlet (D-branes)
dans les théories de cordes topologiques a été abordé par des
techniques d'espace-temps, fondées sur des actions effectives
supersymétriques. La condition de stabilité, obtenue en 1999 par
Minasian, Mariño, Moore et Strominger, est une déformation des
équations de Yang--Mills {\hbox{hermitiennes}}. En présence du
champ de fond antisymétrique $B_{\mu\nu}$, ces équations ont fait
l'objet d'une conjecture en 2003 : elles doivent posséder une
version non-commutative. L'approche par des techniques de surface
d'univers (conditions de bord) trouve un cadre géométrique naturel
dans la géométrie complexe généralisée proposée
récemment par Hitchin.\\

 Les conditions de recollement des différents courants conservés
 (associés aux différentes symétries du probème) induisent des
 contraintes sur la courbure du champ de jauge le long des D-branes
 stables. J'ai établi dans l'article \cite{NCholomorphic} le fait
 que ces contraintes donnent lieu à une condition
 d'holomorphicité sur le fibré associé à une théorie de
 jauge non-commutative. Ceci établit la conjecture citée plus haut
 dans le modèle B, sensible à la géométrie complexe.\\

 Concernant le modèle A, sensible à la géométrie symplectique,
 j'ai établi en collaboration avec Minasian un échange par
 symétrie miroir \cite{GM} entre la trois-forme holomorphe du
 modèle B et la structure symplectique du modèle A, en présence
 des champs de jauge, c'est-à-dire de structures fibrées codées
 par leur caractère de Chern, considéré comme une
 distribution. Les {\hbox{D-branes}} non-lagrangiennes du modèle A,
 avec leur caractère de Chern non-trivial, sont ainsi prises en
 compte par la symétrie miroir entre les deux modèles.  Les
 D-branes coisotropes du modèle A font l'objet de développements
 récents \cite{AldiZaslow,NCKKapustin} suggérant l'intervention de la
 non-commutativité même en l'absence d'holomorphicité. La question de
 l'obtention de ces configurations à partir de nos B-branes stables
 non-commutatives se pose.\\

La multiplication des matrices gamma est apparue dans les couplages de
Ramond--Ramond \cite{HM}, ce qui crée une analogie avec le traitement
des sommes de formes différentielles comme un module de Clifford, et
donne lieu à des modifications des spineurs purs par le secteur de
cordes ouvertes. Cette notion de spineur pur est issue de la géométrie
complexe généralisée, qui trouve une réalisation naturelle en théorie
des cordes et des D-branes. Les développements r\'ecents utilisant les spineurs
purs, par Grassi et Vanhove \cite{Grassi--Vanhove}, vont dans la
direction de la M-théorie topologique à partir de constructions impliquant la limite de la super-particule. D'autre part, la
quantification du problème variationnel de Hitchin a été
récemment abordée par Pestun et Witten\cite{Pestun--Witten}. Les
spineurs purs, que nous avons modifiés au niveau classique,
devraient y jouer un rôle, en for\c{c}ant l'inclusion de champs de jauge dans la fonctionnelle de Hitchin.\\


\appendix

\bibliography{references}

\newpage

\tableofcontents

\end{document}